\documentclass[structabstract]{aa}

\usepackage{graphicx}

\usepackage{txfonts}
\usepackage[utf8]{inputenc}
\usepackage[table]{xcolor}
\usepackage{array, slashbox}
\usepackage{float}
\usepackage{subfigure}
\usepackage{multirow}
\usepackage[]{natbib}
\begin{document}
   \title{Atmospheric characterization of cold exoplanets \\
   using a 1.5-m coronagraphic space telescope}

   \author{A.-L. Maire\inst{1,4} \and R. Galicher\inst{2,3} \and A. Boccaletti\inst{1,4} \and P. Baudoz\inst{1,4} \and J. Schneider\inst{5} \and K. L. Cahoy\inst{6} \and D. M. Stam\inst{7} \and W. A. Traub\inst{8}}
   
   \institute{LESIA, Observatoire de Paris, CNRS, University Pierre et Marie Curie Paris 6 and University Denis Diderot Paris 7, 5 place Jules Janssen, 92195 Meudon, France \\
              \email{anne-lise.maire@obspm.fr}
         \and
             National Research Council Canada, Herzberg Institute of Astrophysics, 5071 West Saanich Road, Victoria, BC V9E 2E7, Canada \\
             \email{raphael.galicher@nrc-cnrc.gc.ca}
          \and   
             D\'ept. de Physique, Universit\'e de Montr\'eal, C.P. 6128, Succ. Centre-Ville, Montr\'eal, QC H3C 3J7, Canada 
         \and
             Groupement d'Int\'er\^et Scientifique Partenariat Haute R\'esolution Angulaire Sol Espace (PHASE) between ONERA, Observatoire de Paris, IPAG, LAM, CNRS and University Denis Diderot Paris 7, France
          \and
             LUTh, Observatoire de Paris, CNRS and University Denis Diderot Paris 7, 5 place Jules Janssen, 92195 Meudon, France
          \and
            Dept. of Aeronautics and Astronautics, MIT, 77 Mass. Ave. 37-367, Cambridge, MA 02139, USA
          \and
             SRON Netherlands Institute for Space Research, Sorbonnelaan 2, 3584 CA Utrecht, The Netherlands
          \and
             JPL, California Institute of Technology, M/S 301-355, 4800 Oak Grove Drive, Pasadena, CA 91109, USA         
             }

   \date{Received 3 February 2012 / Accepted 12 March 2012}

 
  \abstract
   {High-contrast imaging is currently the only available technique for the study of the thermodynamical and compositional properties of exoplanets in long-period orbits, comparable to the range from Venus to Jupiter.
   The SPICES (Spectro-Polarimetric Imaging and Characterization of Exoplanetary Systems) project is a coronagraphic space telescope dedicated to the spectro-polarimetric analysis of gaseous and icy giant planets as well as super-Earths at visible wavelengths.
   So far, studies for high-contrast imaging instruments have mainly focused on technical feasibility because of the challenging planet/star flux ratio of 10$^{-8}$--10$^{-10}$ required at short separations (200~mas or so) to image cold exoplanets.
   However, the main interest of such instruments, namely the analysis of planet atmospheric/surface properties, has remained largely unexplored.}
   {The aim of this paper is to determine which planetary properties SPICES or an equivalent direct imaging mission can measure{, considering} realistic reflected planet spectra and instrument limitation.}
   {We use numerical simulations of the SPICES instrument concept and theoretical planet spectra to carry out this performance study. We also define a criterion on the signal-to-noise ratio of the measured spectrum to determine under which conditions SPICES can retrieve planetary physical properties.}
   {We find that the characterization of the main planetary properties (identification of molecules, effect of metallicity, presence of clouds and type of surfaces) would require a median signal-to-noise ratio of at least 30.
   In the case of a solar-type star $\leq$10~pc, SPICES will be able to study Jupiters and Neptunes up to $\sim$5 and $\sim$2~AU respectively{, because} of the drastic flux decrease with separation. It would also analyze cloud and surface coverage of super-Earths of radius 2.5 Earth radii at 1~AU.
Finally, we determine the potential targets in terms of planet separation, radius and distance for several stellar types. 
For a Sun analog, we show that SPICES could characterize Jupiters (M\,$\geq$\,30 Earth masses) as small as 0.5 Jupiter radii at $\lesssim$2~AU up to 10~pc, and super-Earths at 1--2~AU {for the handful of stars that exist} within 4--5~pc. Potentially, SPICES could perform {analysis of a hypothetical Earth-size planet around $\alpha$~Cen A and B.} 
However, these results depend on the planetary spectra we use{, which} are derived for a few planet parameters assuming a solar-type host star. Grids of model spectra are needed for a further performance analysis. Our results obtained for SPICES are also applicable to other small (1--2~m) coronagraphic space telescopes.}
   {}


\authorrunning{A.-L. Maire et al.}
\titlerunning{Exoplanet characterization using space high-contrast imaging}

   \maketitle
%

\section{Introduction}
\label{intro}
    	The exoplanet field in astrophysics is extremely rich and diverse. From detection to characterization, many techniques are being used or developed to address the fundamental questions about planetary formation and evolution. Exoplanets span a number of categories much larger than the Solar System's planets do.
	Since the first discovery, several unexpected types of exoplanets were found from the hot Jupiters \citep{Mayor1995} which are very close to their host stars ($\leq$0.05~AU) to the population of super-Earths (massive telluric planets) which starts to emerge from radial velocity surveys \citep[][hereafter RV]{Mayor2011} and transit surveys \citep[e.g.,][]{L'eger2009, Charbonneau2009, Batalha2011, Borucki2012}.
	There are so many planet categories already detected even with the detection biases of the current methods (RV, transits, imaging, microlensing) that several instruments/missions will be needed to cover the whole field. Methods such as RV and transits appear to be effective at probing for large close-in exoplanets{, and} current efforts are to expand their sensitivity to longer-period and smaller exoplanets \citep{Udry2007, Seager2010}.
    	The detection and characterization of long-period/wide-orbit ($\gtrsim$1~AU) planets are, however, still difficult to accomplish. On the one hand, it requires long-duration monitoring with very stable instruments to detect them from a RV or transit survey. On the other hand, direct imaging has to tackle the high contrast at small angular separation that exist between the planet and its host star.
	
	An extrapolation of the period distribution of giant planets discovered by RV surveys suggests that a large population of these objects at separations larger than 5~AU still remains to be revealed \citep{Marcy2005}.
    	These planets, in particular the population between 5 and 20~AU, are very important for constraining theoretical mechanisms of planetary formation{, since} it may reside at the boundary between the core-accretion and disk instability regimes \citep{Alibert2011, Boss2011}. When it comes to the atmospheric characterization of these planets, direct imaging will probably be the most productive technique.
	Since 2005 \citep{Chauvin2005, Neuhauser2005}, several massive giant planet candidates were imaged around young stars ($<$200 Myrs), the most emblematic being {the four planets around HR\,8799 \citep{Marois2008, Marois2010} and $\beta$~Pictoris\,b \citep{Lagrange2009, Lagrange2010}. We note that the planetary nature of the Fomalhaut companion \citep{Kalas2008} has been recently questioned \citep{Janson2012}.}
	Following these discoveries, spectra were obtained for a few planets \citep[e.g.,][]{Mohanty2007, Janson2010, Patience2010, Bowler2010}.
	 A first generation of instruments precisely optimized for the detection and spectral characterization at near- and mid-infrared (IR) wavelengths of young giant planets will see first light in the present decade: SPHERE \citep{Beuzit2008}, GPI \citep{Macintosh2008}, HiCIAO \citep{Hodapp2008}, P1640 Phase II \citep{Hinkley2011} and FLAO \citep{Esposito2010} on ground-based telescopes, and JWST \citep{Clampin2010} and SPICA Coronagraph Instrument \citep{Enya2011} in space. In the next decade, planet finders on Extremely Large Telescopes{, such} as EPICS \citep{Kasper2010} and PFI \citep{Macintosh2006a}{, may} offer the ability to observe mature gas giants, ice giants and possibly super-Earths in the near-IR.
	Detailed studies were performed to consider the feasibility of large aperture coronagraphs and large baseline interferometers for the detection of Earth twins from space. These studies identified areas of technological development that need to be first addressed, which will delay the realization of Terrestrial Planet Finder missions until $\sim$2025--2030.
	Meanwhile, observations have demonstrated the extreme diversity of planets{. This led to a growing consensus within the community that we will need to study all planet types in order to have a complete understanding of their} formation and evolution \citep{Schneider2008}.
	To address some parts of these questions{, a} family of small space missions (mainly coronagraphs) have been proposed \citep{Guyon2010b, Trauger2010} for analyses of ice giants and super-Earths.
	The study of these missions has mainly focused on technical feasibility, as it is a challenge to achieve large contrast close to a bright star \citep{Trauger2007, Guyon2010a, Belikov2010}.
	Another area of study is the estimated number of observable exoplanets of a given type \citep{Trauger2010, Guyon2010b}{, based} on assumptions of their density distribution.
	\citet{Cahoy2009} consider signal-to-noise ratios (SNRs) of broad-band (R\,=\,5) photometric measurements assuming planets with grey albedos. The main interest of such missions, the atmospheric characterization, however, has remained highly unexplored so far. \citet{Cahoy2010} analyze colors and coarse spectra (R\,=\,5 and R\,=\,15) of Jupiter and Neptune atmosphere models{, but} without including instrument limitations such as throughput and noise. From the colors of Solar System planets, \citet{Traub2003} suggested that for planets too faint for spectrometry, even coarse colors could help to distinguish between planet types. \citet{Cahoy2010} showed that while a color criterion could not be a means to uniquely distinguish between planet types{, due} to the intertwined contributions of factors such as metallicity and planet-star separation, colors would still provide some constraints on possible planet types.

	Among all of these space coronagraph concept missions, the most recent, SPICES (Spectro-Polarimetric Imaging and Characterization of Exoplanetary Systems) was submitted to the ESA Cosmic Vision call for medium-class missions in 2010 by a consortium of European institutes with American and Japanese participations\footnote{Obs. Paris (LESIA, LUTh, LERMA, GEPI), CEA/SAp, IPAG, LAM, SRON, Univ. Utretch, Obs. Padova, Univ. Exeter, Univ. Cambridge, NASA (JPL, Ames, GSFC), MIT, Univ. Arizona, INTA-CSIC CAB, Obs. Torino, Obs. Geneva, ONERA, UC Berkeley, STScI, CalTech, IFSI Roma, NAOJ, Univ. Hokkaido, Univ. Li\`ege, MPIA, Univ. Kiel and Obs. Vienna with support from Astrium and CNES/PASO.} \citep{Boccaletti2012a}. SPICES has a twofold motivation: 1/ the systematic atmospheric characterization of gas and ice giants as well as super-Earths in the solar neighborhood{, and} 2/ the development and validation of key technologies in order to prepare future direct imaging projects dealing with Earth twins spectral characterization. The science objectives and the technical concept of SPICES are described in \citet{Boccaletti2012a}. The main science driver is the study of planetary systems as a whole for the understanding of planet formation and evolution. {With a maximum imaged field of view of $\sim$13$\arcsec$}, SPICES will focus on targets previously identified using other methods (planets and circumstellar disks){, but} can also detect new planets such as outer planets in known planetary systems and exo-zodiacal disks $<$100~zodis. {A preliminary estimation of the number of characterizable planets gives an order of magnitude of 100 objects for an allocated time of three years over the five years of the mission \citep{Boccaletti2012a}.}
	The main purpose of the instrument is to obtain flux and polarization spectra at visible wavelengths of cold/mature exoplanets, especially those previously discovered by RV surveys \citep{Udry2007} or astrometry with GAIA \citep{Casertano2008}. These surveys will provide the orbital elements and minimum mass of planets, but not the radius which determines the amount of reflected light together with the albedo. It is therefore essential to perform accurate measurements of spectra to possibly distinguish between planet types. 

	In this work, we have developed a numerical simulation to model the instrument concept of SPICES. Under realistic assumptions of noise, instrument performance and reflected exoplanetary spectra, we test the ability to distinguish between spectra of planets differing in surface gravity, atmospheric composition, metallicity, cloud coverage and surface type. We do not consider polarized light in this paper and leave such a study for future work.
	Our primary goal is not to refine the instrument concept given in \citet{Boccaletti2012a}. Firstly, we investigate the exoplanet detection space realistically, and then we may use our results to update the design. Our work will also be beneficial to other space coronagraph proposals.
	In Sect.~\ref{methods}, we describe the numerical model and the assumptions we use for our study. In Sect.~\ref{perf}, we analyze the effects of speckle noise, read-out noise, exozodi and photon noise on the performance. We specify what kinds of planets can be detected in each case. We then study the impact of the spectral type of the host star. In Sect.~\ref{spectrometry}, we define a criterion that gives the required flux accuracy that the instrument has to produce in order to disentangle spectra for similar planet types (Jupiters, Neptunes or super-Earths){, but} with different values of physical parameters (composition, metallicity or cloud and surface coverage). We conclude that our instrument fulfills these constraints for most of the considered cases. In Sect.~\ref{overallperf}, we generalize our results to other stellar types to define a parameter space of potential targets.

\section{Models}
\label{methods}

 \subsection{Numerical model of the instrument}
 \label{nummodel}
\begin{figure}[t]
	\centering
	\includegraphics[width=5.7 cm]{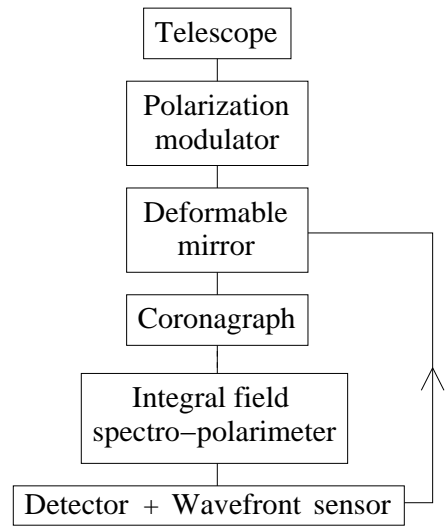}
	\caption{SPICES conceptual baseline.}
	\label{SPICESscheme}       
\end{figure}

 	The instrument concept of SPICES is designed to provide polarimetric and spectroscopic measurements in the 0.45--0.90~$\mu$m range (Fig.~\ref{SPICESscheme} and \citet{Boccaletti2012a}). To achieve high contrasts required to exoplanet characterization, SPICES combines a high-quality off-axis telescope, high-accuracy wavefront control, a broad-band coronagraph with small inner working angle {(IWA; the angular separation of 50\% throughput),} and optical elements and detectors to collect the polarimetric and spectroscopic information. As for the telescope, very low wavefront aberrations have already been demonstrated for GAIA primary mirrors ($\sim$\,8~nm rms on surface\footnote{http://sci.esa.int/science-e/www/object/index.cfm?fobjectid=47688}){, and} SPICES can benefit from the same technology. The wavefront control is achieved with the combination of a focal plane wavefront sensor and a deformable mirror (DM). The wavefront sensing is achieved with a self-coherent camera \citep[SCC,][]{Galicher2008}{, which} is a very simple modification in the coronagraph design.
	The SCC spatially modulates speckles (residual stellar light) that are recorded in the science image (no additional channel) and can accurately retrieve the wavefront errors (phase and amplitude) that induce these speckles. It then drives a DM to correct for them. As the correction is never perfect, the speckle noise is strongly attenuated{, but} residual speckles still remain in the image. The SCC provides a means to calibrate them and extract the companion or disk information{, without} prior information on the spectrum of the latter \citep{Baudoz2006, Galicher2010}. The DM is a 64x64 actuator mirror, as a larger number of actuators increases the field of view which can be corrected.
	A vortex coronagraph {in the focal plane applies} an azimuthal phase ramp to the corrected incident wavefront to cancel the starlight \citep{Mawet2005}. The vortex coronagraph can be made achromatic over a wide spectral bandwidth \citep[$\sim$50\%,][]{Mawet2010a}. This type of coronagraph was successfully used by \citet{Serabyn2010} to re-image the HR$\,$8799 multi-planet system.
	Finally, an integral field spectrograph (IFS) based on a micro-lens array \citep{Antichi2009} allows recording of a dispersed image of the corrected field of view. Once a (x,\,y,\,$\lambda$) data cube is reconstructed from the detector image, the SCC speckle calibration is applied to every spectral channel separately{, therefore} drastically reducing the chromaticity of this device \citep{Galicher2010}.
	In the current design, the beam is divided into two branches{, each} assigned to a state of polarization and half of the spectral band. An achromatic modulator{, at} the very beginning of the instrument{, selects} the linear polarization direction on the sky that is analyzed by the polarizer in each branch.

 \begin{figure*}[ht]         
	\centering
	\includegraphics[trim = 4mm 4mm 4mm 4mm, clip,width=0.26\textwidth]{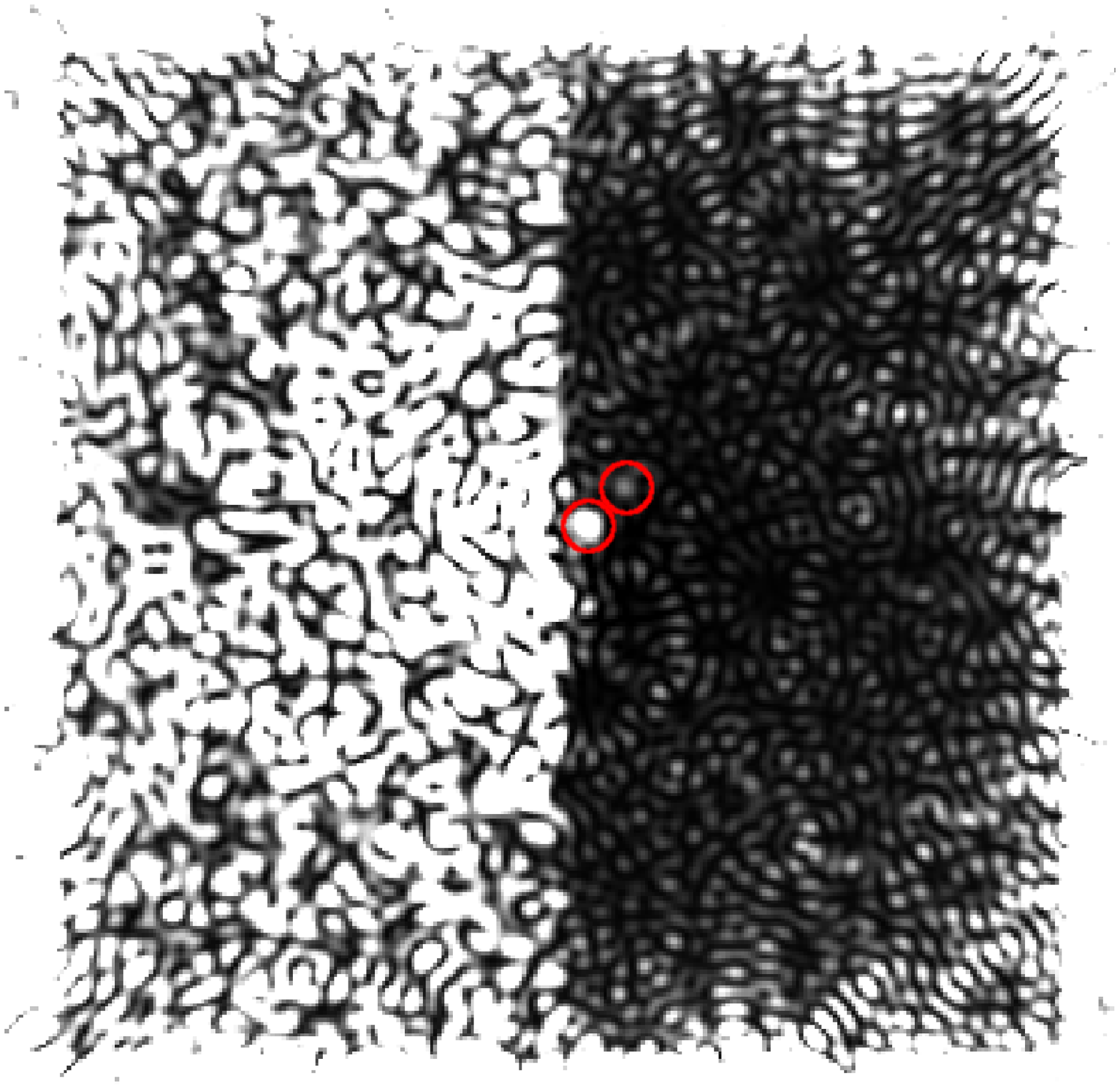}
	\includegraphics[trim = 4mm 4mm 4mm 4mm, clip,width=0.26\textwidth]{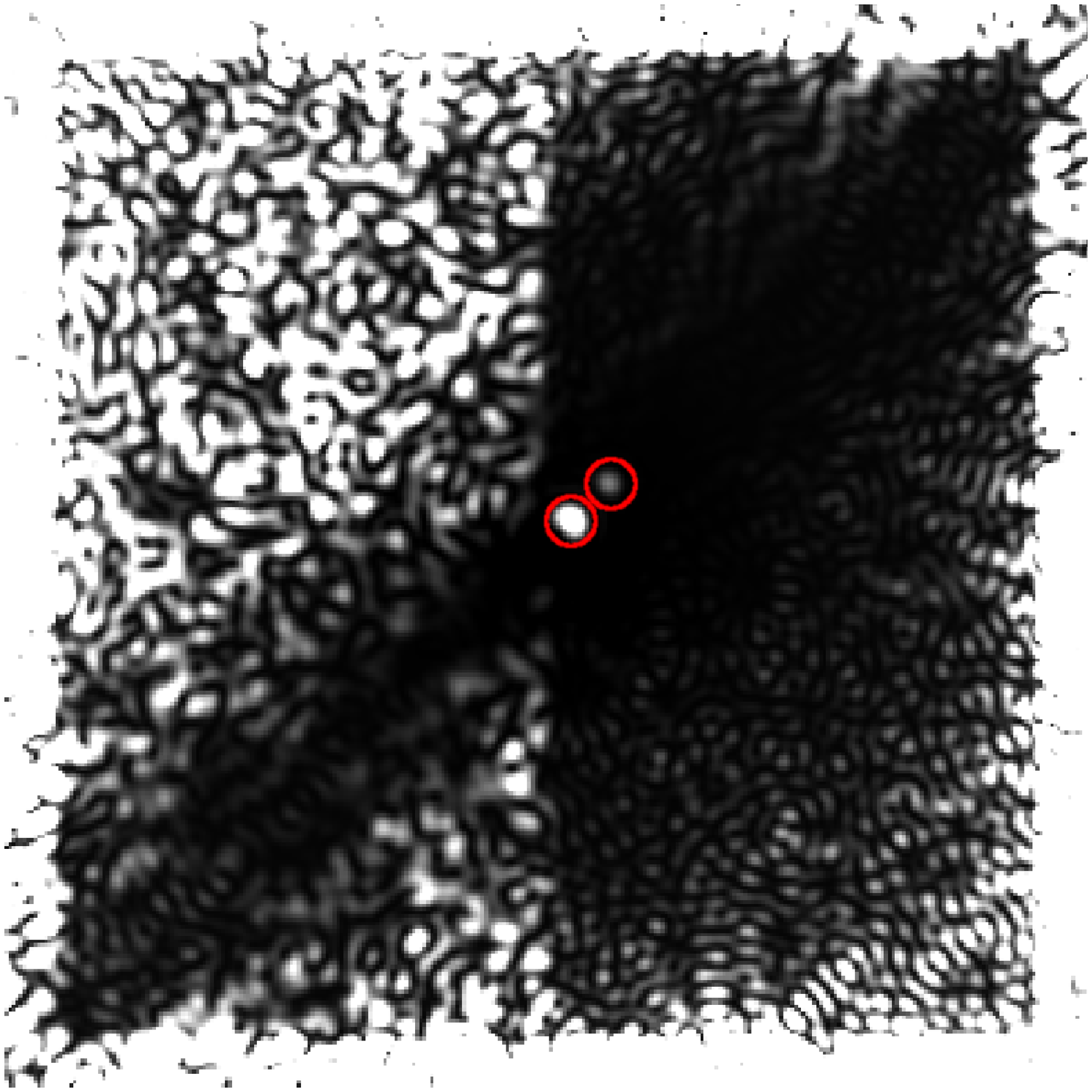}
	\includegraphics[trim = 4mm 4mm 4mm 4mm, clip,width=0.26\textwidth]{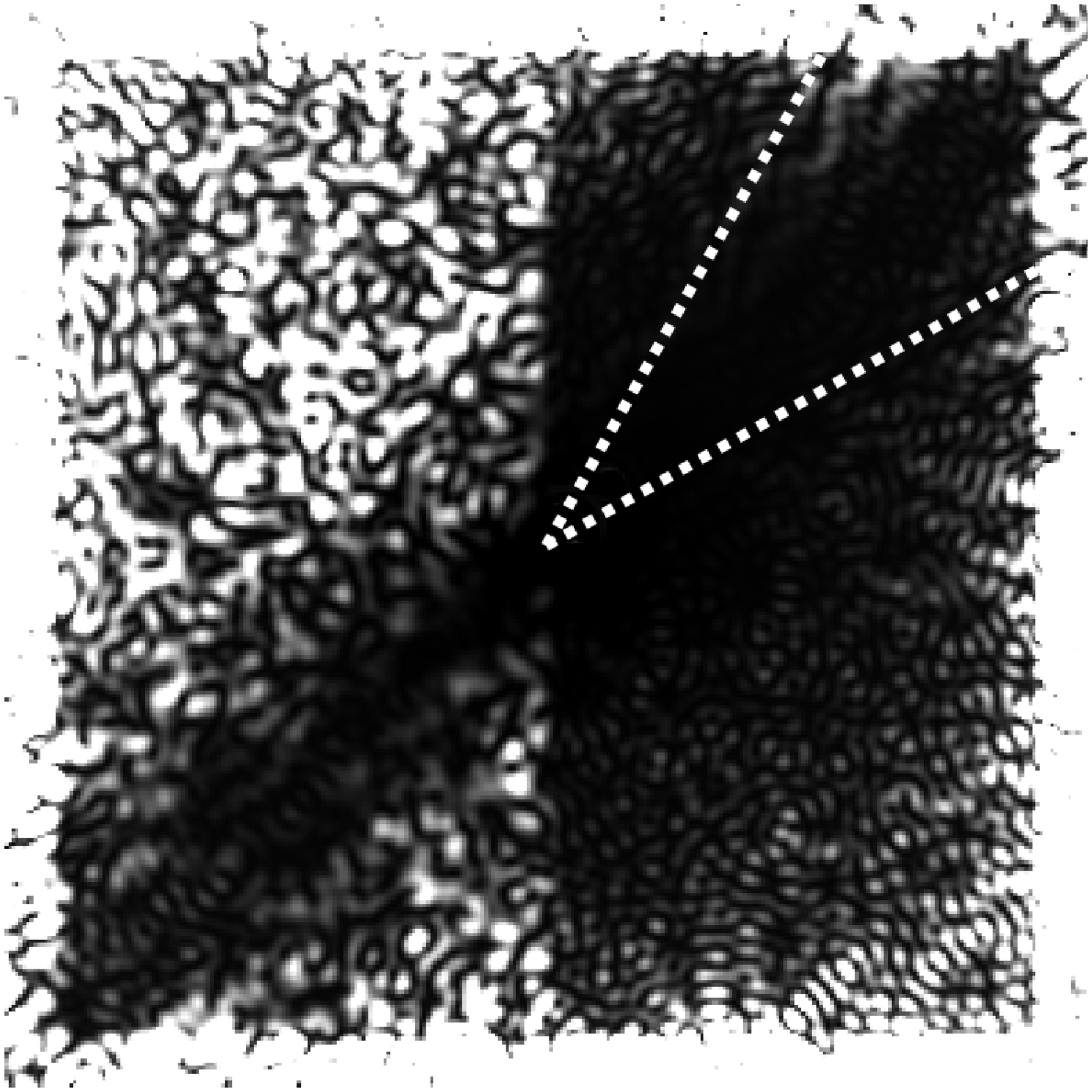}
	\includegraphics[trim = 15mm 18mm 14mm 16mm, clip,width=0.144\textwidth]{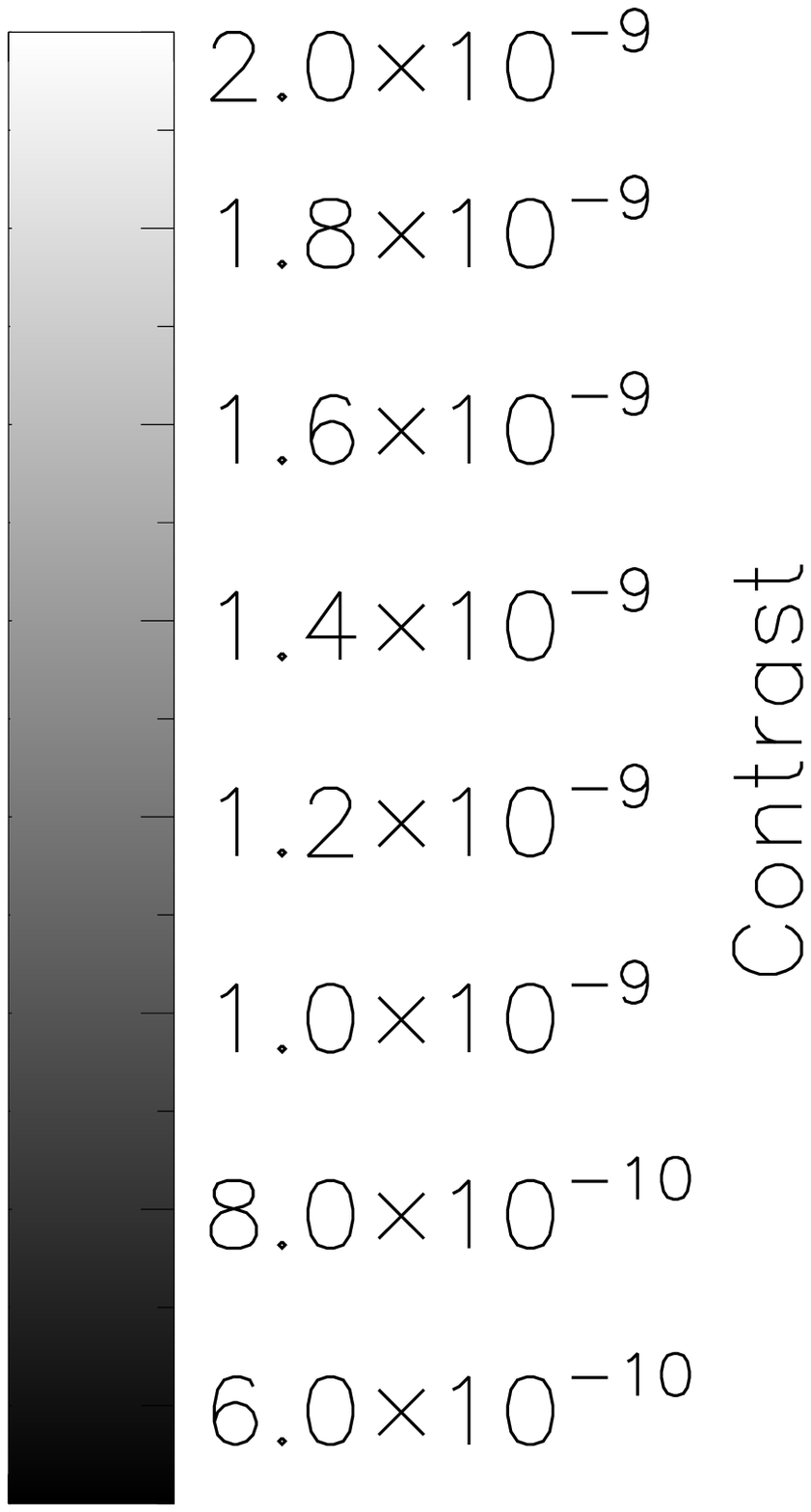}
	\caption{Central part of images without detection noise produced by the simulation{, after} the coronagraph (left) and after the speckle calibration by the self-coherent camera (middle and right). The image size is $\sim$64\,$\times$\,64\,$\rm{(\lambda/D)^2}$ ($\lambda$\,=\,0.675~$\mu$m). In the two left images, there are two jovian planets of contrasts $\sim$10$^{-8}$ and $\sim$10$^{-9}$ (red circles). In the right image, we indicate the calculation area of the profiles shown in Sect.~\ref{perf} with a white dotted line. The intensity scales are linear and identical.}
	\label{images}
\end{figure*}
 
 	We built a numerical code{, written} in the Interactive Data Language (IDL){, to} model the instrument concept and to simulate the SPICES performance. Our code operates in three steps:
 \begin{itemize}
  	\item Step 1: The simulation of non-coronagraphic and coronagraphic image cubes of on-axis (the star) and off-axis (the planet(s)) sources. 
  		The third dimension of the cubes represents the spectral channels. 
  		We assume that the image cube extraction from the IFS data is perfect{, but} correctly account for flat field impact at step 2 (Sect.~\ref{simassumptions}).
   	\item Step 2: The normalization of the star and planet spectra{, and} introduction of noise (photon noise, zodi, exozodi, read-out noise, flat field).
  	\item Step 3: The SCC speckle calibration of the images recorded at step 2 and the measurement of the planet spectra. 
  \end{itemize}
  
  	In Fig.~\ref{images}, we show an example of simulated images {without detection noise} for a single spectral channel ($\lambda$\,=\,0.675~$\mu$m, $\Delta\lambda$\,$\simeq$\,0.013~$\mu$m){, after} the coronagraph (Fig.~\ref{images}, left) and after the speckle calibration (Fig.~\ref{images}, middle and right){. The image field is} limited to the zone corrected by the DM (64\,$\times$\,64\,$(\lambda/D)^2$ which corresponds to $\sim$6\,$\times$\,6\,arcsec$^2$ at $\lambda$\,=\,0.675~$\mu$m). The corrected zone size is set by the linear number of DM actuators \citep{Galicher2010}. As we introduce both amplitude and phase aberrations in the entrance pupil of the instrument and SPICES uses a sole DM, the speckles are corrected only in one half of the field of view (right half in the images).
  	After speckle calibration, the contrast is enhanced as shown in the middle and right images. With the current SCC, the calibration is more efficient in a diagonal because of the chromatism limitation \citep{Galicher2010}. A new version of the technique is under study to calibrate speckles in all directions. In the left and middle images of Fig.~\ref{images}, we added two jovian planets of contrasts $\sim$10$^{-8}$ and $\sim$10$^{-9}$ at 2.2 and 5.4~$\lambda/D$ (red circles). The closest planet is detected in the coronagraphic image (left) with a SNR\,$\simeq$\,25. 
The farthest planet can be seen in this image if its position is known. However, it is not possible to claim a detection in this image (SNR\,$\simeq$\,5){, whereas} it is detected with a SNR\,$\simeq$\,600 after calibration (middle).
 
\subsection{Instrument assumptions}
 \label{simassumptions}
  \begin{table}[t]
 \caption{Assumptions used for the instrument simulations.}
 \begin{center}
 \begin{tabular}{p{4.7cm}l}
 \hline\hline
 Parameter & Value \\
 \hline
 Telescope diameter & 1.5~m \\
 Spectral bandwidth & 0.45--0.90~$\rm{\mu}$m \\
 Spectral resolution ${R}$ ($\lambda$\,=\,0.675~$\mu$m) & 50 \\
 Number of spectral channels & 35 \\
$\Delta\lambda$ per channel & $\sim$0.013~$\mu$m \\
 $\lambda$/D sampling at 0.45~$\mu$m & 3 pixels \\
 Wavefront errors ($\lambda$\,=\,0.675~$\mu$m) & 15~nm rms (f$^{\rm{-3}}$ PSD\tablefootmark{a}) \\ 
 Amplitude aberrations & 0.1\% (f$^{\rm{-1}}$ PSD\tablefootmark{a}) \\ 
 Number of actuators on the DM & 64\,$\times$\,64\\
 Phase estimation by the SCC & perfect \\
 Throughput & 23\% \\
 Quantum efficiency & 70\% \\
 Maximum integration time & 200~h \\
 Read-out noise & 0.2~e- rms per pixel\\
 Full well capacity & 300\,000~e- \\
 Time per single exposure & 1\,000~s \\
 Flat field accuracy & 0.5\% \\
 Zodiacal light & V\,=\,23.1~mag arcsec$^{\rm{-2}}$\\
 Exozodi level & 1~zodi \\
 \hline
 \end{tabular}
 \label{hypotheses}
 \end{center}
 \tablefoot{The parameter values are discussed in Sects.~\ref{nummodel}, \ref{simassumptions} and \ref{noise}. \\
 \tablefoottext{a}{f is the spatial frequency of the optical defect, PSD refers to power spectral density.}}
 \end{table}

	In terms of contrast, SPICES has to reach values as low as 10$^{-9}$ at 2~$\lambda/D$ and 10$^{-10}$ at 4~$\lambda/D$ \citep{Boccaletti2012a} to produce interesting science results. Such performance is achieved assuming the parameters and the requirements on noise that are listed in Table~\ref{hypotheses}. We consider these values in the numerical simulation and discuss some of them in this section.
	{The star is assumed to be perfectly centered onto the coronagraph focal plane mask.} Since the pointing accuracy is a critical aspect, SPICES will include a dedicated procedure for a precise control of the coronagraph alignment at the level of $\sim$0.2~mas \citep{Boccaletti2012a}. {Our simulations showed} that this value keeps the speckle noise at a level of $\sim$10$^{-10}$ between 2 and 4~$\lambda/D${, which} is below the photon noise (Sect.~\ref{noisyperf}).
  	In the optical design of the instrument, the whole spectral band is split into two branches, the first branch covering the 0.45--0.7~$\mu$m band ($\Delta \lambda / \lambda$\,$\simeq$\,43\%) and the second branch the 0.65--0.9~$\mu$m band ($\Delta \lambda / \lambda$\,$\simeq$\,32\,\%). The overlapping is for calibration purposes.
 	We assume that all phase and amplitude aberrations are located in planes conjugated to the instrument pupil and we use the matrix direct Fourier transform \citep{Soummer2007} to propagate the light. {Fresnel propagation effect will be included using} the PROPER library \citep{Krist2007}. The main impact will be a partial modification of the speckle pattern with wavelength \citep{Marois2006a}. This will not impact the SCC wavefront estimation since each spectral channel is treated separately{, but} it will reduce the efficiency of the wavefront correction with a sole DM \citep{Shaklan2006}.
 	{A slightly undersized Lyot stop is used} for the coronagraph (95\% of the pupil diameter).
 	We consider that the vortex coronagraph performance is achromatic for SPICES' spectral bands. Current vortex coronagraphs are limited to contrasts of $\sim$4\,$\times$\,$10^{-8}$ for a 20\% bandwidth{, but} strong efforts have been made in the past few years to develop them in laboratory, test them on the sky and further improve their performance \citep{Serabyn2011}.
	{15 monochromatic images are co-added} to simulate each spectral channel image.
	The SCC requires an oversampling with respect to Shannon's criterion. {The pixel number per spatial resolution element ($\lambda/D$) is 3} at the minimum wavelength{. The} SCC fringes are tilted at 45$^{\circ}$ with respect to the pixel grid.
	We {use} a f$^{\rm{-3}}$ power law (f is the spatial frequency of the optical defect) for the power spectral density (PSD) of the phase aberrations{, since} it quite well reproduces the aberrations measured on the VLT and HST mirrors \citep{Borde2006}. Simulations indicate that the amplitude aberrations will have to be $\lesssim$0.1\% {in order to meet SPICES' requirements if they follow a f$^{\rm{-1}}$ PSD,} but this value will be more stringent if the law exponent is $>$1 \citep[chap.\,IV.2, Fig.~IV.2.2]{Galicher2009}.
	SPICES' optical aberrations are expected to evolve very slowly with time. SPICES will be located at the L2 point{, which} is believed to be a very stable environment{. This} may be confirmed by the GAIA and JWST missions. Assuming such a stable environment, we plan to allocate a significant amount of time {at the beginning of the mission for the purpose of accurately estimating} SPICES' aberrations. Then, {the DM} will compensate for the slow variations. {In our simulation, the SCC} perfectly estimates for the wavefront aberrations (phase and amplitude). 
	The perfectly estimated wavefront is projected onto the 64\,$\times$\,64 DM using the method of energy minimization in the pupil plane \citep{Borde2006}. The DM influence functions are modeled by adapting the formula of \citet{Huang2008} to fit the parameters of a realistic DM.
	Finally, {the numerical noise introduced by the extraction of individual spectra from the IFS to build data cubes is assumed} negligible. 
 
 	In step 2, {blackbody spectra for the star and planetary spectra from \citet{Cahoy2010} and \citet{Stam2008} for the planets are introduced}. {The latter are discussed} in detail in Sect.~\ref{modelatm}. Photon and read-out noise, flat field variations{, and} zodiacal and exo-zodiacal light are accounted for. {The instrument throughput is set to 23\%, considering $\sim$15 optical surfaces of reflectivity 90\% from the primary mirror to the detector, and the quantum efficiency of the detector is 70\%.}
 	Using an algorithm to correct for cosmic ray contamination, \citet{Robberto2009} found that single exposures of 1\,000\,s will keep cosmic-ray induced glitches negligible with respect to a read-out noise level of a few e- rms per pixel. We {adopt} this single exposure for our simulations and we do not account for glitches.
 	{The detector flat field is modeled} as gaussian noise with a mean of 1 and a rms of 0.5\%.
	The IFS spreads the spectrum of an object point on different detector pixels. As a consequence, every pixel of the (x,\,y,\,$\lambda$) cube is affected by a specific flat field. We use the measured values of zodiacal light from \citet{Giavalisco2002}. {An exo-zodiacal disk with a 60$^\circ$-inclination with respect to {a face-on orbit} and a 45$^\circ$-orientation from the horizontal direction is simulated using} the Zodipic algorithm \citep{Kuchner2004}. The inclination value is the statistical median assuming a uniformly random orientation.
	{The orientation corresponds to the SCC fringe. The actual scientific strategy of SPICES requires prior knowledge of the orientation of the planet orbit. This has to be achieved by combining astrometric measurements with RV data. The present and near-future instruments like VLT/PRIMA and GAIA (we note that stars with V\,$<$\,6 are too bright for the latter) can in principle provide the information. We also plan to use a new version of the SCC that could enlarge the high-contrast part of the image and thus, could relax the constraints on the orbital knowledge.}

 \begin{table*}[t]
 \caption{Parameters of the exoplanetary atmosphere models used in this paper.}
 \begin{center}
 \begin{tabular}{c c c c c c}
 \hline\hline
 Planet & Separations (AU) & Radius & Metallicity (solar units) & Atmospheric structure & Surface type \\
 \hline
 Jupiter & 0.8, 2, 5 and 10 & 1~R$_{\rm{J}}$ & 1 and 3 & $-$ & $-$ \\
 Neptune & 0.8, 2, 5 and 10 & 1~R$_{\rm{N}}$ & 10 and 30 & $-$ & $-$ \\
 Super-Earth & 1 & 2.5~R$_{\rm{E}}$ & $-$ & 0, 50 and 100\% clouds & Forest, ocean and forest-ocean mix \\
 \hline
 \end{tabular}
 \label{planetmodels}
 \end{center}
 \tablefoot{R$_{\rm{J}}$, R$_{\rm{N}}$ and R$_{\rm{E}}$ refer to Jupiter, Neptune and Earth radii respectively.}
 \end{table*}

\subsection{Selecting spectral bandwidth and resolution}
\label{bandwidthresol}
\begin{figure}[t]
	\centering
	\includegraphics[width=.47\textwidth]{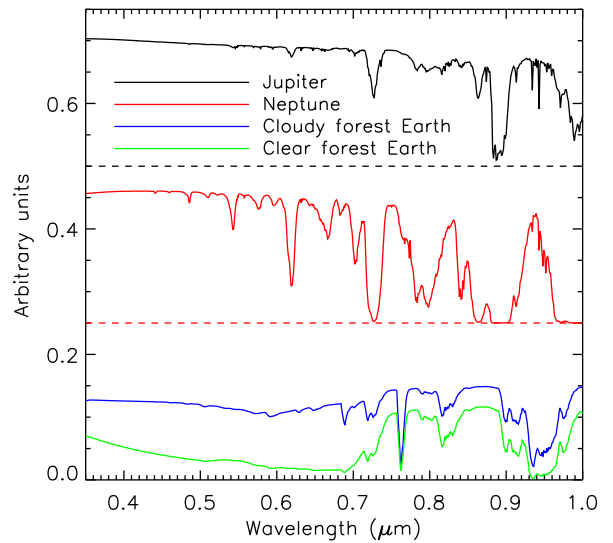}
	\caption{High-resolution albedo spectra of Jupiter, Neptune \citep{Cahoy2010} and two Earth analogs \citep{Stam2008}. The curves of the Jupiter and Neptune are vertically shifted with respect to the actual albedo for the sake of clarity. The dashed horizontal lines indicate the reference position for the Jupiter spectrum (black) and the Neptune spectrum (red).}
	\label{highresmodels}       
\end{figure}

\begin{figure*}[t]
	\centering
	\includegraphics[trim = 5mm 4mm 5mm 5mm, clip,width=0.47\textwidth]{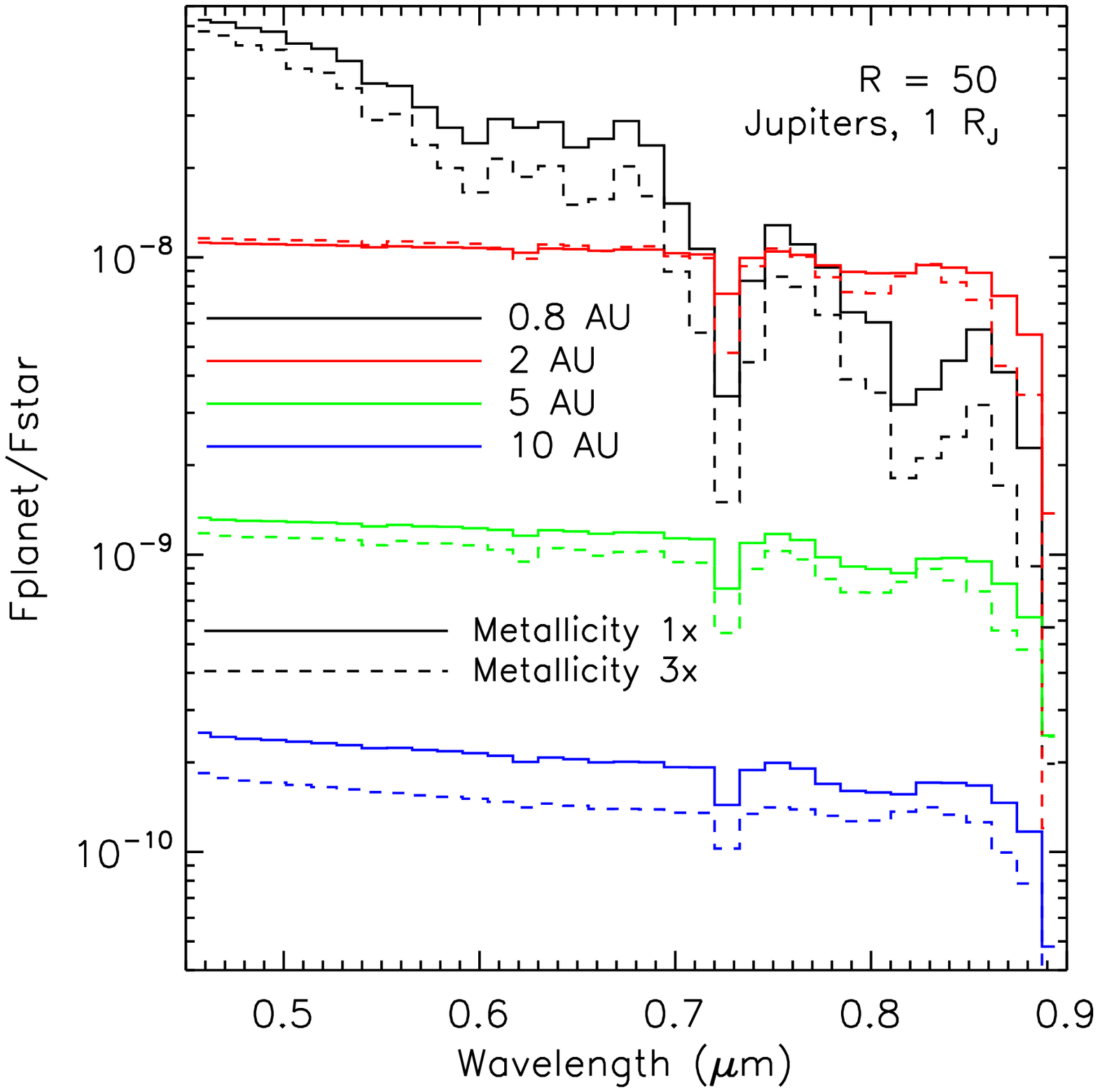}
	\includegraphics[trim = 5mm 4mm 5mm 5mm, clip,width=0.47\textwidth]{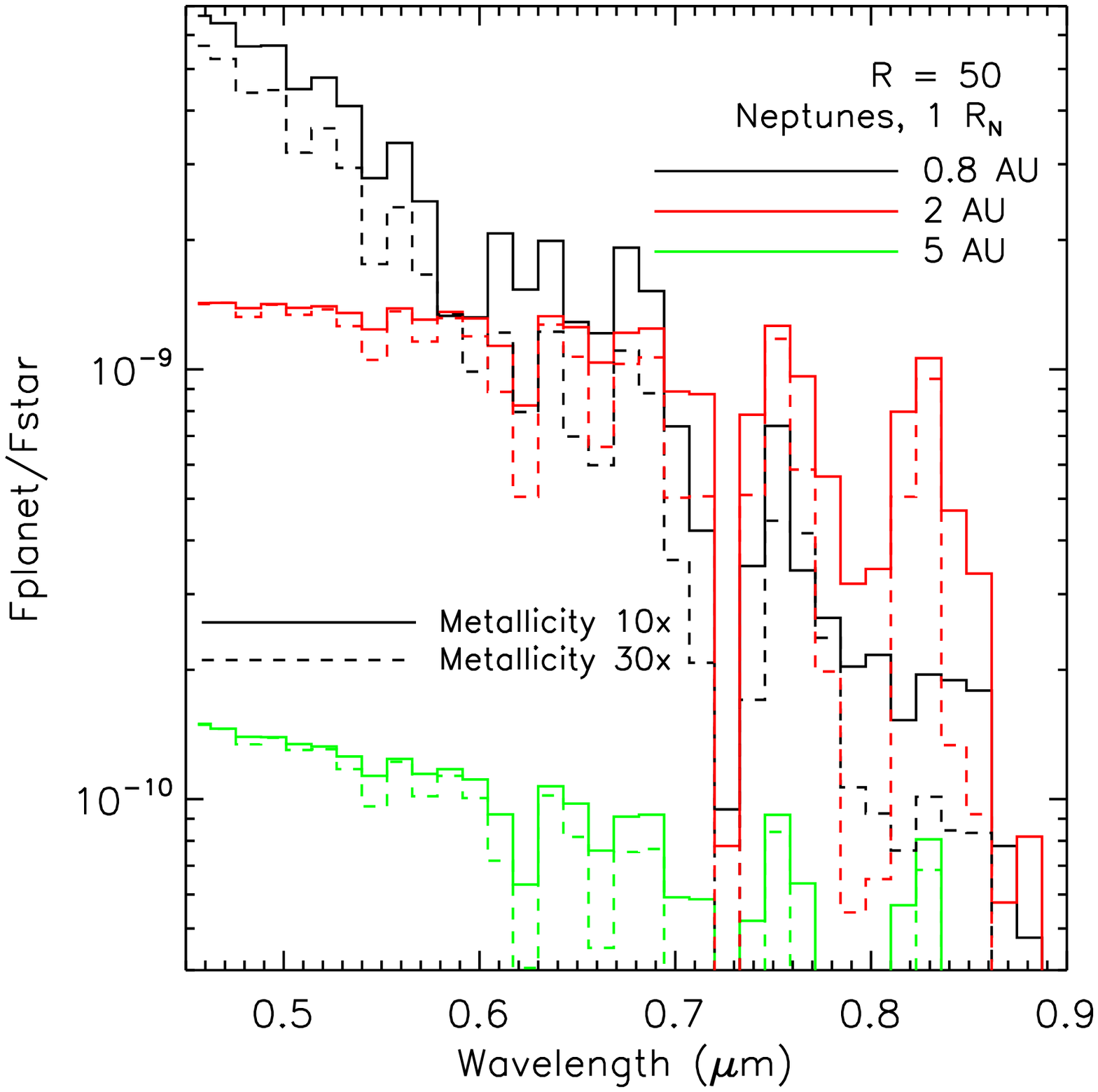}
	\caption{Theoretical models of Jupiter-like (left panel) and Neptune-like (right panel) atmospheres degraded to the resolution of SPICES \citep[models adapted from][]{Cahoy2010}.}
	\label{Cahoymodels}       
\end{figure*}

	In this section we discuss the different parameters we have to take into account for the choice of SPICES' spectral bandwidth and resolution. To determine the spectral bandwidth{, we} first examine theoretical spectra representative of Jupiter, Neptune and terrestrial atmospheres (Fig.~\ref{highresmodels}, giant planet models from \citet{Cahoy2010} and Earth models from \citet{Stam2008}). The spectra are calculated for visible wavelengths: 0.35--1~$\mu$m for \citet{Cahoy2010} and 0.3--1~$\mu$m for \citet{Stam2008}. We note that a 0.45--0.90~$\mu$m bandwidth offers a good compromise and enables to measure Rayleigh scattering at the blue wavelengths as well as the molecular absorption bands in the red part. The trade-off for the long-wavelength cut-off results from a technological limitation{, since} visible detectors have weak efficiencies above 0.90~$\mu$m. Therefore, SPICES will not measure the wide and strong water absorption band at 0.94~$\mu$m in the spectrum of Earth analogs (Fig.~\ref{highresmodels}). Although deep absorption bands are easily identifiable in model spectra, their depth would be hard to measure given the SPICES performance.
	However, other but shallower water bands at 0.72 and 0.82~$\mu$m are also present in planetary spectra. Similarly, giant planets feature a strong methane band at 0.89~$\mu$m at the boundary of SPICES' bandpass, but {there} are weaker bands at, for instance, 0.62, 0.73 and 0.79~$\mu$m. Measuring absorption bands at different wavelengths allows to infer the gas abundances{, if} the cloud top altitudes can be derived from a known gas which is well mixed in the atmosphere \citep[][and references therein]{Stam2008}. Polarimetry combined with flux could also help to break the degeneracy \citep{Stam2004}. For the telluric planets, \citet{DesMarais2002} define the spectral bandwidths and list the molecules that exoplanet missions should address: molecular oxygen (O$_{2}$), ozone (O$_{3}$), water (H$_{2}$O), methane (CH$_{4}$) and carbon dioxide (CO$_{2}$). The spectral range of SPICES (0.45--0.90~$\mu$m) permits to measure all these molecules except CO$_{2}$. In addition to atmospheric gases, \citet{Seager2005} have emphasized the scientific interest of the detection of surface features like the ``red edge'' (the rise of the clear Earth spectrum beyond 0.7~$\mu$m in Fig.~\ref{highresmodels}){. Nevertheless,} several studies showed that this measurement is difficult for the Earth itself \citep[e.g.,][]{Woolf2002, Arnold2002, Montan'es-Rodriguez2005}.
	\citet{Seager2005} note that the ``red edge'' should be detected with molecular oxygen to be sure that it is related to vegetation{, because} minerals may present a similar feature but at different wavelengths.
	From the comparison of widths of SPICES' spectral channels and of molecular bands, we find that a spectral resolution of at least 50 is required to identify the main bands of the spectra of giant planets as well as super-Earths (Tables~\ref{giantmolecules} and \ref{SEmolecules} and Figs.~\ref{Cahoymodels} and \ref{Stammodels}). Our analysis confirms previous results \citep{Schneider2009}.

\subsection{Planetary atmosphere models}
\label{modelatm}
	The following sections \ref{jupneptmodels} and \ref{earthmodels} describe the models we use in our simulation. In this paper, we consider two sets of models, one for giant planets and one for telluric planets. \citet{Cahoy2010}'s models cover a large range of star-planet separations and planet metallicities. \citet{Stam2008}'s models assume different planet surfaces and atmospheric structures (cloudy and clear atmospheres). Table~\ref{planetmodels} summarizes the main parameters we consider for these models. We set a maximum mass of 10 Earth masses (M$_{\rm{E}}$) for the super-Earths, which corresponds to a maximum radius of 2.5\,Earth radii (R$_{\rm{E}}$) from the mass-radius relation of \citet{Grasset2009}{. Recall that Stam's} atmosphere models assume an Earth-like planet.

	From the degraded spectra of Figs.~\ref{Cahoymodels} and \ref{Stammodels} (see Sect.~\ref{bandwidthresol} for the discussion of the spectral resolution), the planet/star contrast is expected to be $\sim$\,10$^{-8}$--10$^{-10}$ for Jupiters and $\sim$\,10$^{-9}$--10$^{-11}$ for Neptunes and super-Earths{, depending} on the separation from the star and on the wavelength. Note that \citet{Cahoy2010} and \citet{Stam2008} present geometric albedo spectra{, while} we plot contrast curves using the following formula:
\begin{equation}
C(\lambda) = A(\lambda,\alpha) \; \frac{R_p^2}{a^2} 
\label{contrastG2}
\end{equation}
where $\lambda$ is the wavelength, $C(\lambda)$ is the planet/star contrast spectrum, $A(\lambda,\alpha)$ is the planet albedo spectrum at phase angle $\alpha$, $R_p$ is the planet radius and $a$ is the star-planet separation. Phase angle is the angle star-planet-observer: when the planet is behind the star $\alpha$\,=\,0$^\circ${, and} when the planet is in front of the star $\alpha$\,=\,180$^\circ$. In this work, we only consider the case of a planet at its maximum elongation from the star, $\alpha$\,=\,90$^\circ$. The flux reflected by a planet depends on both its albedo and its radius (Eq.~(\ref{contrastG2})). To determine the albedo from observations, we need to independently estimate for the radius. We can use theoretical mass-radius relations \citep[e.g.,][]{Fortney2007, Baraffe2008, Grasset2009}. The polarimetric capabilities of SPICES could also help to establish the planetary properties without prior information about the planet's radius \citep{Stam2004, Stam2008}. This will be subject for future work. When we will study SPICES' ability to retrieve the planet properties from measured spectra in Sect.~\ref{spectrometry}{, we} assume that the radius is known.

\subsubsection{Theoretical spectra of Jupiter and Neptune analogs}
\label{jupneptmodels}
\begin{table}[t]
 \caption{List of molecules observable at R\,=\,50 in the Jupiter and Neptune models of Fig.~\ref{Cahoymodels}.}
 \begin{center}
 \begin{tabular}{c c l}
 \hline\hline
 Molecule & Approximate $\rm{\lambda}$ ($\rm{\mu}$m) & Note \\
 \hline
 CH$_{\rm{4}}$ & 0.54 \\
 CH$_{\rm{4}}$ & 0.62 \\
 H$_{\rm{2}}$O & 0.65 & for 0.8-AU models \\
 CH$_{\rm{4}}$ & 0.66 \\
 CH$_{\rm{4}}$ & 0.73 \\
 CH$_{\rm{4}}$ & 0.79 \\
 H$_{\rm{2}}$O & 0.82 & for 0.8-AU models \\
 CH$_{\rm{4}}$ & 0.84 \\
 CH$_{\rm{4}}$ & 0.86  \\
 \end{tabular}
 \label{giantmolecules}
 \end{center}
 \end{table}
	{We use the models of \citet{Cahoy2010}, who} calculate atmospheric structures of old ($\sim$4.5\,Gyr) Jupiter and Neptune analogs in radiative equilibrium with the radiation of a solar-type host star at separations of 0.8, 2, 5 and 10~AU, for different metallicities and for optical wavelengths (0.35 to 1~$\mu$m). This range matches the separations of the planets that small space coronagraphs can potentially observe. Table~\ref{giantmolecules} lists the main spectral bands observable with a spectral resolution of 50. Several theoretical spectra of Jupiter and Neptune analogs are shown in Fig.~\ref{Cahoymodels}. Unlike young planets{, which} are dominated by thermal radiation, the star-planet separation drastically alters the structure and composition of mature planetary atmospheres. Therefore, a simple scaling of the amount of reflected light with distance is not sufficient to model realistic spectra and to derive the actual performance of a mission like SPICES.
	If planets are too warm for any molecules to condense into clouds{, their} spectra are dominated by Rayleigh scattering{. This is illustrated for the case of a separation of 0.8~AU in Fig.~\ref{Cahoymodels}}. At 2~AU, bright water clouds form and dominate the atmospheric opacity all over the spectrum. At 5~AU, ammonia clouds form above the water clouds. At 10~AU, the same clouds form but at a deeper pressure level{, and} Rayleigh scattering again dominates the reflected flux at short wavelengths. The planet is $\sim$4 times fainter than the Jupiter at 5~AU{, as} expected in the case where the flux decrease follows an inverse square power law of separation. For all cloudy planets (2, 5 and 10~AU), the clouds are optically thick.
	The Neptune spectra exhibit the same but stronger absorption bands as the Jupiter spectra. Increasing the planet metallicity usually decreases its albedo{. The exception is the 2-AU Jupiter model for} short wavelengths because for this case the water clouds are high in the atmosphere and thick. \citet{Cahoy2010} note that the metallicity increase produces larger differences between the Jupiter spectra than between the Neptune spectra for separations of 5 and 10~AU. Methane bands dominate the spectra over all the bandwidth (e.g., 0.62, 0.73, 0.79 and 0.89~$\mu$m). Their depths depend on the nature of the light-scattering particles (gases, clouds, aerosols).

\subsubsection{Theoretical spectra of telluric planets}
\label{earthmodels}
\begin{figure}[t]
	\centering
	\includegraphics[trim = 4mm 4mm 4mm 4mm, clip, width=.47\textwidth]{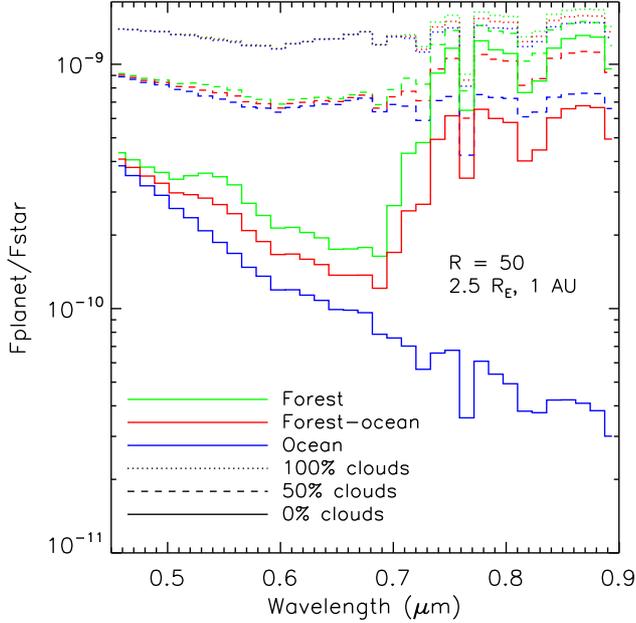}
	\caption{Same as Fig.~\ref{Cahoymodels} but for the terrestrial atmosphere models of \citet{Stam2008}.}
	\label{Stammodels}       
\end{figure}

\begin{table}[t]
 \caption{Same as Table~\ref{giantmolecules} but for the terrestrial planet models of Fig.~\ref{Stammodels}.}
 \begin{center}
 \begin{tabular}{c c l}
 \hline\hline
 Molecule & Approximate $\rm{\lambda}$ ($\rm{\mu}$m) & Note \\
 \hline
 O$_{\rm{3}}$ & 0.5 $-$ 0.7 & the Chappuis band\\
 O$_{\rm{2}}$ & 0.69 & B-band\\
 H$_{\rm{2}}$O & 0.72 \\
 O$_{\rm{2}}$ & 0.76 & A-band\\
 H$_{\rm{2}}$O & 0.82 \\
 \end{tabular}
 \label{SEmolecules}
 \end{center}
 \end{table}
	In this section{, we} summarize the main features of the telluric planet models. \citet{Stam2008} uses a pressure-temperature vertical profile of the Earth to derive flux and polarization spectra for several surface and cloud coverages and for a spectral range between 0.3 and 1~$\mu$m. We recall that we do not consider polarization effects in this paper and that these models are relevant for a separation of 1~AU around a solar-type star. \citet{Stam2008} considers atmospheres with and without a horizontal homogeneous water cloud layer{, and} surfaces completely covered by either forest or black ocean with a Fresnel reflecting interface (Fig.~\ref{Stammodels}). The water clouds are optically thick and located in the troposphere. The atmospheric absorbers are water, molecular oxygen and ozone (Table~\ref{SEmolecules}). The model albedo of vegetation presents two main features: a local maximum between 0.5 and 0.6~$\mu$m{, which} is due to two absorption bands of chlorophyll at 0.45 and 0.67~$\mu$m{, and} the increase of the albedo beyond 0.7~$\mu$m due to the ``red edge'' (Sect.~\ref{bandwidthresol}). The appearance of the ``red edge" in the planet spectra depends strongly on the cloud thickness and coverage, but it still alters the spectrum shape for a partial cloud coverage despite their large optical thickness (Fig.~\ref{Stammodels}, see the spectra of the 50\% cloudy ocean and 50\% cloudy forest planets in blue and green dashed lines respectively). At short wavelengths Rayleigh scattering dominates while at long wavelengths scattering by clouds is the most important process.
	\citet{Stam2008} uses a weighted sum of homogeneous models to simulate a quasi horizontally inhomogeneous model representative of the Earth (70\% of the surface covered by ocean and 30\% by forest) with different cloud coverages. In this paper, we use different weighted sums for simulating three cloud coverages (0, 50 and 100\%) and three surface types (forest, forest-ocean equally mix and ocean){, as} indicated in Table~\ref{planetmodels}.

\subsubsection{Planet contrasts as a function of the stellar type}
\label{contraststellartypes}
	The atmospheric structures of the models were calculated by assuming $\sim$4.5-Gyr planets in radiative equilibrium with the flux of a Sun-like parent star. These models can be transposed to other stellar types assuming flux conservation with the standard formula:
\begin{equation}
4 \pi \, R_p^2 \, \sigma \,  T_{eq}^4 \; = \; (1 - A_{B}) \, \pi \, R_p^2 \, \frac{L_{\star}}{4 \, \pi \, a^2}
\label{eqrad}
\end{equation}
where $R_p$ is the planet radius, $\sigma$ is Stefan's constant, $T_{eq}$ the planet atmosphere equilibrium temperature, $A_{B}$ the planet Bond albedo and $L_{\star}$ the host star luminosity.
This formula does not account for the effects of the wavelength dependence of the star emission on the atmosphere \citep{Marley1999, Fortney2007}.

\begin{table}[t]
 \caption{Star-planet separations for giant planets extrapolated to other stellar types assuming Eq.~(\ref{sep}). The values for the solar-type star are those modeled by \citet{Cahoy2010}.}
 \begin{center}
 \begin{tabular}{c c c c c c}
 \hline\hline
 Spectral type & Luminosity (L$_{\sun}$) & \multicolumn{4}{c}{Separations (AU)}\\
 \hline
  A0 & 28 & 4.2 & 10.6 & 26.5 & 53.0 \\
  F0 & 4.8 & 1.8 & 4.4 & 11.0 & 21.9 \\
  \rowcolor[gray]{0.8} G2 & 1 & 0.8 & 2 & 5 & 10 \\
  K0 & 0.45 & 0.5 & 1.3 & 3.4 & 6.7 \\
  M0 & 0.09 & 0.24 & 0.6 & 1.5 & 3 \\
 \hline 
 \end{tabular}
 \label{sepstartype}
 \end{center}
 \end{table}

	Because we are using models with discrete values (especially separations and stellar luminosity){, we} cannot extrapolate the planet spectra to any separations around any stars. Instead, we calculate the correspondence between separations and stellar luminosity{, considering} $T_{eq}$ and $A_{B}$ only depend on the incident stellar flux at the planet position $L_{\star}/a^2$. Therefore, Eq.~(\ref{eqrad}) becomes:
\begin{equation}
L_{\star} \propto a^2 \quad \Rightarrow \quad a_{S_{p}} \; = \; a_{G2} \, \sqrt{L_{S_{p}}}
\label{sep}
\end{equation}
where $a_{S_{p}}$ is the star-planet separation for a star of spectral type $S_{p}$, $a_{G2}$ the star-planet separation for a G2 star and $L_{S_{p}}$ the star luminosity in solar units.
	For example, a Jupiter at 2~AU from a solar-like star would have the same atmospheric structure as a Jupiter at $\sim$10.5~AU from an A0 star. Table~\ref{sepstartype} gives the correspondences for different star-planet separations and stellar types.

	Substituting Eq.~(\ref{sep}) into Eq.~(\ref{contrastG2}){, we} obtain the contrast of a planet around a host star of type $S_p$:
\begin{equation}
C(\lambda) = A(\lambda,\alpha) \; \frac{R_{{p}}^2}{a_{G2}^2 \, L_{S_{p}}}
\label{contrast}
\end{equation}
	We use Eqs.~(\ref{sep}) and (\ref{contrast}) to derive the star-planet separations and contrasts in Sect.~\ref{perf}. While this calculation is a fairly good estimation for the giant and cloudy telluric planet models, it is less accurate for the clear terrestrial models. For stars cooler than the Sun, \citet{Wolstencroft2002} suggest that the ``red edge'' could indeed be shifted towards wavelengths redder than 1~$\mu$m{, if} the photon number involved in the photosynthesis processes is greater than for the mechanism operating on Earth. If this hypothesis is verified, the ``red edge'' will be outside the bandwidth covered by SPICES and undetectable. However, \citet{Kiang2007} warn that a theory predicting the ``red edge'' wavelength for a given stellar type assuming the same mechanism as on Earth is still missing.
	
\section{Performance in detection}
\label{perf}

The instrument model presented in Sect.~\ref{nummodel} provides an estimation of the achievable contrast map in the field of view for each spectral channel. In this section{, we} estimate the average contrast that is reached in the darkest area of the field of view (area contained by the dotted lines in Fig.~\ref{images}) as a function of the angular separation from the central star. {As we explained in Sect.~\ref{simassumptions}, we assume that the planet orbital parameters are known and its position can be matched with the orientation of the corrected area.}
	
\subsection{Impact of speckle noise}
\label{specklenoise}	

\begin{figure}[t]
	\centering
	\includegraphics[width=.47\textwidth]{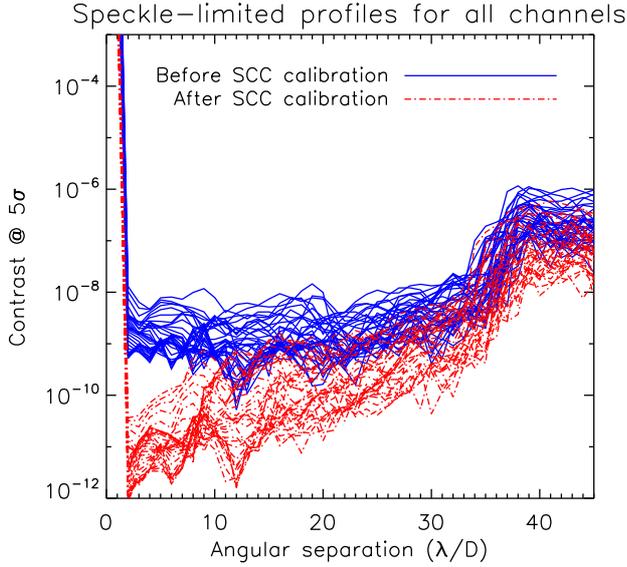}
	\caption{5-$\rm{\sigma}$ detection profiles of the instrument contrast achieved by SPICES for all spectral channels{, before} (blue solid lines) and after (red dashed lines) the SCC speckle calibration.}
	\label{perfnonoiseallchan}       
	\end{figure}

	Figure~\ref{perfnonoiseallchan} shows SPICES' contrast curves against angular separation ($\lambda_0$\,=\,0.675~$\mu$m) for all spectral channels{, before} and after the speckle calibration by the SCC. We see that the speckle subtraction is necessary to reach the requirement of $\sim$$10^{-10}$ at a few $\lambda/D$ (Sect.~\ref{simassumptions}). The wavelength dispersion of the performance is due to the phase aberration dependence on wavelength ($\propto$\,$\lambda^{-1}$) and the SCC calibration dependence on spectral resolution \citep{Galicher2010}. Recall that we set the same bandwidth for all channels so spectral resolution increases with wavelength. The steep increase of the detection limit around 32~$\lambda/D$ corresponds to the DM cut-off spatial frequency. This cut-off and the more efficient SCC calibration at small separations explain the degradation with angular separation.

\begin{figure}[t]
	\centering
	\includegraphics[width=.47\textwidth]{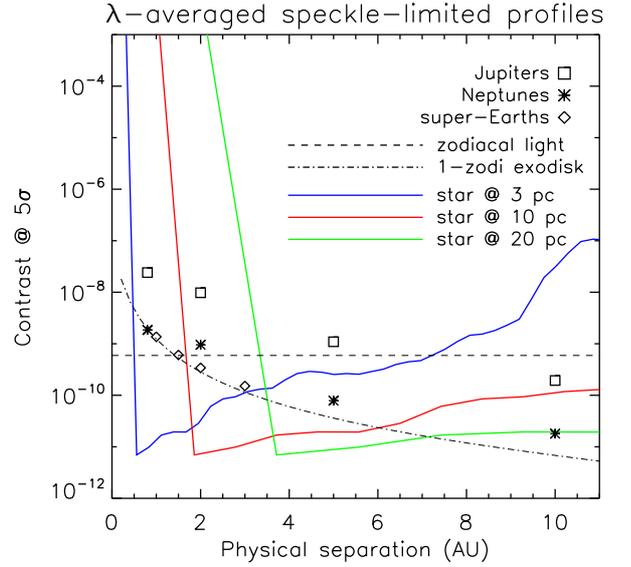}
	\caption{5-$\rm{\sigma}$ detection profiles of the instrument contrast averaged over the SPICES bandpass (solid lines) and compared to averaged planet contrasts calculated for a solar-type star (symbols).
	{For comparison, we also plot the mean contrast of the zodiacal light (horizontal dashed line) and a 1-zodi exodisk (dot-dashed curve), both for a star at 10~pc}.}
	\label{perfnonoise}        
\end{figure}

	To have a clear and simple view of the performance, we plot radial profiles averaged on all the spectral channels against the physical separation in AU for three star distances in Fig.~\ref{perfnonoise}.
	The performance is limited at short separations by the coronagraph IWA {($\sim$2~$\lambda/D$, Sect.~\ref{nummodel})}, and at large separations by the size of the DM corrected area (cut-off at 32~$\lambda/D$). As we express the x-axis in AU, the contrast curve scales with the star distance. We also plot the positions of Jupiter of 1 solar metallicity and 1 Jupiter radius (R$_{\rm{J}}$), Neptune of 10 solar metallicities and 1 Neptune radius (R$_{\rm{N}}$), and 2.5-R$_{\rm{E}}$ cloudy planets. We assume that the super-Earth properties do not evolve with separation for the considered range. We recall that \citet{Stam2008} uses a temperature-pressure profile of the Earth so the model is relevant for a separation of 1~AU. For stars at 20~pc, the farthest Jupiters (5 and 10~AU) and Neptunes (5~AU) are detected with SNR\,$>$\,5. The 10-AU Neptune is below the curve (SNR\,$\simeq$\,4.5). For a 10-pc star, the same planets are still detected as well as planets as close as 2~AU. For the closest star we consider (3~pc), planets as close as 1~AU are very well detected and characterized (Sect.~\ref{spectrometry}). In this case, the 10-AU Jupiter and 5-AU Neptune are not detected with the current instrument design because of the speckle calibration degradation with angular separation. Solutions may exist to improve the detection at these large angular separations ($\gtrsim$18~$\lambda/D$) such as spectral deconvolution \citep{Sparks2002}.

	In addition, we plot the averaged contrast levels for zodiacal light and a 1-zodi exodisk {(star distance of 10~pc)}. {Recall that they increase as the square of the star distance.} Although the instrument concept can reduce speckle noise below a contrast of 10$^{\rm{-10}}$, the final performance is limited by these extended background sources. The considered exo-zodiacal disk limitation is at the level of Neptunes and super-Earths. Therefore, to obtain a correct estimation of planet fluxes, the zodiacal and exo-zodiacal contributions must be carefully calibrated and removed.
	The exo-zodiacal disk intensity has been identified as critical for Earth-twin detection{. To address this question,} exo-disk surveys have been recommended to prepare target lists of faint exo-disks for space nulling interferometers like the Terrestrial Planet Finder Interferometer and Darwin \citep{Lunine2008, Coud'eduForesto2010, Hatzes2010}.
	In the case of SPICES, the problematic is different as we do {single-aperture} imaging. The exact procedure to account for zodiacal and exo-zodiacal contributions remains to be defined although it is a data reduction issue, which is beyond the scope of this paper.
	In the following, we consider the model distribution of both zodiacal and exo-zodiacal intensities can be subtracted from the data{. The} photon noise of these contributions may still limit the contrast performance (Sect.~\ref{noise}).

\subsection{Impact of detection noise}
\label{noise}
\begin{figure}[t]
	\centering
	\includegraphics[width=.47\textwidth]{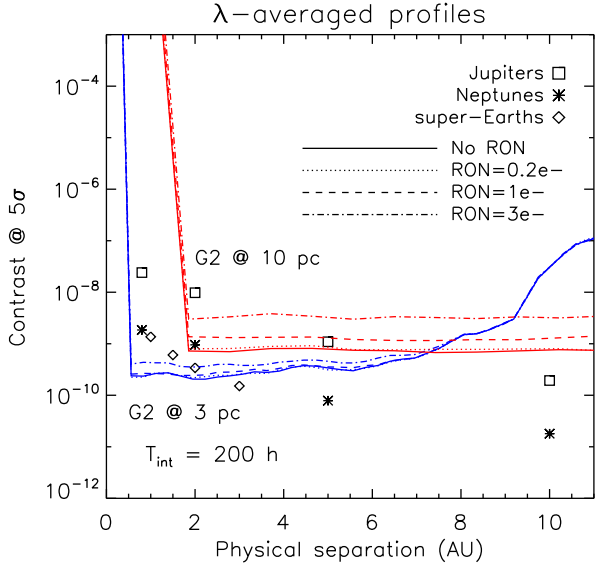}
	\caption{5-$\rm{\sigma}$ detection profiles accounting for the photon noise of G2 star at 3 (blue lines) and 10~pc (red lines){, and} for read-out noise levels from 0 to 3~e- rms per pixel.}
	\label{ron}       
\end{figure}
	Due to a small primary mirror (typically 1.5~m in diameter) together with a spectral resolution R\,=\,50, space coronagraphs like SPICES
will be limited by photon noise (from the stellar background and the planet) or read-out noise for most of the targets{, because} very long exposures would be required to reach the instrument limitation at $\sim$\,$10^{\rm{-10}}$. For instance, $\sim$10\,000~h are needed to achieve $\sim$$10^{\rm{-10}}$ at 5\,$\sigma$ for a G2 star at 10~pc.
From this exposure time and assuming that the noise follows the photon noise behavior (proportional to the square root of the photon number){, we estimate the total integration time required to detect a super-Earth at SNR\,=\,5, for typical values of distance and contrast accessible to SPICES. It is 200~h for a planet of contrast 2.5\,$\times$\,10$^{\rm{-10}}$ at 2~AU around a solar-type star at 5~pc.} No detailed study of the maximum integration time per target has been carried out to date for SPICES. In this paper, we set the maximum integration time per target to 200~h{. This} is a good trade-off between achieving high contrasts and observing a large number of targets during the mission.

	We present the impact of photon and read-out noise on SPICES' performance {for the case of a G2 star at 3 and 10~pc in Fig.~\ref{ron}}. We note that the read-out noise is a major limitation for the furthest star but not for the closest. This is due to the fact that the full well capacity of the detector (Table~\ref{hypotheses}) is not filled after a 1\,000-s exposure in both cases. The number of single exposures and read-out noise level are thus the same{, but} the photon count is greater for the closest star. We note that {the dozen of stars located within 3~pc have types later than G (the exception being Sirius). Thus,} the read-out noise will not be a fundamental limitation for close stars. We base the read-out noise requirement on the farthest star and set its value to 0.2~e- rms per pixel. Electron multiplying CCDs can achieve such a low read-out noise{, and} a large set of devices have been qualified for space during the GAIA preparation \citep{Smith2006}.

\begin{figure}[t]
	\centering
	\includegraphics[width=.47\textwidth]{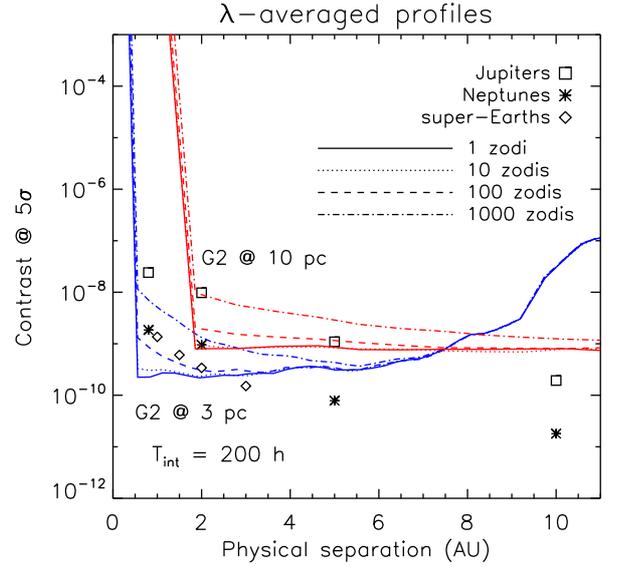}
	\caption{5-$\rm{\sigma}$ detection profiles for a Sun analog at 3 and 10~pc and exo-zodiacal intensities ranging from 1 to 1\,000~zodis{, assuming} a 0.2~e-/pixel rms read-out noise and photon noise.}
	\label{exozodilevel}       
\end{figure}

	As indicated in the previous section, an exo-disk can prevent the detection of faint planets if its photon noise becomes too important. To help to prepare a target list, we estimate the exozodi level that may hamper the detection of SPICES' targets. Figure~\ref{exozodilevel} presents the performance for different exozodi levels and two distances of a solar-type star. {The read-out noise is set to 0.2~e- rms per pixel. We assume that} the exo-disks have no structure and can be subtracted out from the data to the precision imposed by photon noise. 
	We find that the exo-disk photon noise does not significantly limit the performance up to 10 zodis{, but} begins preventing the Neptune and super-Earth detection when larger than 100~zodis. 
	For exposures shorter than 200~h (more dominant photon noise), the acceptable exozodi level is lower.

\begin{figure}[t]
	\centering
	\includegraphics[trim = 5mm 4mm 5mm 12mm, clip, height=0.375\textwidth]{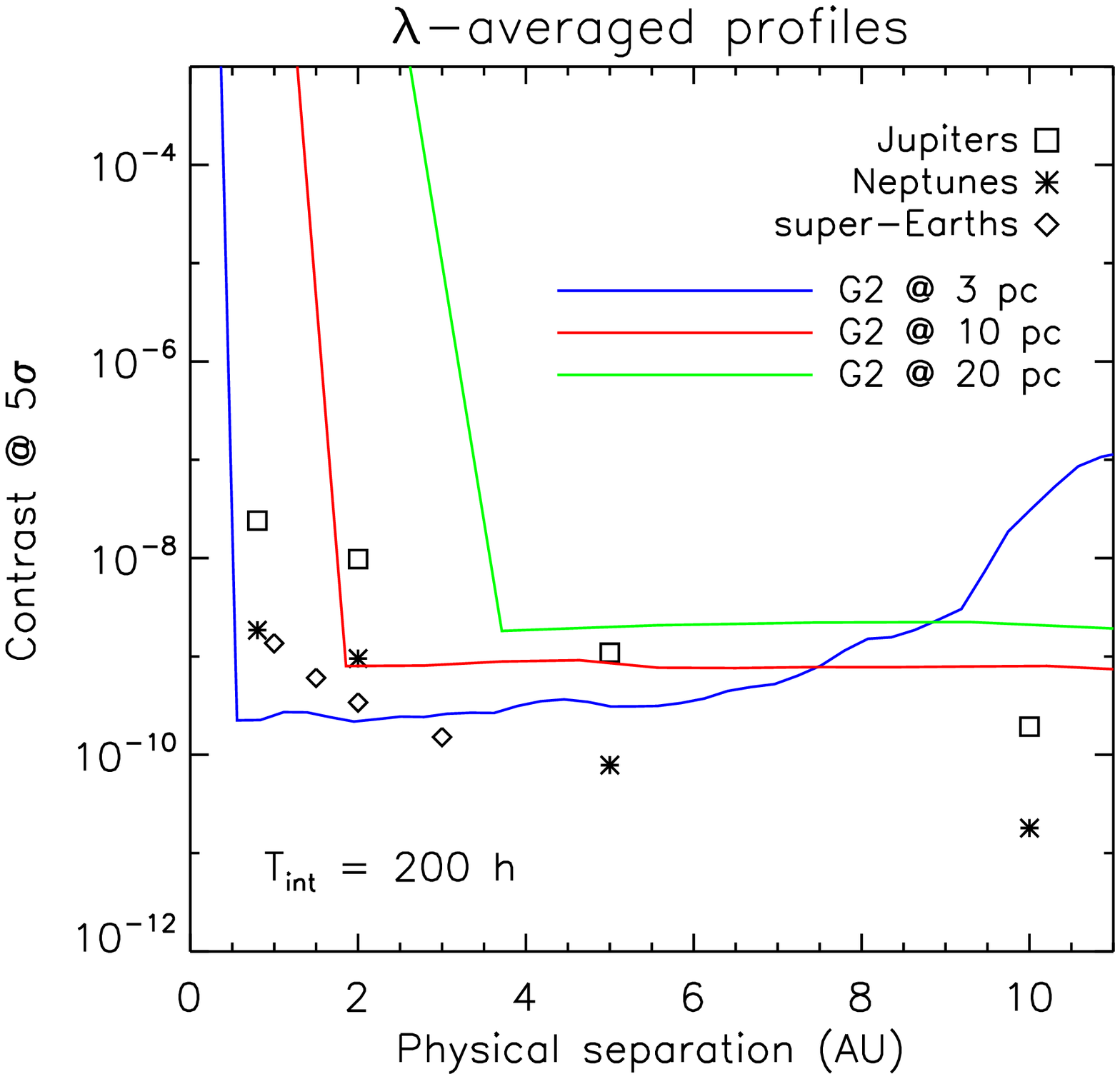} 
	\includegraphics[trim = 5mm 4mm 5mm 12mm, clip, height=0.375\textwidth]{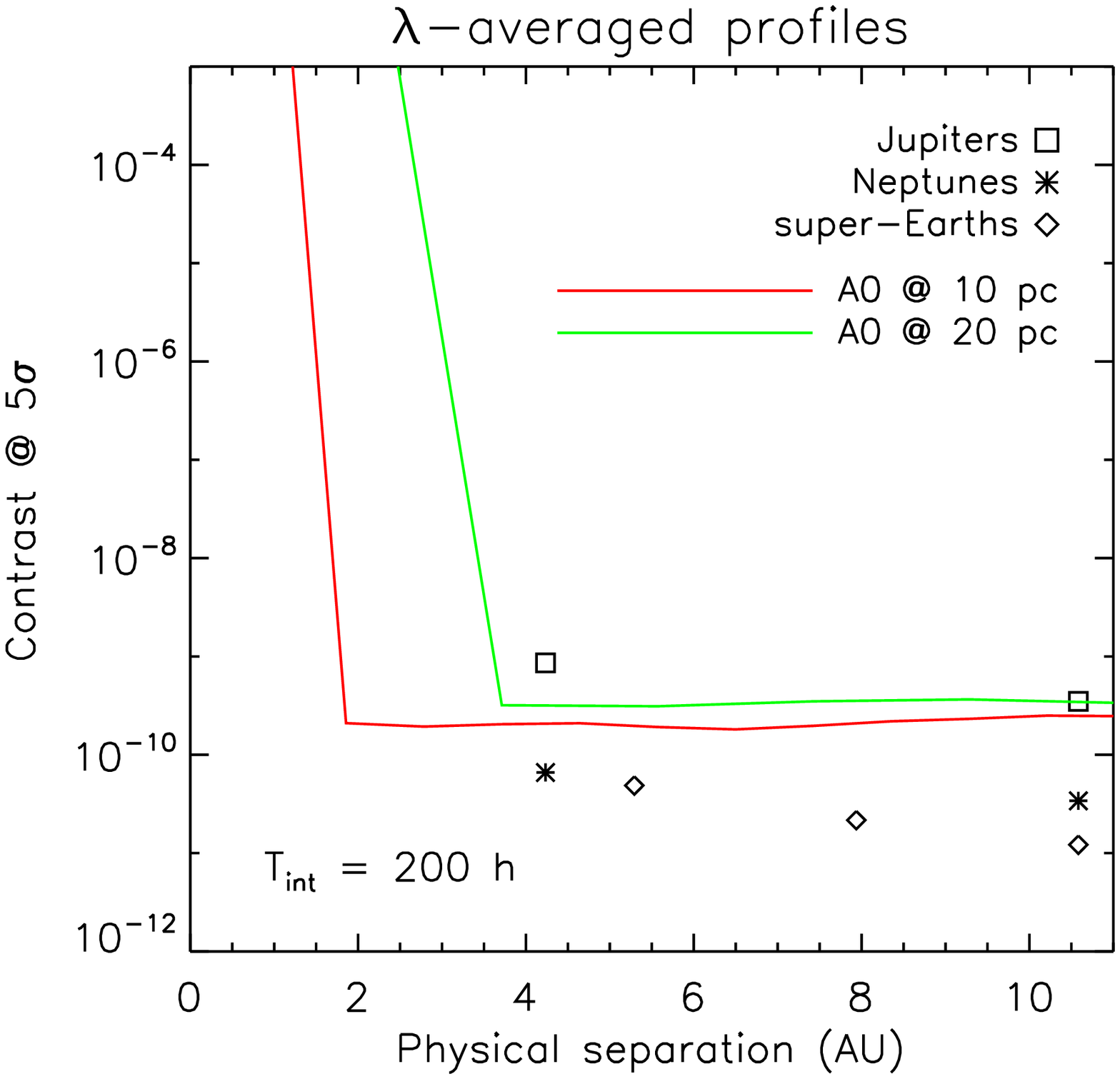} 
	\includegraphics[trim = 5mm 4mm 5mm 12mm, clip, height=0.375\textwidth]{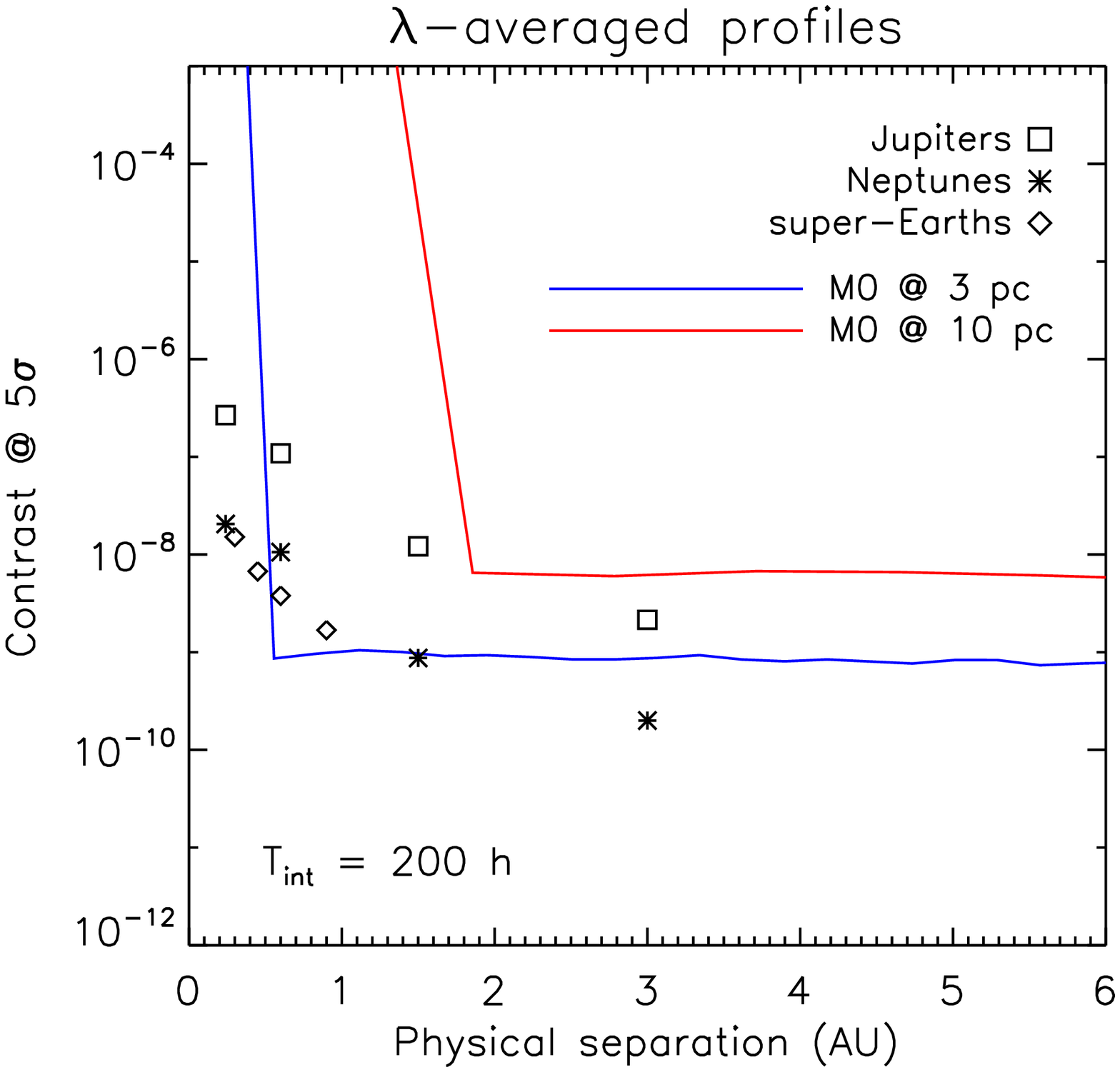} 
	\caption{5-$\rm{\sigma}$ detection profiles for a 200-h exposure compared to averaged planet contrasts for G2 (top), A0 (middle) and M0 (bottom) type stars and several star distances.}
	\label{noisyprof}       
\end{figure}

\subsection{Impact of stellar type}
\label{noisyperf}
	Assuming a generic exposure time of 200 hours and an exo-zodiacal disk of 1 zodi, we test the detectability of planets for several stellar types. Figure~\ref{noisyprof} presents the 5-$\sigma$ detection levels for three stars of type G2, A0 and M0 and several distances (3, 10, and 20~pc). We consider the star-planet separations given in Table~\ref{sepstartype} (they scaled by L$_{\star}^{1/2}$){, and} we apply Eq.~(\ref{contrast}) to find the wavelength-averaged planet contrasts. For a G2 star (Fig.~\ref{noisyprof}, top), SPICES can potentially access jovian planets, icy giants and super-Earths with separations smaller than $\sim$6, $\sim$3 and $\sim$2~AU respectively. No planet is detected at distances larger than $\sim$10~pc. The detectability of the closest planets (0.8 to 2~AU) will be limited by the coronagraph IWA ($\sim$2~$\lambda/D$) for stars at 4 to 10~pc.
	As for an A0 parent star (Fig.~\ref{noisyprof}, middle), Jupiter-like planets are not detected when further than 10~AU for stars within 20~pc. 
	The separation ranges considered by \citet{Cahoy2010} and \citet{Stam2008} in the case of a G2 star do not allow to study planets at separations shorter than $\sim$4~AU around an A0 star{. Nonetheless,} we can roughly estimate that Neptune analogs and super-Earths can be detected in the range 2--4~AU for stars closer than 10~pc. There are no A stars within 5~pc (except for the Sirius binary system) so we do not plot the 3-pc curve.
	Finally, the very close M stars (Fig.~\ref{noisyprof}, bottom) at 3--5~pc are of great interest for detecting Jupiter-like planets in the 0.5--4~AU range as well as super-Earths in close orbits at 0.5--1~AU.

	From the current exoplanet database \citep{Schneider2011}, we assess that only a handful of known extrasolar planets match the limitations described here. However, we note that RV surveys are not complete in the case of early and late stellar types \citep[A and M in particular,][]{Udry2007} and at long periods (a few AUs) even for nearby stars.

\section{Performance in spectrometry}
\label{spectrometry}
 
 In this section, we analyze {the SPICES performance more in detail, by estimating SNRs of the measured planetary spectra}. The objective is to set the constraints on the SNR to allow the differentiation between planetary models: impact of physical star-planet separation and metallicity for the Jupiter and Neptune analogs (Sects.~\ref{Jupstudy} and \ref{Neptstudy}){, and} variations due to cloud and surface coverage for the rocky planets (Sect.~\ref{SEstudy}).
 
\subsection{Criterion of characterization}
\label{spectroparam}
	We first explain our characterization criterion. The underlying question is to know if a measured spectrum $S$ is reproduced by one of two model spectra noted $M_i$ ($i$ refering to the model index, $i$\,=\,1,2). This depends on the noise of the measured spectrum $N=S$/SNR. We define the following criteria of comparison of $S$ to $M_i$:
\begin{equation}
  crit_i\,=\,\mathrm{median}_{\lambda} \left( \frac{S(\lambda)\,-\,M_i(\lambda)}{N(\lambda)} \right)
  \label{definecrit}
\end{equation}
where the median is calculated over the spectral channels. We choose the median because it accounts well for the overall quality of a spectrum.
	For a given measurement ($S$, $N$), the model which best matches the measured spectrum gives the lowest criterion value. Substituting the definition of $N$ to express $crit_i$ as a function of SNR and assuming the latter is nearly constant on the measured spectrum, we can write:
\begin{equation}
crit_i\,=\, \mathrm{median}_{\lambda} \left( \frac{S(\lambda)\,-\,M_i(\lambda)}{S(\lambda)} \right) \times \rm{SNR}
\label{criti}
\end{equation}
	In this paper, we use theoretical models to assess the instrument performance and to set the SNR required to measure differences between them ($S$\,=\,$M_1$ and $M_i$\,=\,$M_2$). We consider that the models are differentiated when their difference is 10 times above the measured noise so $crit_i$\,=\,10. Inverting Eq.~(\ref{criti}) and calling SNR$_{\rm{r}}$ the value of the required SNR, we finally obtain:
\begin{equation}
\mathrm{SNR}_\mathrm{r} \,=\,10\,\times \, \frac{1}{\mathrm{median}_{\lambda} \left( \frac{M_1(\lambda)\,-\,M_2(\lambda)}{M_1(\lambda)} \right)}
\label{criterion}
\end{equation}
	As an example, when we study the metallicity (Sects.~\ref{Jupstudy} and \ref{Neptstudy}), $M_1$ will be the lowest metallicity model and $M_2$ the highest metallicity model. We consider different spectral bandwidths to determine SNR$_{\rm{r}}$ (Eq.~(\ref{criterion})) when analyzing different planetary properties.
	For instance, metallicity strenghtens the bands (Fig.~\ref{Cahoymodels}){, and} cloud and surface coverage alter the spectrum at blue and red wavelengths respectively (Fig.~\ref{Stammodels}). We give the calculated SNR$_{\rm{r}}$ in Table~\ref{tabsnr} and we specify the spectral channels in the last column. We stress the point that the SNR$_{\rm{r}}$ values {correspond to} the spectrum of the brightest planet considered for each analyzed property: for instance, the low-metallicity planets for the Jupiters and Neptunes (Sects.~\ref{Jupstudy} and \ref{Neptstudy}) and the cloudy planets for the super-Earths (Sect.~\ref{SEstudy}).

\begin{table}[t]
 \caption{Values of SNR$_{\rm{r}}$ derived from Eq.~(\ref{criterion}).}
 \begin{center}
 \begin{tabular}{l l c l}
 \hline\hline
 Planet & Parameter & SNR$_{\rm{r}}$ & Note \\
 \hline
 Jupiter & 0.8/2~AU & 15 & \\
 Jupiter 0.8~AU & metallicity 1/3x & 30 \\
 Jupiter 2~AU & metallicity 1/3x & 30 & CH$_{\rm{4}}$ bands \\
 Jupiter 5~AU & metallicity 1/3x & 30 & CH$_{\rm{4}}$ bands \\
 Neptune & 0.8/2~AU & 15 & \\
 Neptune 0.8~AU & metallicity 10/30x & 30 \\
 Neptune 2~AU & metallicity 10/30x & 25 & CH$_{\rm{4}}$ bands \\
 Forest Earth & 0/50/100\% clouds & 25 & blue channels \\
 Ocean Earth & 0/50/100\% clouds & 25 & blue channels \\
 Clear Earth & 0/50/100\% forests & 12 & red channels \\
 50\% cloudy Earth & 0/50/100\% forests & 30 & red channels \\
 Cloudy Earth & 0/50/100\% forests & 220 & red channels \\
 \hline 
 \end{tabular}
 \label{tabsnr}
 \end{center}
 \end{table}
		
\begin{table}[t]
 \caption{Maximum star distance at which SPICES resolves the planet separation at the central wavelength of the bandwidth.}
 \begin{center}
 \begin{tabular}{c c}
 \hline\hline
 Planet separation (AU) & Star distance (pc) \\
 \hline
 0.8 & 4 \\
 1 & 5 \\
 2 & 10 \\
 5 & 25 \\
 \hline 
 \end{tabular}
 \label{septodistsun}
 \end{center}
 \end{table}
 
	In the remainder of this section, we study the ability of SPICES to disentangle planetary models. For each planet separation, we consider the distance at which the star-planet system is resolved at quadrature (Table~\ref{septodistsun}) and we derive the exposure time to achieve the SNR$_{\rm{r}}$ values quoted in Table~\ref{tabsnr}. We restrain the study to the case of a solar-type star and we assume a maximum exposure time of 200~hours ($\sim$8~days).
We perform our simulations for five independent realizations of speckle pattern. We then average our results to minimize the impact of an optimistic or pessimistic speckle pattern. To save computing time{, we} use the same five speckle patterns for all planet cases{, although} we randomly change the photon and read-out noise.
We assume the planet position to be perfectly known. We integrate the planet flux within apertures of diameter 1~$\lambda/D$ for each spectral channel. This corresponds to the full width at half maximum of the point spread function.
The 1-$\sigma$ error bars shown in the plots account for the variation of both speckle and noise realizations.

\subsection{Jupiter models}
\label{Jupstudy}
\begin{figure}[t]
	\centering
	\includegraphics[width=.47\textwidth]{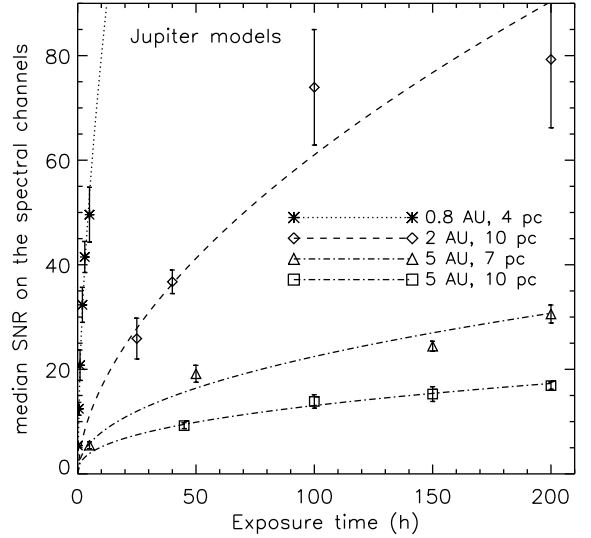}
	\caption{Evolution of the median value of SNR$_{\rm{m}}$ (see text) as a function of exposure time for Jupiter planets (symbols). The curves are power-law fits. We plot 1-$\sigma$ error bars.}
	\label{Jupsnrtime}       
\end{figure}

	 We consider the models of Jupiter analogs described in Sect.~\ref{jupneptmodels} for several separations and metallicities. Figure~\ref{Jupsnrtime} shows the evolution with exposure time of the median SNR measured from the simulated data (SNR$_{\rm{m}}$). We represent each data point with its corresponding 1-$\sigma$ error bar. The SNR$_{\rm{m}}$ dependence on exposure time may change from one observed planet to another as a function of the planet intensity and location in the diffraction pattern of the host star. We fit power-law curves since we expect SNR$_{\rm{m}}$ to be proportional to the square root of the integration time{, if} the dominant noise is the photon noise{, or} SNR$_{\rm{m}}$ to be constant{, if} it is the speckle noise.
	 We find that all exponents are close to 0.5{, which} corresponds to the case of photon noise limitation. {SNR$_{\rm{m}}$ rapidly increases with time for the brightest Jupiter models (separations of 0.8 and 2~AU), while the growth is slower for the faintest model (5~AU)}. We use Fig.~\ref{Jupsnrtime} to derive all the exposure times given in this section.
	
\begin{figure}[t]
	\centering
	\includegraphics[width=.47\textwidth]{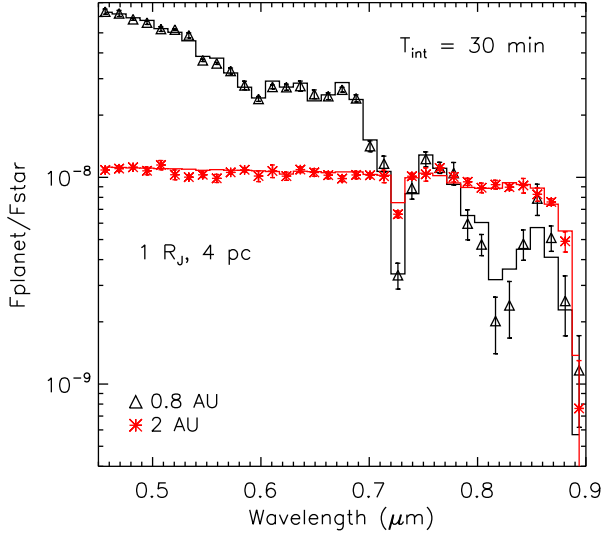}
	\caption{Spectra of Jupiters at 0.8 and 2~AU as they would be measured by SPICES (symbols) and the corresponding model spectra (curves).}
	\label{slopeJup}       
\end{figure}

	As long as astrometric measurements are not available, the orbital inclination is unknown for non-transiting planets detected by RV. We recall that we use an inclination of 60$^\circ$ which is the statistical median for random orientations (Sect.~\ref{simassumptions}). Although direct imaging is able to put constraints on this parameter, it requires {several images in practice} covering the orbit and high SNRs. The fact that planets are easier to observe at quadrature is also not particularly favorable for a precise determination of inclination and then mass. Therefore, in the case of a single {observation, the physical separation of the planet to the star is poorly constrained, particularly since eccentricity could also be significant \citep{Udry2007}.} We can thus confuse a giant planet close to its star with a large planet at large separation if their projected separation is the same on the image. In addition, for an eccentric orbit, a planet's albedo can depend strongly on the orbital position: the planet can be almost cloud-free near perihelion and covered by clouds near aphelion. The spectroscopic characterization could help to break degeneracies in these parameters{, if} the spectral differences are large enough to be detected. Considering the theoretical models, giant planet spectra mostly differ in the blue{, where} Rayleigh scattering dominates for a planet at 0.8~AU. The application of the criterion defined in Eq.~(\ref{criterion}) to giant planet spectra indicates that SNR$_{\rm{r}}$\,$\simeq$\,15 permits to distinguish between the atmospheres of two giants at 0.8 and 2~AU respectively (Table~\ref{tabsnr}). This performance is achieved in $\sim$30~min for a distance of 4~pc (Fig.~\ref{Jupsnrtime}){, which} corresponds to the upper limit at which a separation of 0.8~AU is accessible to SPICES (Table~\ref{septodistsun}). We plot the two spectra as they would be measured by the instrument with 1-$\sigma$ error bars as well as the corresponding models in Fig.~\ref{slopeJup}. As expected, the blue half of the bandwidth is the region where the two spectra can be distinguished with no ambiguity. SPICES will be able to measure Rayleigh scattering and estimate the star-planet separation. {However, we note that these measurements would be possible for a few stars only because of the small angular resolution.}

\begin{figure}[t]
	\centering
	\includegraphics[trim = 8mm 4mm 6mm 10mm, clip, height=0.363\textwidth]{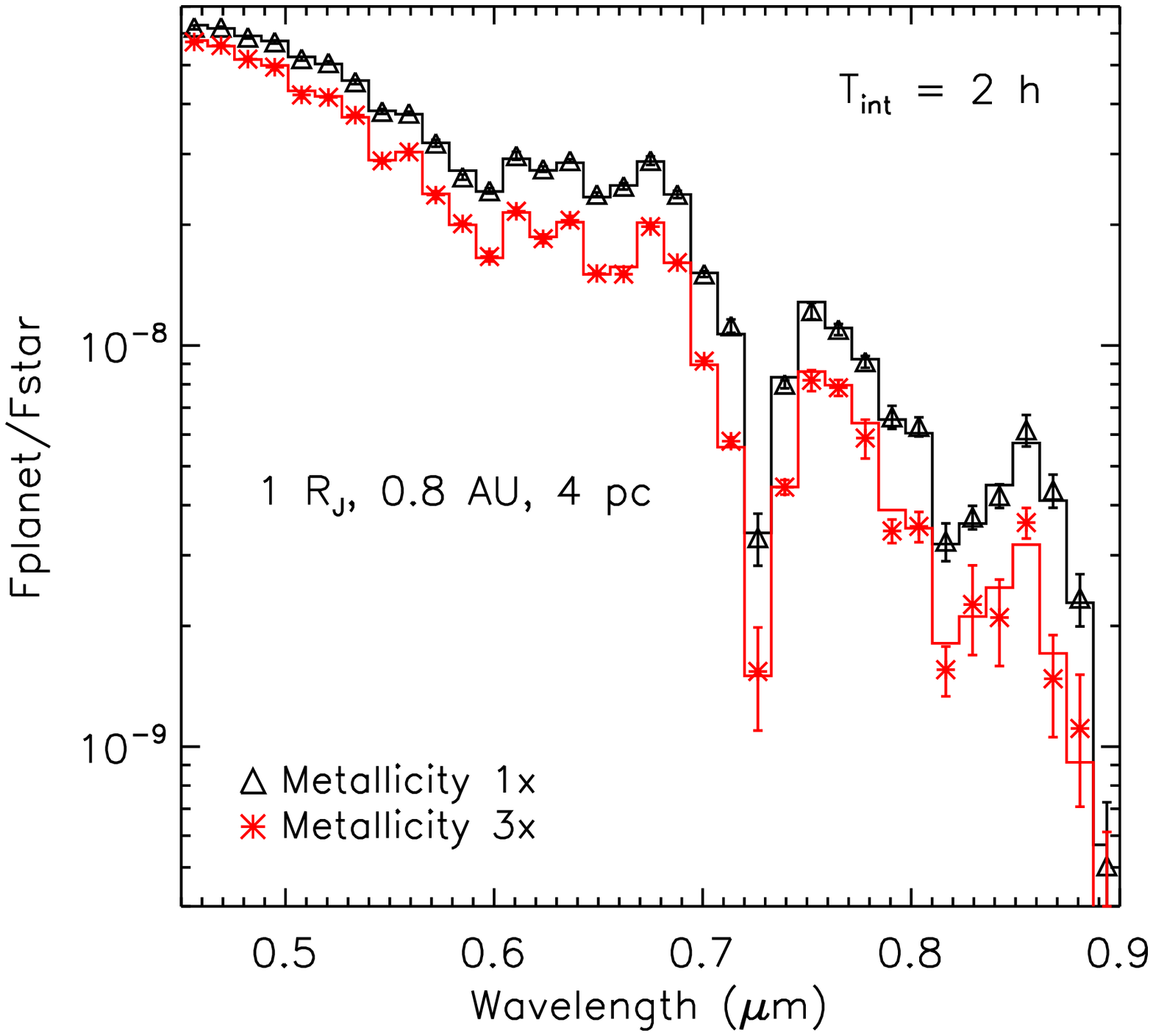} 
	\includegraphics[trim = 8mm 4mm 6mm 10mm, clip, height=0.363\textwidth]{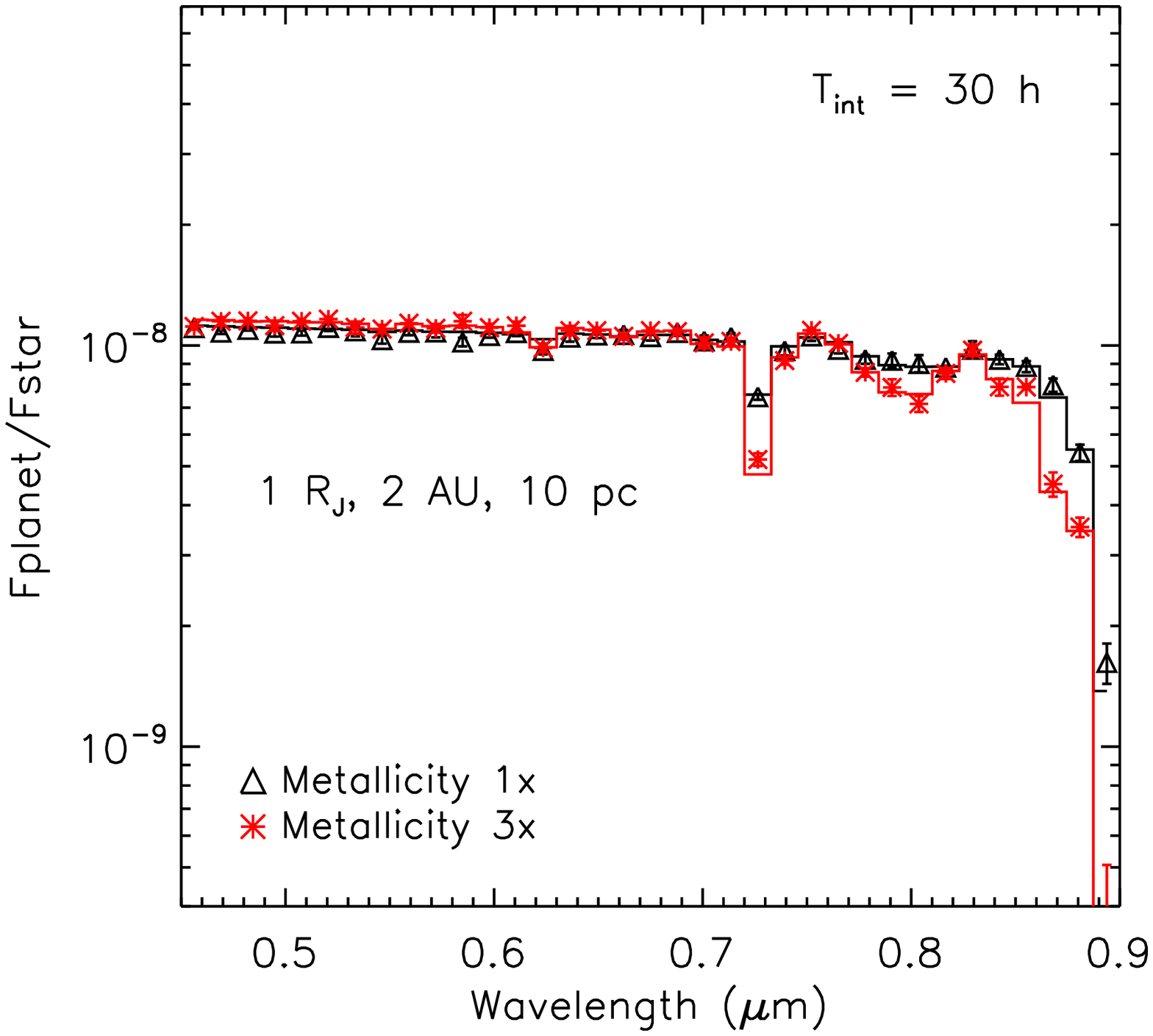} 
	\includegraphics[trim = 8mm 4mm 6mm 10mm, clip, height=0.363\textwidth]{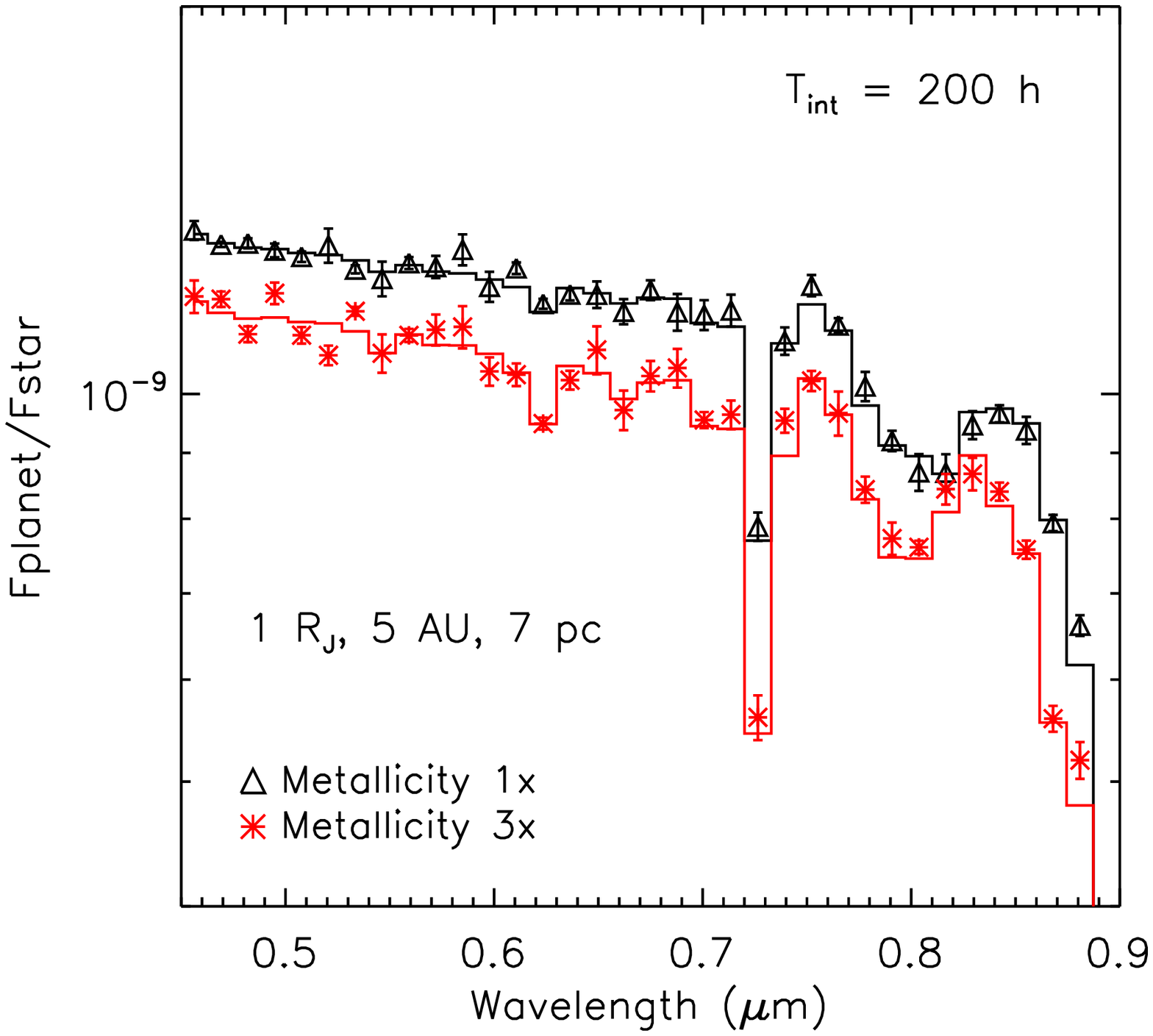}
	\caption{Measured (symbols) and theoretical (lines) spectra of the 0.8-AU (top), 2-AU (middle) and 5-AU (bottom) Jupiter models for 1 and 3 times the solar metallicity. The planet/star contrast scale is identical to Fig.~\ref{slopeJup} for comparison except for the bottom panel.}
	\label{Jupmetalspices}       
\end{figure}

	We now study SPICES' ability to measure {the spectral differences between Jupiter models with 1 and 3 times the solar metallicity for star-planet separations of 0.8, 2 and 5~AU at levels an order of magnitude above the noise}. As indicated in Table~\ref{tabsnr}{, this} requires SNR$_{\rm{r}}$\,$\simeq$\,30 for all separations. For the 0.8-AU Jupiters{, this value is achieved within the distance} for which our instrument can resolve such a planet ($\leq$4~pc). In particular, {an integration time of 2~h satisfies the criterion for a star at 4~pc,} and the differences between the measured spectra are 10 times larger than the noise over the blue half of the spectra (Fig.~\ref{Jupmetalspices}, top panel).
	Recall that the error bars shown in the figures are at 1\,$\sigma$. Similarly, the SNR criterion is satisfied for the 2-AU planets at the {maximum} distance of 10~pc for an exposure time of $\sim$30~h. Metallicity effects are larger in the methane bands for this case and we focus on the 0.73-$\mu$m methane band and the blue edge of the 0.89-$\mu$m deep band to disentangle the spectra (Fig.~\ref{Jupmetalspices}, middle panel).
	For a 5-AU Jupiter, the two metallicity cases are distinguished in 200~h at a distance of 7~pc in the methane bands at 0.62 and 0.73~$\mu$m and on the blue edge of the deep feature at 0.89~$\mu$m (Fig.~\ref{Jupmetalspices}, bottom panel). We note that for the 0.8- and 5-AU cases, the metallicity effects can mimic a radius variation by shifting the whole spectrum{, while} for the 2-AU case they alter the flux specifically in the absorption bands. We conclude that for resolved systems, SPICES will be able to analyze metallicity enhancements as small as a factor of 3 for all 0.8- and 2-AU Jupiter targets around solar-type stars. As planets at 5~AU are fainter, they will be accessible only for G2 stars within 7~pc{, considering the maximum exposure time of $\sim$200~h.}

\begin{figure}[t]
	\centering
	\includegraphics[width=.47\textwidth]{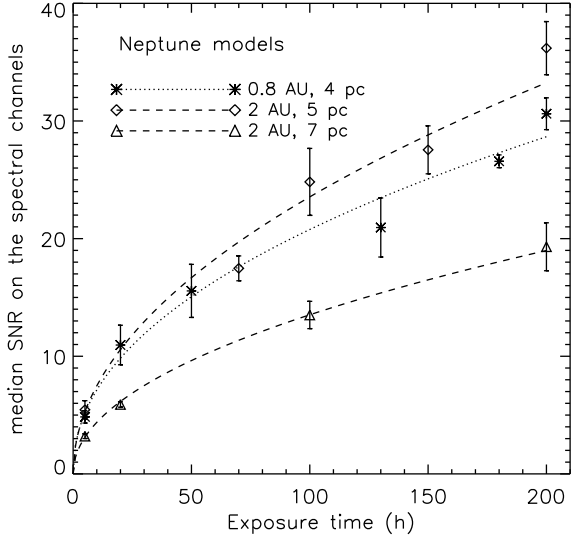}
	\caption{Same as Fig.~\ref{Jupsnrtime} but for the Neptune models of Sect.~\ref{Neptstudy}.}
	\label{Neptsnrtime}       
\end{figure}

  \begin{figure}[t]
	\centering
	\includegraphics[width=.47\textwidth]{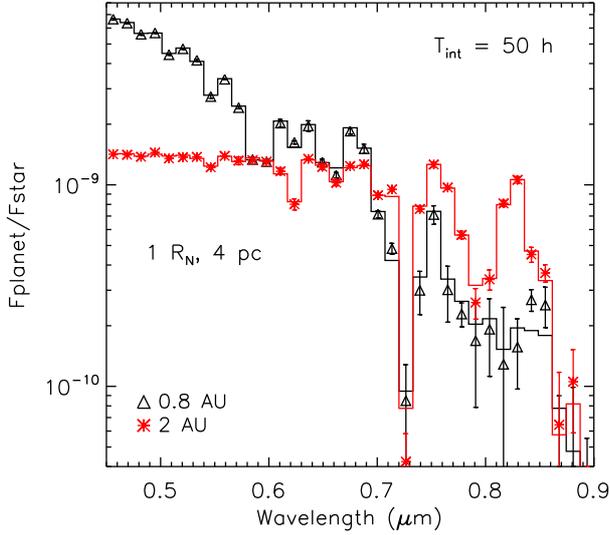}
	\caption{Same as Fig.~\ref{slopeJup} but for the 0.8- and 2-AU Neptunes.}
	\label{slopeNept}       
\end{figure} 

\subsection{Neptune models}
\label{Neptstudy} 
 
\begin{figure}[t]
	\centering
	\includegraphics[width=0.47\textwidth]{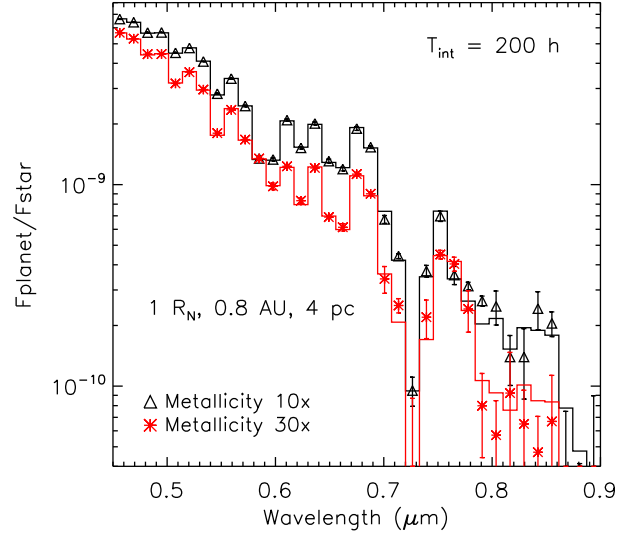} 
	\includegraphics[width=0.47\textwidth]{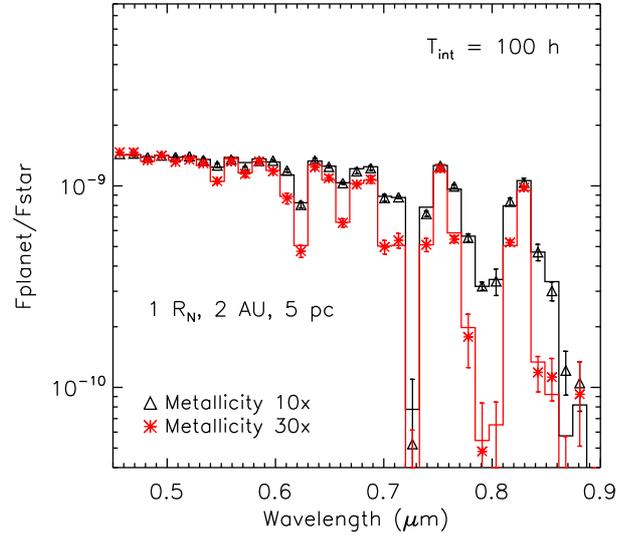}
	\caption{Same as Fig.~\ref{Jupmetalspices} but for the measurement of metallicity effects between 10 and 30 times the solar metallicity of the 0.8-AU (top) and 2-AU (bottom) Neptunes. The vertical scale is identical to Fig.~\ref{slopeNept} for comparison.}
	\label{Neptmetalspices}       
\end{figure}   
 
	Figure~\ref{Neptsnrtime} is similar to Fig.~\ref{Jupsnrtime} but for the Neptune analogs and is used to derive the integration times mentioned below.
	According to Table~\ref{tabsnr}, the distinction between the 0.8- and 2-AU planets requires SNR$_{\rm{r}}$\,$\sim$\,15 in the blue part of the spectral range. Similarly to the {Jupiter spectra, the spectrum of} very close-in Neptunes will feature a negative spectral slope{, due} to Rayleigh scattering{, which} noticeably differs from the nearly flat spectrum of a farther planet. For the maximum distance of 4~pc at which a 0.8-AU Neptune is angularly resolved by SPICES, this value is reached in $\sim$50~h. Therefore, the same analysis can be performed for closer solar-type targets {(as we explained in the previous section, Rayleigh scattering would be measurable for a few objects only)}. The simulated measurements for the 4-pc Neptunes at 0.8 and 2~AU are plotted in Fig.~\ref{slopeNept}. The 0.8- and 2-AU Neptune {spectral} differences can be measured for $\lambda$\,$<$\,0.58~$\mu$m{, whereas} the {spectral measurements are degenerate} for redder wavelengths given the noise level.
 
  	We then test if SPICES can distinguish metallicity effects between Neptunes of 10 and 30 times the solar metallicity (Table~\ref{planetmodels}). We recall that these values are those studied by \citet{Cahoy2010}. {The required SNR$_{\rm{r}}$ is $\sim$30 over the full spectral range for disentangling the 0.8-AU spectra.} It is $\sim$25 in the methane bands for the 2-AU planet spectra (Table~\ref{tabsnr}).
	For the 0.8-AU planets{, a} $\sim$200-h exposure is requested to measure metallicity variations {for distances} as far as 4~pc. The spectral differences are detected over the blue channels up to $\sim$0.65~$\mu$m (Fig.~\ref{Neptmetalspices}, top panel).
	For a separation of 2~AU, the bottom panel of Fig.~\ref{Neptmetalspices} shows that the metallicity signatures mainly impact the methane bands at 0.62 and 0.66~$\mu$m as well as on the edges of the 0.79-$\mu$m deep band. We find that SPICES can distinguish the spectra for stars within $\sim$6~pc.  

\subsection{2.5-R$_{E}$ planet models}
\label{SEstudy}

  \begin{figure}[t]
	\centering
	\includegraphics[width=.47\textwidth]{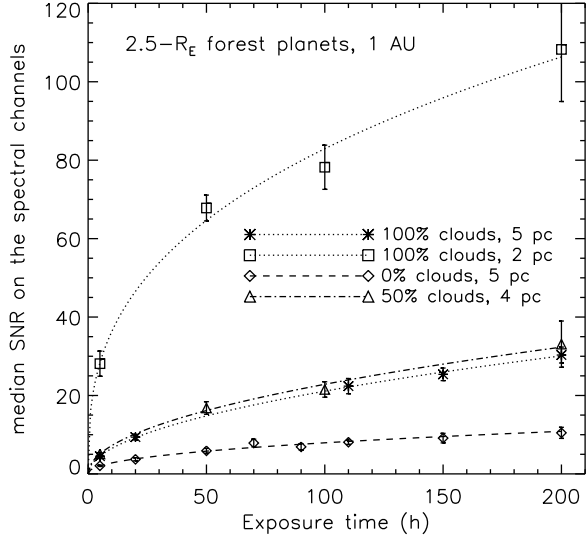}
	\caption{Same as Fig.~\ref{Jupsnrtime} but for the 2.5-R$_{\rm{E}}$ planet models discussed in Sect.~\ref{SEstudy}.}
	\label{SEsnrtime}       
\end{figure} 

 \begin{figure}[t]
	\centering
	\includegraphics[width=.47\textwidth]{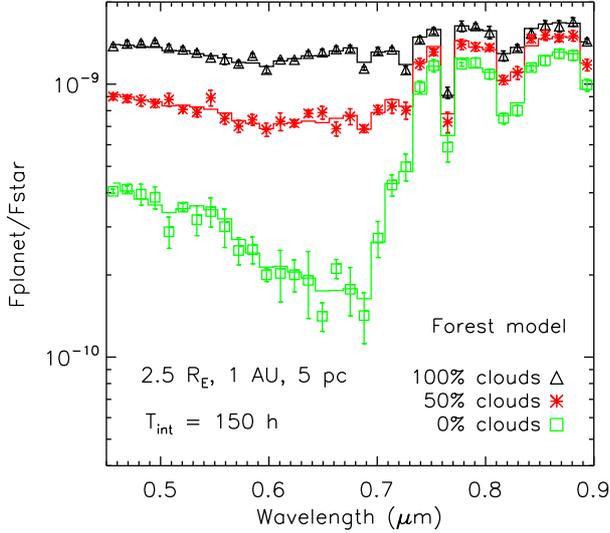}
	\caption{Measured (symbols) and model (lines) spectra of 2.5-R$_{\rm{E}}$ forest planets at 1~AU for cloud coverage of 0, 50 and 100\%.}
	\label{SEcloudcover}       
\end{figure}

	In this section, we analyze SPICES' capability to measure the properties of super-Earths, and in particular the effects of cloud and surface coverage for planets at 1~AU (their parameters are given in Table~\ref{planetmodels}). Figure~\ref{SEsnrtime} represents the SNR$_{\rm{m}}$ evolution as a function of the exposure time for several models of planets discussed in this section. The labels ``clear'' and ``cloudy'' refer to the models with 0\% clouds and 100\% clouds respectively. The star distance is limited to 5~pc (Table~\ref{septodistsun}) to allow the planet to be angularly separated from the star. {A few G-type stars would thus be accessible to SPICES.}

\begin{figure}[t]
	\centering
	\includegraphics[width=0.47\textwidth]{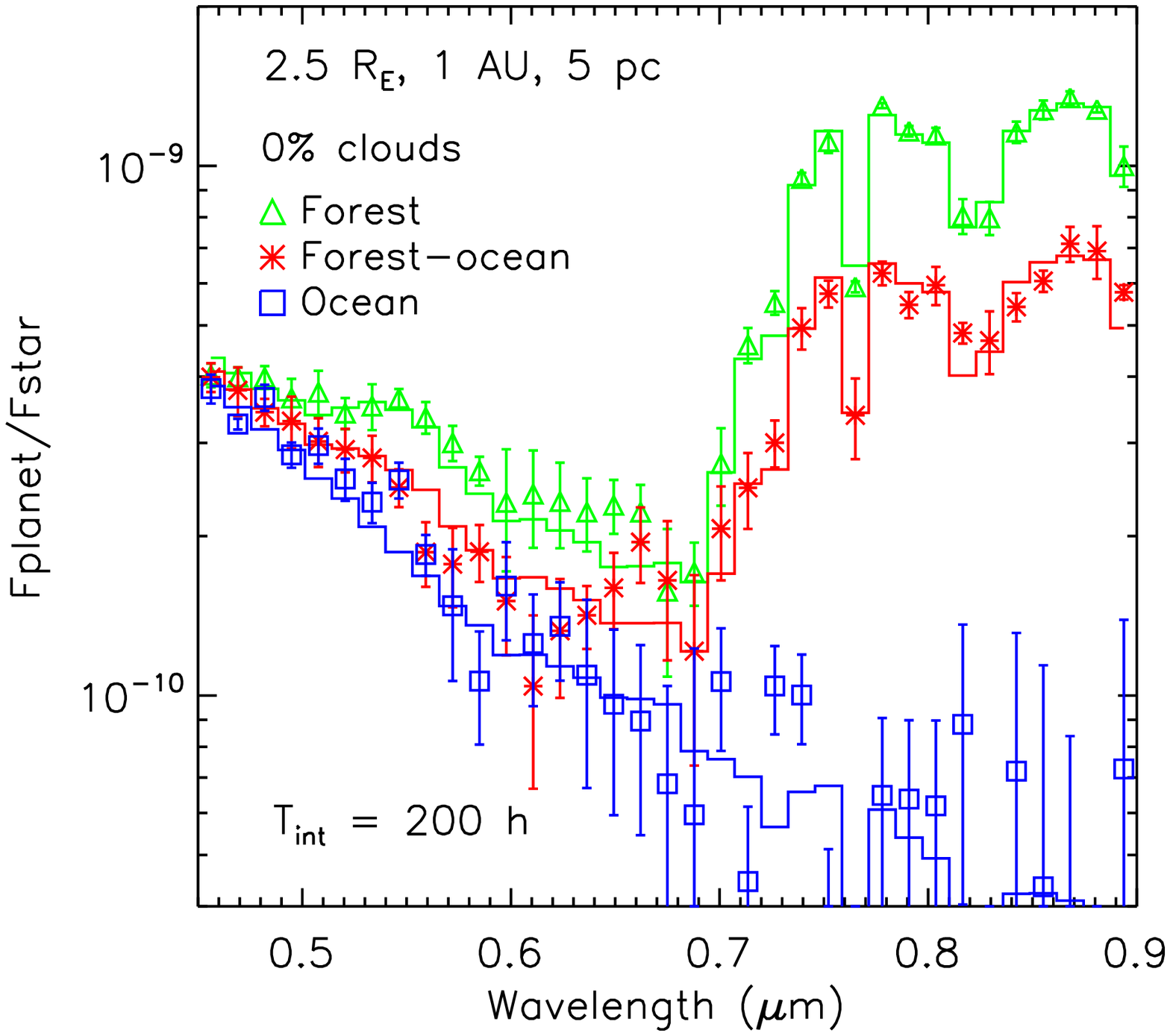} 
	\includegraphics[width=0.47\textwidth]{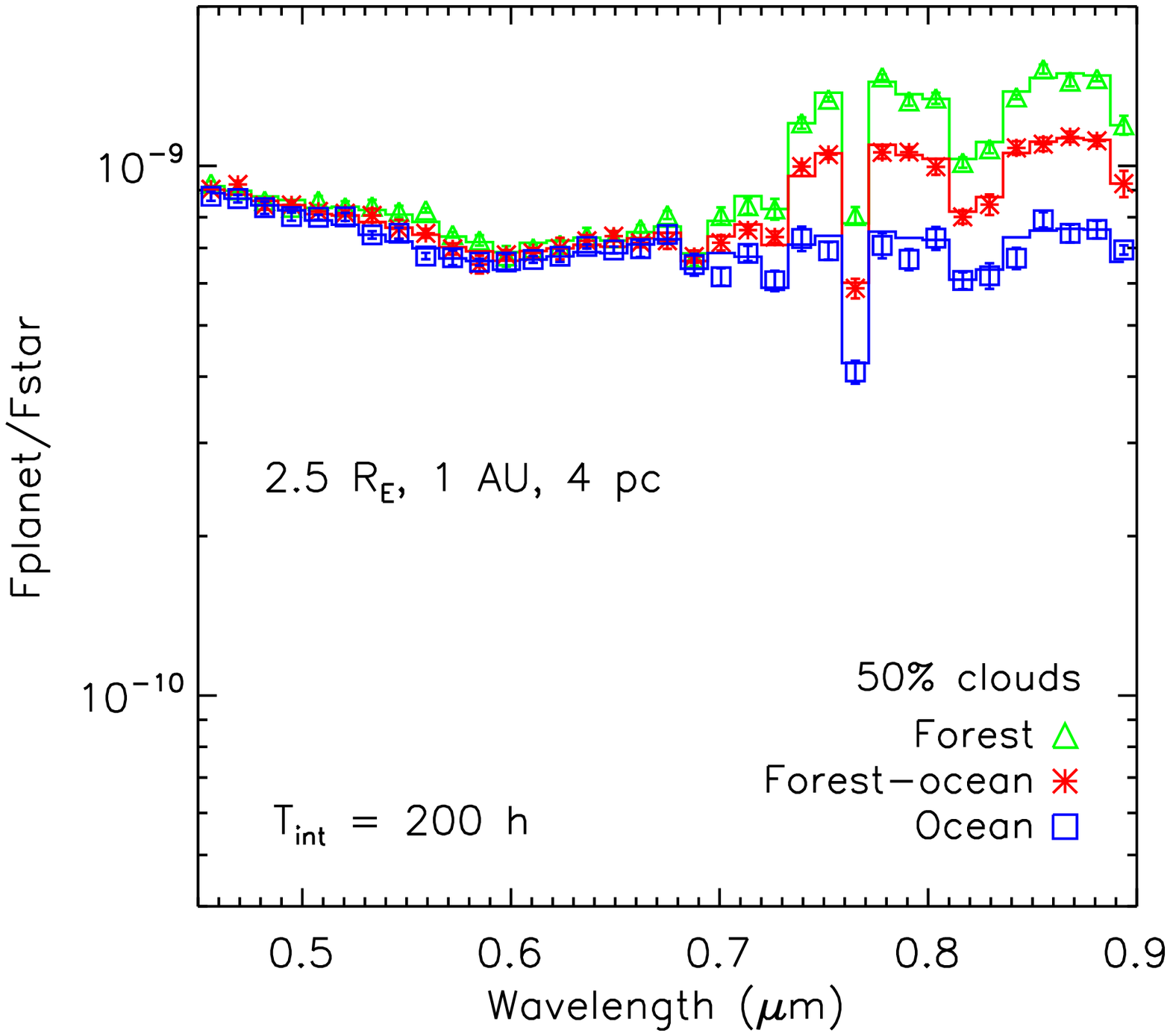}
	\caption{Measured spectra of 2.5-R$_{\rm{E}}$ planets for the clear (top) and 50\% cloudy (bottom) models with surfaces composed of forest, 50\% forest-50\% ocean and ocean with the corresponding theoretical spectra. The vertical scale is identical to Fig.~\ref{SEcloudcover} for comparison.}
	\label{SEsurfcover}       
\end{figure}

	We first investigate the impact of cloud coverage for both forest and ocean surfaces. The influence of clouds is more important in the blue (Fig.~\ref{SEcloudcover}{, for} surfaces entirely covered with forest), because of the strong ``red edge'' reflection in the red, and SNR$_{\rm{r}}$\,$\sim$\,25 (Table~\ref{tabsnr}) is needed to measure the difference between 0, 50 and 100\% clouds. At the maximum distance of 5~pc{, this} performance is met in $\sim$150~h for a super-Earth entirely covered with forest around a G2 star (Figs.~\ref{SEsnrtime} and \ref{SEcloudcover}). Systems closer than 5~pc will also be accessible for the characterization of the cloud coverage. We do not compute the exposure time for surfaces covered by ocean{, but} it will be similar since SNR$_{\rm{r}}$\,$\sim$\,25 (Table~\ref{tabsnr}) and the cloudy spectra are nearly identical for both surfaces (green and blue dotted lines in Fig.~\ref{Stammodels}). Figure~\ref{SEcloudcover} also shows that the main atmospheric gases (O$_2$ and H$_2$O) are quite well retrieved whatever the cloud coverage (SNR$_{\rm{m}}$ $>$\,5 and $>$\,15 respectively){. The} broad ozone signature is mostly detectable when the cloud coverage is large (SNR$_{\rm{m}}$\,$\simeq$\,23). We recall that the band wavelengths are given in Table~\ref{SEmolecules}.
	
	Information about the planet surface can be obtained for moderate cloud coverage{, because} it produces noticeable signatures (Fig.~\ref{SEsurfcover}). The differences between surface types are larger for a clear atmosphere{, especially} in the half red part of the spectral range. We consider three generic cases of {planets with an ocean, an equally mixed surface of ocean and forest, and a forest.} To separate these cases, SNR$_{\rm{r}}\,$$\sim$\,12 (Table~\ref{tabsnr}) is required on the forest model and SNR$_{\rm{m}}$\,$>$\,12 is achieved in a $\sim$200-h observation for the worst case of a G2 star at 5~pc (Fig.~\ref{SEsurfcover}, top panel). SPICES can distinguish these three cases for any terrestrial planets on a 1-AU orbit within 5~pc. If we consider 50\% cloud coverage, surface effects are more difficult to distinguish and would require SNR$_{\rm{r}}$\,$\sim$\,30 in the red part (Table~\ref{tabsnr}). The exposure time exceeds the limit of 200~h for a target at 5~pc{, which rather limits} the sample to 4~pc (Fig.~\ref{SEsurfcover}, bottom panel). Molecular oxygen and water absorptions as well as the ``red edge'' can still be measured. 
On the contrary, 100\% cloud coverage definitely prevents the identification of surfaces since it would require SNR$_{\rm{r}}$\,$\sim$\,220{. This} performance that is out of reach of a small telescope like SPICES in a reasonable amount of time. In the favorable case of a Sun-like star at 2~pc ($\alpha$~Cen A is the sole known case), the instrument achieves SNR$_{\rm{m}}$\,$\sim$\,110 {and} allow to distinguish between cloudy planets totally covered with ocean and forest respectively.

\section{Potential targets}
\label{overallperf}
\begin{table}[t]
 	\caption{Maximum star distance at which SPICES resolves the planet separation at the central wavelength of the bandwidth in the case of a M0 host star.}
	\begin{center}
 	\begin{tabular}{c c}
	 \hline\hline
 	Planet separation (AU) & Star distance (pc) \\
	\hline
 	0.24 & 1.2 \\
 	0.3 & 1.5 \\
 	0.6 & 3 \\
 	1.5 & 7.5 \\
 	3 & 15 \\
 	\hline 
 	\end{tabular}
 	\label{septodistM}
 	\end{center}
 \end{table}
 
 \begin{figure}[t]
	\centering
	\includegraphics[trim = 8mm 4mm 6mm 12mm, clip, height=0.349\textwidth]{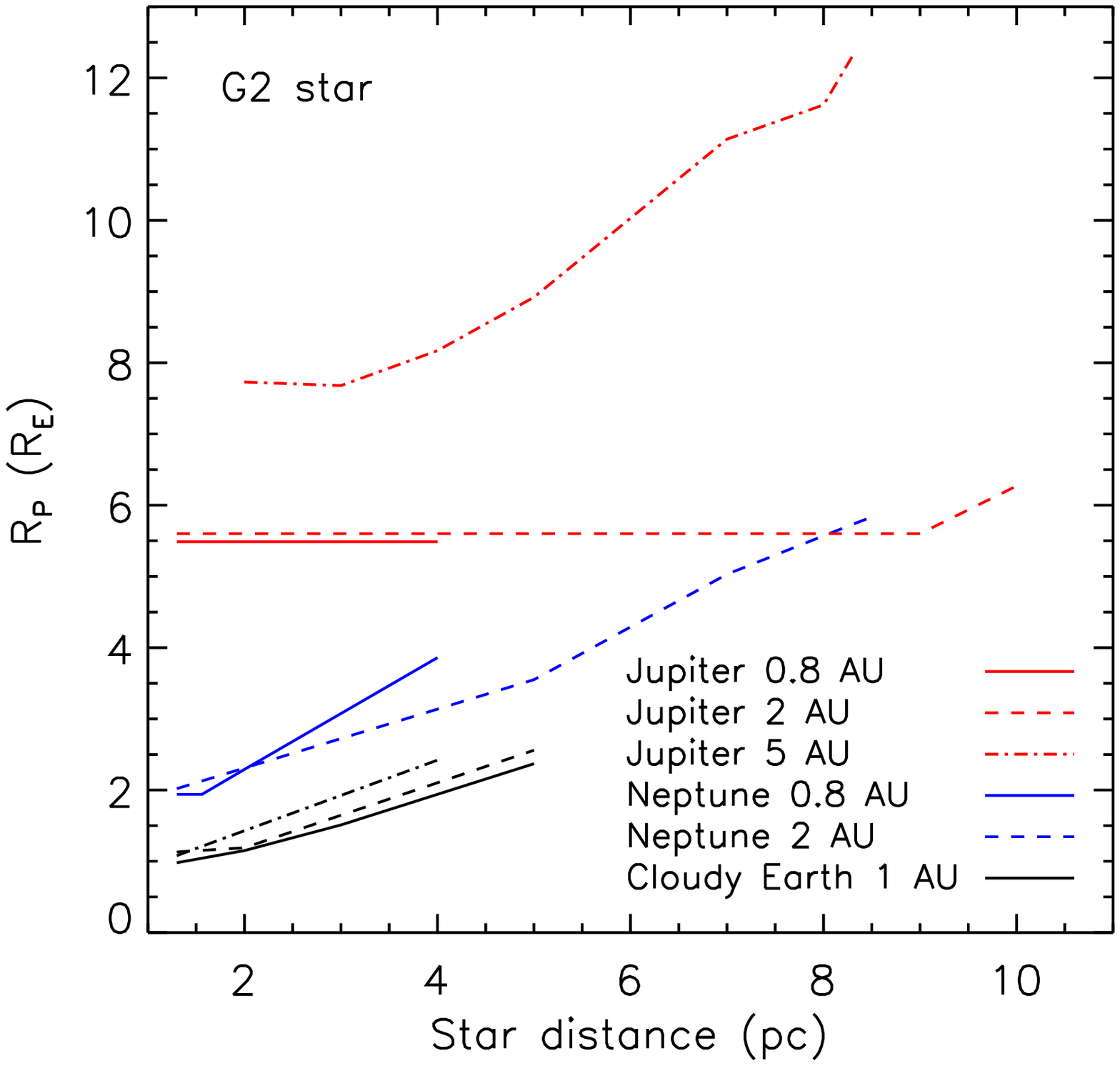} 
	\includegraphics[trim = 8mm 4mm 6mm 12mm, clip, height=0.349\textwidth]{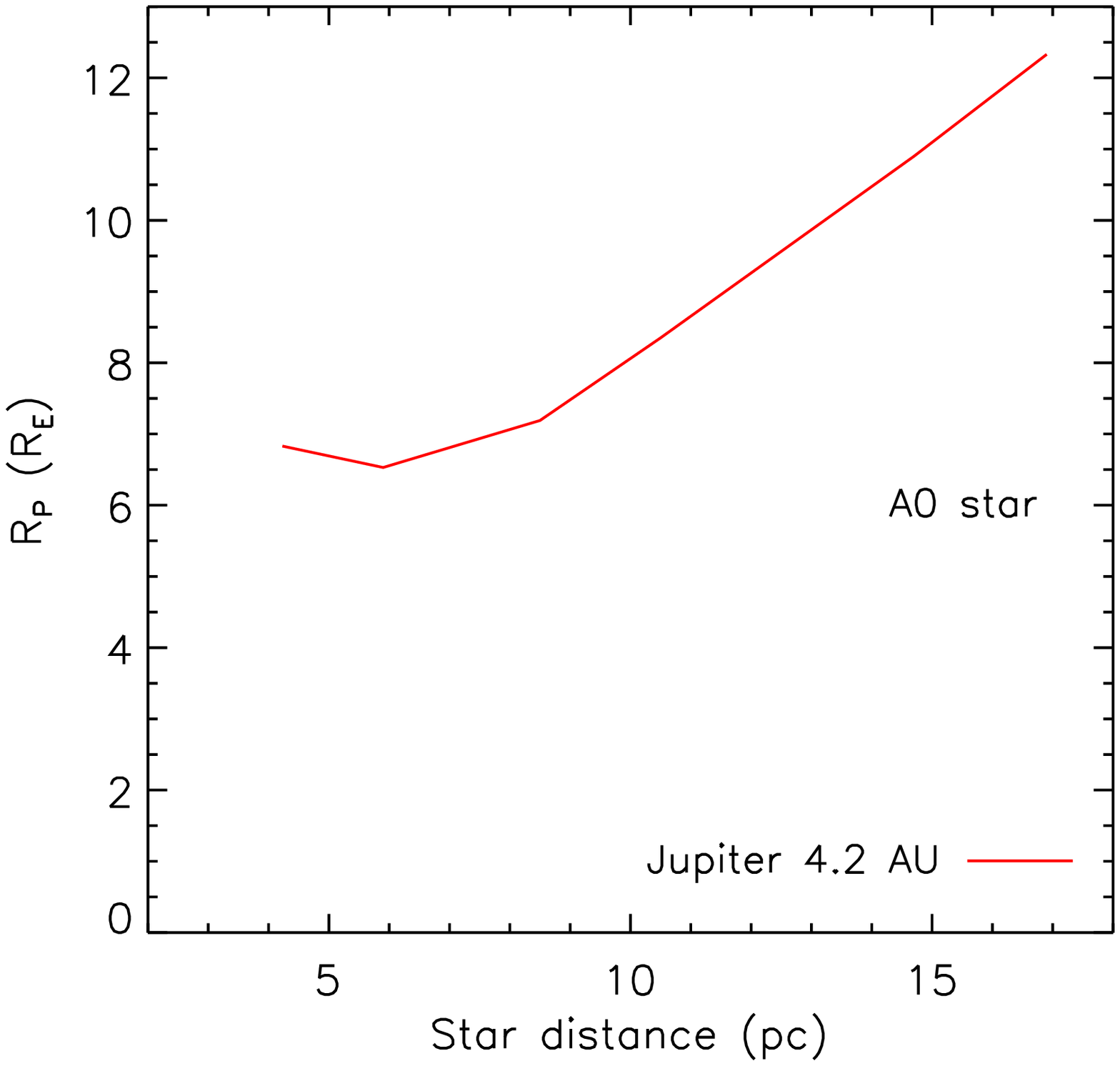} 
	\includegraphics[trim = 8mm 4mm 6mm 12mm, clip, height=0.349\textwidth]{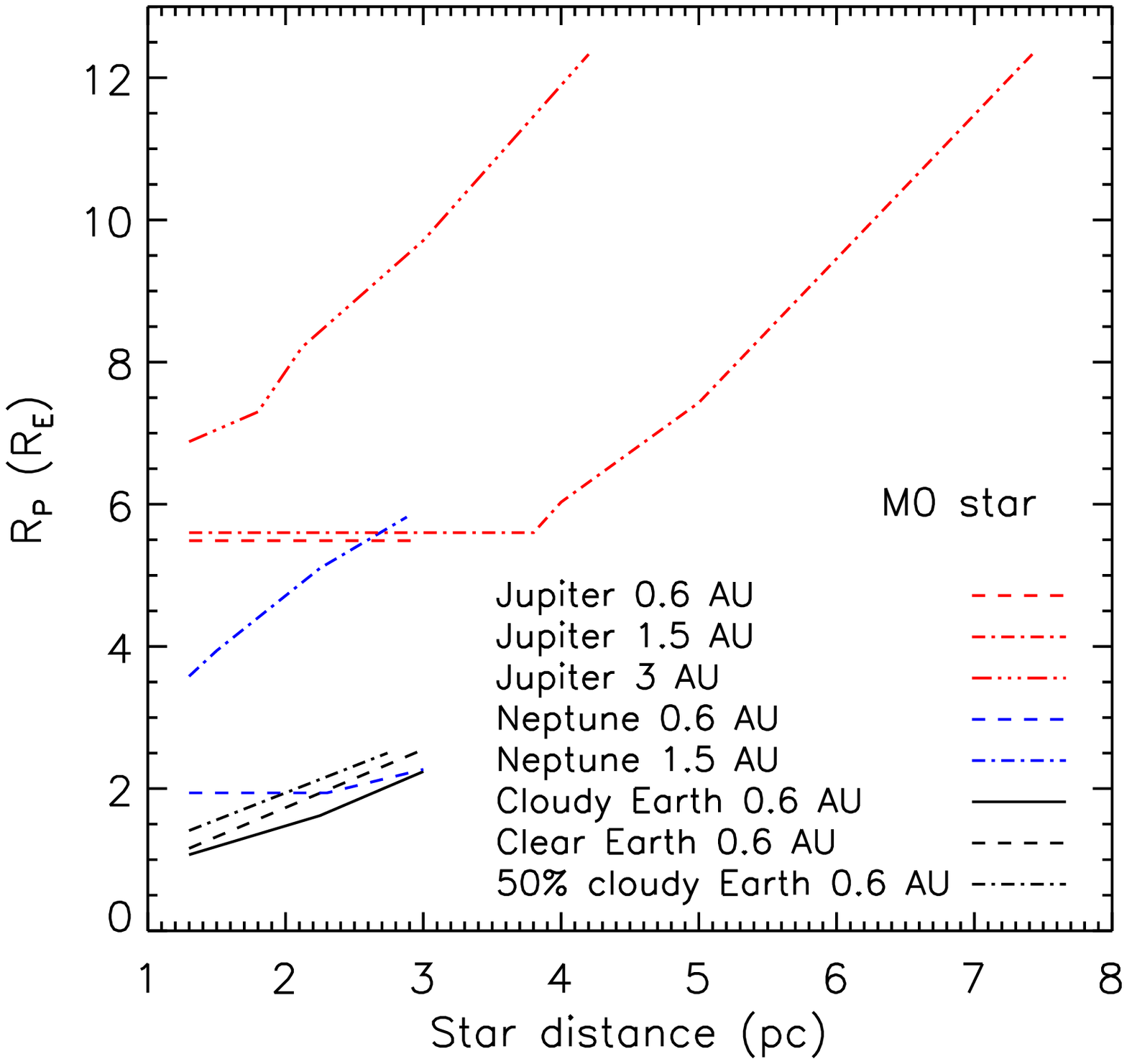}
	\caption{Minimum radius of model planets for which characterization is possible (see text) for different star types: G2 (top), A0 (middle) and M0 (bottom). We slightly offset for clarity the curves of the 0.8-AU Jupiter (top) and 0.6-AU Jupiter (bottom). Note that for lack of space, we do not indicate the labels for all Earths in the top panel{, but} they are the same as those of the bottom panel except for the planet separation.}
	\label{sumperfspectro}       
\end{figure}

\begin{table*}[t]
 \caption{SPICES performance summary for exoplanet spectroscopic characterization. We strongly encourage the reader to refer to the rest of the paper to access the assumptions and the detail of the studies.}
 \begin{center}
 \begin{tabular}{c l c c c c}
 \hline\hline
 Planet & Probed property & \multicolumn{3}{c}{Stellar type} & Sections \\
 & & \multicolumn{1}{c}{A0} & \multicolumn{1}{c}{G2} & \multicolumn{1}{c}{M0} \\
 \hline
 \multirow{6}{*}{Jupiter}
 & \multirow{3}{3.7 cm}{Rayleigh scattering}
 & 1~AU\,$\leq$\,a\,$\leq$\,4.2~AU & 0.25~AU\,$\leq$\,a\,$\leq$\,0.8~AU &\multirow{3}{*}{$-$} & \multirow{6}{*}{\ref{jupneptmodels}, \ref{Jupstudy}, and \ref{overallperf}} \\
 & & 5~pc\,$\leq$\,d\,$\leq$\,{17}~pc & 1.3~pc\,$\leq$\,d\,$\leq$\,4~pc & \\
 \vspace{0.13cm}
 & & 0.5~R$_{\rm{J}}$\,$\leq$\,R$_{\rm{p}}$\,$\leq$\,1.1~R$_{\rm{J}}$ & 0.5~R$_{\rm{J}}$\,$\leq$\,R$_{\rm{p}}$\,$\leq$\,1.1~R$_{\rm{J}}$ & \\
 & \multirow{3}{3.7 cm}{Solar metallicity 1x/3x}
 & 1~AU\,$\leq$\,a\,$\leq$\,4.2~AU & 0.25~AU\,$\leq$\,a\,$\leq$\,5~AU & 0.25~AU\,$\leq$\,a\,$\leq$\,3~AU \\
 & & 5~pc\,$\leq$\,d\,$\leq$\,{17}~pc & 1.3~pc\,$\leq$\,d\,$\leq$\,10~pc & 1.3~pc\,$\leq$\,d\,$\leq$\,7.5~pc  \\
 & & 0.6~R$_{\rm{J}}$\,$\leq$\,R$_{\rm{p}}$\,$\leq$\,1.1~R$_{\rm{J}}$ & 0.5~R$_{\rm{J}}$\,$\leq$\,R$_{\rm{p}}$\,$\leq$\,1.1~R$_{\rm{J}}$ & 0.5~R$_{\rm{J}}$\,$\leq$\,R$_{\rm{p}}$\,$\leq$\,1.1~R$_{\rm{J}}$ \\
 \hline
  \multirow{6}{*}{Neptune}
 & \multirow{3}{3.7 cm}{Rayleigh scattering}
 & \multirow{3}{*}{$-$} & 0.25~AU\,$\leq$\,a\,$\leq$\,0.8~AU & \multirow{3}{*}{$-$} & \multirow{6}{*}{\ref{jupneptmodels}, \ref{Neptstudy}, and \ref{overallperf}}\\
 & & & 1.3~pc\,$\leq$\,d\,$\leq$\,4~pc & \\
  \vspace{0.13cm}
 & & & 0.5~R$_{\rm{N}}$\,$\leq$\,R$_{\rm{p}}$\,$\leq$\,1.5~R$_{\rm{N}}$ & \\
 & \multirow{3}{3.7 cm}{Solar metallicity 10x/30x}
 & \multirow{3}{*}{$-$} & 0.25~AU\,$\leq$\,a\,$\leq$\,2~AU & 0.25~AU\,$\leq$\,a\,$\leq$\,{1.5}~AU \\
 & & & 1.3~pc\,$\leq$\,d\,$\leq$\,{8.5}~pc & 1.3~pc\,$\leq$\,d\,$\leq$\,3~pc \\
 & & & 0.5~R$_{\rm{N}}$\,$\leq$\,R$_{\rm{p}}$\,$\leq$\,1.5~R$_{\rm{N}}$ & 0.5~R$_{\rm{N}}$\,$\leq$\,R$_{\rm{p}}$\,$\leq$\,1.5~R$_{\rm{N}}$ \\
   \hline
 \multirow{9}{*}{Earth}
 & \multirow{3}{3.7 cm}{Cloud coverage 0/50/100\%}
 & \multirow{3}{*}{$-$} & 0.25~AU\,$\leq$\,a\,$\leq$\,1~AU & 0.25~AU\,$\leq$\,a\,$\leq$\,0.6~AU & \multirow{9}{*}{\ref{earthmodels}, \ref{SEstudy}, and \ref{overallperf}}\\
 & & & 1.3~pc\,$\leq$\,d\,$\leq$\,5~pc & 1.3~pc\,$\leq$\,d\,$\leq$\,3~pc \\
  \vspace{0.13cm}
 & & & 1~R$_{\rm{E}}$\,$\leq$\,R$_{\rm{p}}$\,$\leq$\,2.5~R$_{\rm{E}}$ & 1.1~R$_{\rm{E}}$\,$\leq$\,R$_{\rm{p}}$\,$\leq$\,2.5~R$_{\rm{E}}$ \\
 & \multirow{3}{3.7 cm}{Forest coverage 0/50/100\% clouds\,=\,0\%}
 & \multirow{3}{*}{$-$} & 0.25~AU\,$\leq$\,a\,$\leq$\,1~AU & 0.25~AU\,$\leq$\,a\,$\leq$\,0.6~AU \\
 & & & 1.3~pc\,$\leq$\,d\,$\leq$\,5~pc & 1.3~pc\,$\leq$\,d\,$\leq$\,3~pc \\
  \vspace{0.13cm}
 & & & 1.1~R$_{\rm{E}}$\,$\leq$\,R$_{\rm{p}}$\,$\leq$\,2.5~R$_{\rm{E}}$ & 1.2~R$_{\rm{E}}$\,$\leq$\,R$_{\rm{p}}$\,$\leq$\,2.5~R$_{\rm{E}}$ \\
 & \multirow{3}{3.7 cm}{Forest coverage 0/50/100\% clouds\,=\,50\%}
 & \multirow{3}{*}{$-$} & 0.25~AU\,$\leq$\,a\,$\leq$\,1~AU & 0.25~AU\,$\leq$\,a\,$\leq$\,0.6~AU \\
 & & & 1.3~pc\,$\leq$\,d\,$\leq$\,4~pc & 1.3~pc\,$\leq$\,d\,$\leq$\,2.8~pc \\
 & & & 1.1~R$_{\rm{E}}$\,$\leq$\,R$_{\rm{p}}$\,$\leq$\,2.5~R$_{\rm{E}}$ & 1.4~R$_{\rm{E}}$\,$\leq$\,R$_{\rm{p}}$\,$\leq$\,2.5~R$_{\rm{E}}$ \\
 \hline 
 \end{tabular}
 \label{tabsummary}
 \end{center}
 \end{table*}

	In this section, we extend the results obtained above in order to determine the minimum planet radius accessible to SPICES for each planet category studied in the previous section. We also study the volume of the planet sample which can be characterized (metallicity, cloud coverage, surface type) with SPICES. The models used are the gas giants with metallicity 1, the ice giants with metallicity 10{, the cloudy forest Earths for the cloud coverage measurement, and the forest Earths with} 0 and 50\% clouds for the surface type analysis. We consider that the planet flux increases as the square of its radius and that the atmosphere composition and structure remain the same. We set the following values of SNR$_{\rm{r}}$: 30 for all Jupiters and Neptunes and 25, 12 and 30 respectively for the Earths (Table~\ref{tabsnr}). We assume a maximum exposure time of 200~h and three host stars of type G2, A0 and M0. We carry out this study for the four models of Jupiter and Neptune at 0.8, 2, 5 and 10~AU and the Earth model at 1~AU. We recall that these separations are relevant for a G2 star and the corresponding separations for A0 and M0 stars are given in Table~\ref{sepstartype}. We consider the following radius ranges for the planets: 
\begin{itemize}
	\item R$_{\rm{p}}$$\,\leq\,$2.5~R$_{\rm{E}}$ for the Earths \citep{Grasset2009}.
	\item 0.5~R$_{\rm{N}}$\,$\leq\,$R$_{\rm{p}}$$\,\leq\,$1.5~R$_{\rm{N}}$ for the Neptunes. The lower limit is the radius for the maximum mass we consider for the Earths (10~M$_{\rm{E}}$). The upper value corresponds to a maximum mass of 30~M$_{\rm{E}}$ in the mass-radius relation of \citet{Fortney2007}.
	\item 0.5~R$_{\rm{J}}$\,$\leq\,$R$_{\rm{p}}$$\,\leq\,$1.1~R$_{\rm{J}}$ for the Jupiters. We derive the lower value from the upper mass limit we assume for the Neptunes. The upper limit corresponds to the maximum radius of a 4.5-Gyr planet at 1~AU in \citet{Fortney2007}. We note that transit measurements find potentially larger planets (up to 1.4~R$_{\rm{J}}$){, but very close to the star. The inflated radius of these planets could be explained by additional mechanisms to stellar radiation \citep{Fortney2010}.}
\end{itemize}

	Figure~\ref{sumperfspectro} gives the minimum planet radius (in R$_{\rm{E}}$ units) for which SPICES is able to characterize the metallicity of giants or the cloud and forest coverage of terrestrial planets. The colors refer to the planet type (Jupiter, Neptune or Earth) and the line styles to the planet model (separation for the giants and cloud coverage for the Earths).
	
	For G2 stars (Fig.~\ref{sumperfspectro}, top panel), SPICES reaches the lowest radius we consider (0.5~R$_{\rm{J}}$\,$\sim$\,5.5~R$_{\rm{E}}$) for the 0.8- and 2-AU Jupiters{, for} all stars within respectively 4 and 10~pc. {About twenty G stars could be probed for Jupiters at 2~AU, while a few stars could be searched for Jupiters at 0.8~AU.} For a Jupiter at 5~AU, the upper limit in radius (1.1~R$_{\rm{J}}$\,$\sim$\,12~R$_{\rm{E}}$) is reached when the star is at 8.5~pc{. The} radius linearly decreases as the star distance decreases down to 8~R$_{\rm{E}}$ at 4~pc (flux proportional to the square ratio of the planet radius to the star distance). At shorter distances, the deviations from the linearity are due to the speckle noise. When the star distance decreases, the angular separation of the star-planet system increases and the level of the speckle background increases, especially towards the edges of the dark hole (Fig.~\ref{perfnonoiseallchan}).
	As for Neptune-like planets, the minimum radius scales nearly linearly with distance from $\sim$ 2 to 4~R$_{\rm{E}}$ at 0.8~AU{, and} from 2 to 6~R$_{\rm{E}}$ at 2~AU. SPICES can characterize cloudy and clear telluric planets at 1~AU around {a few} G2 stars up to 5~pc (IWA limitation){. Potentially, it is} able to reach Earth-size planets for very close stars like $\alpha$~Cen A ($\sim$\,1.3~pc).
	
	For A0 stars (middle panel), the only planet SPICES can study among the planetary atmosphere models we consider is the cloud-free Jupiter. We already noted the lack of models for separations smaller than $\sim$4~AU (Sect.~\ref{noisyperf}). The shortest separation available is 4.2~AU (Table~\ref{sepstartype}), which is angularly resolved by SPICES when the star is closer than 21~pc. For this planet, the upper limit of the detectable radius roughly follows a linear law when the star distance increases from 9 to 17~pc. {We note that a dozen of A stars are located within the upper limit of the range. For the closest} stars, the detector read-out noise impacts the performance.

	If we focus on M0 stars (bottom panel), we note that the cloud-free giant planets (0.24~AU, Table~\ref{sepstartype}) are inaccessible due to the IWA (Table~\ref{septodistM}). For the Jupiters at 0.6 and 1.5~AU (Table~\ref{sepstartype}), we determine that SPICES allows the analysis of the smallest radius up to 3 and 4~pc respectively. The radius upper limit is achieved at $\sim$7.5 and $\sim$4~pc for separations of 1.5 and 3~AU respectively. {Jupiters at 1.5~AU could be accessible for $\sim$100 M stars.} For the 3-AU Jupiter case, the minimum radius that is detectable decreases as a linear function towards short distances.
	The minimum radius (0.5--0.6~R$_{\rm{N}}$) is feasible for 0.6-AU Neptunes (water clouds) at distances $\leq$\,3~pc. {SPICES can also access icy giants as far as 1.5~AU within $\sim$3~pc ($\sim$10 M stars satisfy this constraint).} Finally, telluric planets at 0.3~AU (the luminosity-scaled distance equivalent to 1~AU from a Sun-like star{, Table~\ref{sepstartype}}) are proven to be difficult to detect with SPICES and only reachable within 1.5~pc (Table~\ref{septodistM}). However, since M stars are of interest in terms of contrast{, we extrapolate the luminosity of the model at 0.3~AU out to 0.6~AU, using} a simple inverse square power law. This is obviously not rigorous, as the atmosphere characteristics would change{, but} it gives a rough estimation for one of the most challenging science cases of the mission. With such an assumption, we find SPICES can characterize telluric planets with radii as small as 1--2~R$_{\rm{E}}$ within 3~pc.

Table~\ref{tabsummary} summarizes SPICES' spectrophotometric performance for all planets and stellar types we considered in terms of star-planet separation (a), star distance (d), and planet radius (R$_{\rm{p}}$) ranges. As explained in Sect.~\ref{jupneptmodels}, exoplanetary atmosphere albedo can drastically change with the star-planet separation. The minimum separation at which SPICES can detect a planet thus depends on the coronagraph IWA, the closest observable star for each spectral type and the planetary atmosphere albedo. Because of the lack of atmosphere models for close-in planets, we cannot derive a precise lower limit for the star-planet separation. However, we give a value that corresponds to the IWA limitation for the closest stars.
For the star distance range, the lower limit is set to either the distance of the closest star of the considered spectral type{, or} the distance below which the planet is fainter than the speckle noise. The upper limit is either the maximum distance at which the planet separation is resolved by SPICES{, or} the distance beyond which the planet is fainter than the photon noise.

\section{Conclusions}
	In this paper{, we} presented an analysis and an estimation of the performance of SPICES, a small coronagraphic space mission operating in the visible. From numerical simulations of the instrument concept given in \citet{Boccaletti2012a}, we first determined that the contrast reached by the instrument meets the top-level requirements ($\sim$10$^{-9}$ at 2~$\lambda/D$ and $\sim$10$^{-10}$ at 4~$\lambda/D$).
	Then, we analyzed the impact of different sources of noise: detector read-out noise, exo-zodiacal intensity and photon noise considering several stellar types. We confirmed previous analyses that exodisks might be a major limitation for the characterization of faint planets{, like} Neptunes and super-Earths{, because} it biases the planet photometry (for disk intensity $\gtrsim$1~zodi) and adds photon noise ($\gtrsim$a few 10~zodis). Exodisk intensity calibration and target selection will be necessary to reduce this limitation. Using planet models calculated for a Sun-like star and assuming flux conservation, we found that the brightest stars (types AF) would not be the most favorable targets for SPICES.
	For instance, only planets with clear atmospheres will be detectable around A0 stars within $\sim$20~pc. On the contrary, stars of types GKM will allow the detection of a large variety of planets as far as $\sim$12~pc for a G2 star and $\sim$7.5~pc for a M0 star: cloud-free, water-cloud and ammonia-cloud Jupiters, cloud-free and water-cloud Neptunes and telluric planets. After this general study of contrast performance, we focused on spectrometric abilities on planets around a solar-type star. We defined a criterion on the SNR of the measured spectra to determine if SPICES or an analog space mission could probe several planetary properties {by} disentangling their spectra (clouds and metallicity for giant planets, cloud and surface coverage for telluric planets). The results are very encouraging since SPICES could characterize the metallicity of Jupiters and Neptunes up to 10~pc and 6~pc respectively for a G2 star. The cloud and surface coverage of a super-Earth orbiting a Sun analog would also be characterized for systems closer than 5~pc. For A0 stars, the instrument {could} study the metallicity of cloud-free Jupiters closer than 17~pc. For M0 stars, cloudy Jupiters and Neptunes {would} be accessible at distances $\leq$7.5~pc and $\leq$3~pc respectively, and super-Earths at distances $\leq$3~pc. These results would give a potential target list of $\sim$300 stars. We also estimated the minimum radius for the planets SPICES could detect.
	
	We emphasize that our study is also useful for other space coronagraph studies currently under development (Sect.~\ref{intro}). Indeed, as far as the authors of this paper are aware of, no study has investigated in detail the spectrophotometric abilities of such missions to retrieve physical parameters of realistic planet spectra. An important point we noted during our study is the need for grids of planetary spectra{, as it has been done} for NIR planet-finders on 8--10~m telescopes \citep[e.g.,][]{Chabrier2000, Baraffe2003, Burrows2006, Fortney2008}. For self-luminous planets, the problem is a bit simplified with respect to mature planets{, because} the emergent spectrum is independent of the separation from the star and the stellar irradiation for separations $\gtrsim$1~AU. Reflected {spectral} models are usually limited to solar-type stars and are derived for a few points of the planet parameter space (mass, separation, metallicity, age). To thoroughly estimate instrument performance{, we} will need spectra for large ranges of planet types (gas and ice giants{, and} super-Earths), separations, stellar types, etc.

	In this paper, we also consider areas for future work that will refine our results. We will include Fresnel propagation in our instrument model to study the impact of out-of pupil aberrations on the performance. Phase and amplitude aberrations will evolve with the wavelength. Their calibration will be as good as they currently are{, because} our focal plane wavefront sensor can estimate both phase and amplitude aberrations in each channel of the IFS independently. The DM correction may be slightly degraded and an optimization of the design may be required to reach a contrast of 10$^{-10}$. We also plan to study the polarimetric performance of SPICES{, as} we expect that the combination of flux and polarization measurements can remove degeneracies that arise when retrieving planet properties from flux measurements alone. Circumstellar disks are another science case to study (dust distribution, rings, planet gaps). Finally, another point of great interest would be to apply the methods we used here to a larger telescope like the Terrestrial Planet Finder Coronagraph \citep{Levine2009}. {Indeed, one of the main results of our work is that the optimal targets of SPICES would be at separations of $\sim$0.8--2~AU, because of the drastic decrease of the reflected flux. For these cases, the performance is limited by the angular resolution of the telescope and not the photon noise. The choice of the telescope diameter was anterior to this study and resulted from a trade-off between the science objectives and the budget allocated to a medium-class mission. We now consider to submit a larger (2.5--3~m) instrument as a large-class mission when a call for proposals will be issued. Besides these considerations, such a telescope} will access a larger volume of target stars and will be less sensitive to zodiacal and exo-zodiacal contributions for planet spectra measurement \citep{Traub2010}.

\begin{acknowledgements}
   The authors wish to thank the SPICES team members for their work in the definition of the science cases and the instrument concept.
   {We also thank the referee for his/her constructive comments on the manuscript.}
   This research has made use of the SIMBAD database, operated at CDS, Strasbourg, France.
   A.-L. M. is supported through a doctoral fellowship from the Minist\`ere de l'\'Education Nationale, de la Recherche et de la Technologie.
\end{acknowledgements}

\bibliographystyle{aa}
\bibliography{biblio.bib}

\begin{thebibliography}{88}
\expandafter\ifx\csname natexlab\endcsname\relax\def\natexlab#1{#1}\fi

\bibitem[{{Alibert} {et~al.}(2011){Alibert}, {Mordasini}, \&
  {Benz}}]{Alibert2011}
{Alibert}, Y., {Mordasini}, C., \& {Benz}, W. 2011, A\&A, 526, A63+

\bibitem[{{Antichi} {et~al.}(2009){Antichi}, {Dohlen}, {Gratton}, {Mesa},
  {Claudi}, {Giro}, {Boccaletti}, {Mouillet}, {Puget}, \&
  {Beuzit}}]{Antichi2009}
{Antichi}, J., {Dohlen}, K., {Gratton}, R.~G., {et~al.} 2009, ApJ, 695, 1042

\bibitem[{{Arnold} {et~al.}(2002){Arnold}, {Gillet}, {Lardi{\`e}re}, {Riaud},
  \& {Schneider}}]{Arnold2002}
{Arnold}, L., {Gillet}, S., {Lardi{\`e}re}, O., {Riaud}, P., \& {Schneider}, J.
  2002, A\&A, 392, 231

\bibitem[{{Baraffe} {et~al.}(2008){Baraffe}, {Chabrier}, \&
  {Barman}}]{Baraffe2008}
{Baraffe}, I., {Chabrier}, G., \& {Barman}, T. 2008, A\&A, 482, 315

\bibitem[{{Baraffe} {et~al.}(2003){Baraffe}, {Chabrier}, {Barman}, {Allard}, \&
  {Hauschildt}}]{Baraffe2003}
{Baraffe}, I., {Chabrier}, G., {Barman}, T.~S., {Allard}, F., \& {Hauschildt},
  P.~H. 2003, A\&A, 402, 701

\bibitem[{{Batalha} {et~al.}(2011){Batalha}, {Borucki}, {Bryson}, {Buchhave},
  {Caldwell}, {Christensen-Dalsgaard}, {Ciardi}, {Dunham}, {Fressin},
  {Gautier}, {Gilliland}, {Haas}, {Howell}, {Jenkins}, {Kjeldsen}, {Koch},
  {Latham}, {Lissauer}, {Marcy}, {Rowe}, {Sasselov}, {Seager}, {Steffen},
  {Torres}, {Basri}, {Brown}, {Charbonneau}, {Christiansen}, {Clarke},
  {Cochran}, {Dupree}, {Fabrycky}, {Fischer}, {Ford}, {Fortney}, {Girouard},
  {Holman}, {Johnson}, {Isaacson}, {Klaus}, {Machalek}, {Moorehead},
  {Morehead}, {Ragozzine}, {Tenenbaum}, {Twicken}, {Quinn}, {VanCleve},
  {Walkowicz}, {Welsh}, {Devore}, \& {Gould}}]{Batalha2011}
{Batalha}, N.~M., {Borucki}, W.~J., {Bryson}, S.~T., {et~al.} 2011, ApJ, 729,
  27

\bibitem[{{Baudoz} {et~al.}(2006){Baudoz}, {Boccaletti}, {Baudrand}, \&
  {Rouan}}]{Baudoz2006}
{Baudoz}, P., {Boccaletti}, A., {Baudrand}, J., \& {Rouan}, D. 2006, in IAU
  Colloq., Vol. 200, Direct Imaging of Exoplanets: Science \& Techniques, ed.
  C.~{Aime} \& F.~{Vakili}, 553

\bibitem[{{Belikov} {et~al.}(2010){Belikov}, {Pluzhnik}, {Connelley},
  {Witteborn}, {Greene}, {Lynch}, {Zell}, \& {Guyon}}]{Belikov2010}
{Belikov}, R., {Pluzhnik}, E., {Connelley}, M.~S., {et~al.} 2010, in Society of
  Photo-Optical Instrumentation Engineers (SPIE) Conf. Series, Vol. 7731,
  77312D

\bibitem[{{Beuzit} {et~al.}(2008){Beuzit}, {Feldt}, {Dohlen}, {Mouillet},
  {Puget}, {Wildi}, {Abe}, {Antichi}, {Baruffolo}, {Baudoz}, {Boccaletti},
  {Carbillet}, {Charton}, {Claudi}, {Downing}, {Fabron}, {Feautrier},
  {Fedrigo}, {Fusco}, {Gach}, {Gratton}, {Henning}, {Hubin}, {Joos}, {Kasper},
  {Langlois}, {Lenzen}, {Moutou}, {Pavlov}, {Petit}, {Pragt}, {Rabou}, {Rigal},
  {Roelfsema}, {Rousset}, {Saisse}, {Schmid}, {Stadler}, {Thalmann}, {Turatto},
  {Udry}, {Vakili}, \& {Waters}}]{Beuzit2008}
{Beuzit}, J.-L., {Feldt}, M., {Dohlen}, K., {et~al.} 2008, in SPIE Conf.
  Series, Vol. 7014, 701418

\bibitem[{{Boccaletti} {et~al.}(2012){Boccaletti}, {Schneider}, {Traub},
  {Lagage}, {Stam}, {Gratton}, {Trauger}, {Cahoy}, {Snik}, {Galicher}, \&
  {Reess}}]{Boccaletti2012a}
{Boccaletti}, A., {Schneider}, J., {Traub}, W.~A., {et~al.} 2012, Exp. Astron.,
  [arXiv::1203.0507]

\bibitem[{{Bord{\'e}} \& {Traub}(2006)}]{Borde2006}
{Bord{\'e}}, P.~J. \& {Traub}, W.~A. 2006, ApJ, 638, 488

\bibitem[{{Borucki} {et~al.}(2012){Borucki}, {Koch}, {Batalha}, {Bryson},
  {Rowe}, {Fressin}, {Torres}, {Caldwell}, {Christensen-Dalsgaard}, {Cochran},
  {DeVore}, {Gautier}, {Geary}, {Gilliland}, {Gould}, {Howell}, {Jenkins},
  {Latham}, {Lissauer}, {Marcy}, {Sasselov}, {Boss}, {Charbonneau}, {Ciardi},
  {Kaltenegger}, {Doyle}, {Dupree}, {Ford}, {Fortney}, {Holman}, {Steffen},
  {Mullally}, {Still}, {Tarter}, {Ballard}, {Buchhave}, {Carter},
  {Christiansen}, {Demory}, {D{\'e}sert}, {Dressing}, {Endl}, {Fabrycky},
  {Fischer}, {Haas}, {Henze}, {Horch}, {Howard}, {Isaacson}, {Kjeldsen},
  {Johnson}, {Klaus}, {Kolodziejczak}, {Barclay}, {Li}, {Meibom}, {Prsa},
  {Quinn}, {Quintana}, {Robertson}, {Sherry}, {Shporer}, {Tenenbaum},
  {Thompson}, {Twicken}, {Van Cleve}, {Welsh}, {Basu}, {Chaplin}, {Miglio},
  {Kawaler}, {Arentoft}, {Stello}, {Metcalfe}, {Verner}, {Karoff}, {Lundkvist},
  {Lund}, {Handberg}, {Elsworth}, {Hekker}, {Huber}, {Bedding}, \&
  {Rapin}}]{Borucki2012}
{Borucki}, W.~J., {Koch}, D.~G., {Batalha}, N., {et~al.} 2012, ApJ, 745, 120

\bibitem[{{Boss}(2011)}]{Boss2011}
{Boss}, A.~P. 2011, ApJ, 731, 74

\bibitem[{{Bowler} {et~al.}(2010){Bowler}, {Liu}, {Dupuy}, \&
  {Cushing}}]{Bowler2010}
{Bowler}, B.~P., {Liu}, M.~C., {Dupuy}, T.~J., \& {Cushing}, M.~C. 2010, ApJ,
  723, 850

\bibitem[{{Burrows} {et~al.}(2006){Burrows}, {Sudarsky}, \&
  {Hubeny}}]{Burrows2006}
{Burrows}, A., {Sudarsky}, D., \& {Hubeny}, I. 2006, ApJ, 640, 1063

\bibitem[{{Cahoy} {et~al.}(2009){Cahoy}, {Guyon}, {Schneider}, {Marley},
  {Belikov}, {Meyer}, {Ridgway}, {Traub}, \& {Woolf}}]{Cahoy2009}
{Cahoy}, K., {Guyon}, O., {Schneider}, G., {et~al.} 2009, in SPIE Conf. Series,
  Vol. 7440, 74400G

\bibitem[{{Cahoy} {et~al.}(2010){Cahoy}, {Marley}, \& {Fortney}}]{Cahoy2010}
{Cahoy}, K.~L., {Marley}, M.~S., \& {Fortney}, J.~J. 2010, ApJ, 724, 189

\bibitem[{{Casertano} {et~al.}(2008){Casertano}, {Lattanzi}, {Sozzetti},
  {Spagna}, {Jancart}, {Morbidelli}, {Pannunzio}, {Pourbaix}, \&
  {Queloz}}]{Casertano2008}
{Casertano}, S., {Lattanzi}, M.~G., {Sozzetti}, A., {et~al.} 2008, A\&A, 482,
  699

\bibitem[{{Chabrier} {et~al.}(2000){Chabrier}, {Baraffe}, {Allard}, \&
  {Hauschildt}}]{Chabrier2000}
{Chabrier}, G., {Baraffe}, I., {Allard}, F., \& {Hauschildt}, P. 2000, ApJ,
  542, 464

\bibitem[{{Charbonneau} {et~al.}(2009){Charbonneau}, {Berta}, {Irwin}, {Burke},
  {Nutzman}, {Buchhave}, {Lovis}, {Bonfils}, {Latham}, {Udry}, {Murray-Clay},
  {Holman}, {Falco}, {Winn}, {Queloz}, {Pepe}, {Mayor}, {Delfosse}, \&
  {Forveille}}]{Charbonneau2009}
{Charbonneau}, D., {Berta}, Z.~K., {Irwin}, J., {et~al.} 2009, Nature, 462, 891

\bibitem[{{Chauvin} {et~al.}(2005){Chauvin}, {Lagrange}, {Dumas}, {Zuckerman},
  {Mouillet}, {Song}, {Beuzit}, \& {Lowrance}}]{Chauvin2005}
{Chauvin}, G., {Lagrange}, A.-M., {Dumas}, C., {et~al.} 2005, A\&A, 438, L25

\bibitem[{{Clampin}(2010)}]{Clampin2010}
{Clampin}, M. 2010, in ASP Conf. Series, Vol. 430, Pathways Towards Habitable
  Planets, ed. V.~{Coud{\'e} du Foresto}, D.~M. {Gelino}, \& I.~{Ribas}, 167--+

\bibitem[{{Coud{\'e} du Foresto} {et~al.}(2010){Coud{\'e} du Foresto}, {Absil},
  {Beaulieu}, {Beichman}, {Boccaletti}, {Chakraborty}, \& {et
  al.}}]{Coud'eduForesto2010}
{Coud{\'e} du Foresto}, V., {Absil}, O., {Beaulieu}, J.-P., {et~al.} 2010, Blue
  Dots report, http://www.blue-dots.net

\bibitem[{{Des Marais} {et~al.}(2002){Des Marais}, {Harwit}, {Jucks},
  {Kasting}, {Lin}, {Lunine}, {Schneider}, {Seager}, {Traub}, \&
  {Woolf}}]{DesMarais2002}
{Des Marais}, D.~J., {Harwit}, M.~O., {Jucks}, K.~W., {et~al.} 2002,
  Astrobiology, 2, 153

\bibitem[{{Enya} {et~al.}(2011){Enya}, {Kotani}, {Haze}, {Aono}, {Nakagawa},
  {Matsuhara}, {Kataza}, {Wada}, {Kawada}, {Fujiwara}, {Mita}, {Takeuchi},
  {Komatsu}, {Sakai}, {Uchida}, {Mitani}, {Yamawaki}, {Miyata}, {Sako},
  {Nakamura}, {Asano}, {Yamashita}, {Narita}, {Matsuo}, {Tamura}, {Nishikawa},
  {Kokubo}, {Hayano}, {Oya}, {Fukagawa}, {Shibai}, {Baba}, {Murakami}, {Itoh},
  {Honda}, {Okamoto}, {Ida}, {Takami}, {Abe}, {Guyon}, {Bierden}, \&
  {Yamamuro}}]{Enya2011}
{Enya}, K., {Kotani}, T., {Haze}, K., {et~al.} 2011, AdSpR, 48, 323

\bibitem[{{Esposito} {et~al.}(2010){Esposito}, {Riccardi}, {Fini}, {Puglisi},
  {Pinna}, {Xompero}, {Briguglio}, {Quir{\'o}s-Pacheco}, {Stefanini}, {Guerra},
  {Busoni}, {Tozzi}, {Pieralli}, {Agapito}, {Brusa-Zappellini}, {Demers},
  {Brynnel}, {Arcidiacono}, \& {Salinari}}]{Esposito2010}
{Esposito}, S., {Riccardi}, A., {Fini}, L., {et~al.} 2010, in SPIE Conf.
  Series, Vol. 7736, 773609

\bibitem[{{Fortney} {et~al.}(2010){Fortney}, {Baraffe}, \&
  {Militzer}}]{Fortney2010}
{Fortney}, J.~J., {Baraffe}, I., \& {Militzer}, B. 2010, Exoplanets, ed.
  S.~Seager (Tucson, AZ: University of Arizona Press), 397--418,
  [arXiv:0911.3154]

\bibitem[{{Fortney} {et~al.}(2007){Fortney}, {Marley}, \&
  {Barnes}}]{Fortney2007}
{Fortney}, J.~J., {Marley}, M.~S., \& {Barnes}, J.~W. 2007, ApJ, 659, 1661

\bibitem[{{Fortney} {et~al.}(2008){Fortney}, {Marley}, {Saumon}, \&
  {Lodders}}]{Fortney2008}
{Fortney}, J.~J., {Marley}, M.~S., {Saumon}, D., \& {Lodders}, K. 2008, ApJ,
  683, 1104

\bibitem[{{Galicher}(2009)}]{Galicher2009}
{Galicher}, R. 2009, PhD thesis, University Denis Diderot Paris 7

\bibitem[{{Galicher} {et~al.}(2008){Galicher}, {Baudoz}, \&
  {Rousset}}]{Galicher2008}
{Galicher}, R., {Baudoz}, P., \& {Rousset}, G. 2008, A\&A, 488, L9

\bibitem[{{Galicher} {et~al.}(2010){Galicher}, {Baudoz}, {Rousset}, {Totems},
  \& {Mas}}]{Galicher2010}
{Galicher}, R., {Baudoz}, P., {Rousset}, G., {Totems}, J., \& {Mas}, M. 2010,
  A\&A, 509, A31+

\bibitem[{{Giavalisco} {et~al.}(2002){Giavalisco}, {Sahu}, \&
  {Bohlin}}]{Giavalisco2002}
{Giavalisco}, M., {Sahu}, K., \& {Bohlin}, R.~C. 2002, New Estimates of the Sky
  Background for the HST Exposure Time Calculator, STScI Instrument Science
  Report WFC3-ISR 2002 - 02

\bibitem[{{Grasset} {et~al.}(2009){Grasset}, {Schneider}, \&
  {Sotin}}]{Grasset2009}
{Grasset}, O., {Schneider}, J., \& {Sotin}, C. 2009, ApJ, 693, 722

\bibitem[{{Guyon} {et~al.}(2010a){Guyon}, {Pluzhnik}, {Martinache}, {Totems},
  {Tanaka}, {Matsuo}, {Blain}, \& {Belikov}}]{Guyon2010a}
{Guyon}, O., {Pluzhnik}, E., {Martinache}, F., {et~al.} 2010a, PASP, 122, 71

\bibitem[{{Guyon} {et~al.}(2010b){Guyon}, {Shaklan}, {Levine}, {Cahoy},
  {Tenerelli}, {Belikov}, \& {Kern}}]{Guyon2010b}
{Guyon}, O., {Shaklan}, S., {Levine}, M., {et~al.} 2010b, in SPIE Conf. Series,
  Vol. 7731, 773129

\bibitem[{{Hatzes} {et~al.}(2010){Hatzes}, {Boccaletti}, {Dvorak}, {Micela},
  {Morbidelli}, {Quirrenbach}, {Rauer}, {Selsis}, {Tinetti}, \&
  {Udry}}]{Hatzes2010}
{Hatzes}, A., {Boccaletti}, A., {Dvorak}, R., {et~al.} 2010, Exoplanet Research
  Advisory Team Report, http://sci.esa.int/eprat

\bibitem[{{Hinkley} {et~al.}(2011){Hinkley}, {Oppenheimer}, {Zimmerman},
  {Brenner}, {Parry}, {Crepp}, {Vasisht}, {Ligon}, {King}, {Soummer},
  {Sivaramakrishnan}, {Beichman}, {Shao}, {Roberts}, {Bouchez}, {Dekany},
  {Pueyo}, {Roberts}, {Lockhart}, {Zhai}, {Shelton}, \&
  {Burruss}}]{Hinkley2011}
{Hinkley}, S., {Oppenheimer}, B.~R., {Zimmerman}, N., {et~al.} 2011, PASP, 123,
  74

\bibitem[{{Hodapp} {et~al.}(2008){Hodapp}, {Suzuki}, {Tamura}, {Abe}, {Suto},
  {Kandori}, {Morino}, {Nishimura}, {Takami}, {Guyon}, {Jacobson},
  {Stahlberger}, {Yamada}, {Shelton}, {Hashimoto}, {Tavrov}, {Nishikawa},
  {Ukita}, {Izumiura}, {Hayashi}, {Nakajima}, {Yamada}, \&
  {Usuda}}]{Hodapp2008}
{Hodapp}, K.~W., {Suzuki}, R., {Tamura}, M., {et~al.} 2008, in SPIE Conf.
  Series, Vol. 7014, 701419

\bibitem[{{Huang} {et~al.}(2008){Huang}, {Rao}, \& {Jiang}}]{Huang2008}
{Huang}, L., {Rao}, C., \& {Jiang}, W. 2008, Opt. Express, 16, 108

\bibitem[{{Janson} {et~al.}(2010){Janson}, {Bergfors}, {Goto}, {Brandner}, \&
  {Lafreni{\`e}re}}]{Janson2010}
{Janson}, M., {Bergfors}, C., {Goto}, M., {Brandner}, W., \& {Lafreni{\`e}re},
  D. 2010, ApJL, 710, L35

\bibitem[{{Janson} {et~al.}(2012){Janson}, {Carson}, {Lafreni{\`e}re},
  {Spiegel}, {Bent}, \& {Wong}}]{Janson2012}
{Janson}, M., {Carson}, J.~C., {Lafreni{\`e}re}, D., {et~al.} 2012, ApJ, 747,
  116

\bibitem[{{Kalas} {et~al.}(2008){Kalas}, {Graham}, {Chiang}, {Fitzgerald},
  {Clampin}, {Kite}, {Stapelfeldt}, {Marois}, \& {Krist}}]{Kalas2008}
{Kalas}, P., {Graham}, J.~R., {Chiang}, E., {et~al.} 2008, Science, 322, 1345

\bibitem[{{Kasper} {et~al.}(2010){Kasper}, {Beuzit}, {Verinaud}, {Gratton},
  {Kerber}, {Yaitskova}, {Boccaletti}, {Thatte}, {Schmid}, {Keller}, {Baudoz},
  {Abe}, {Aller-Carpentier}, {Antichi}, {Bonavita}, {Dohlen}, {Fedrigo},
  {Hanenburg}, {Hubin}, {Jager}, {Korkiakoski}, {Martinez}, {Mesa}, {Preis},
  {Rabou}, {Roelfsema}, {Salter}, {Tecza}, \& {Venema}}]{Kasper2010}
{Kasper}, M., {Beuzit}, J.-L., {Verinaud}, C., {et~al.} 2010, in SPIE Conf.
  Series, Vol. 7735, 77352E

\bibitem[{{Kiang} {et~al.}(2007){Kiang}, {Segura}, {Tinetti}, {Govindjee},
  {Blankenship}, {Cohen}, {Siefert}, {Crisp}, \& {Meadows}}]{Kiang2007}
{Kiang}, N.~Y., {Segura}, A., {Tinetti}, G., {et~al.} 2007, Astrobiology, 7,
  252

\bibitem[{{Krist}(2007)}]{Krist2007}
{Krist}, J.~E. 2007, in SPIE Conf. Series, Vol. 6675, 66750P

\bibitem[{{Kuchner}(2004)}]{Kuchner2004}
{Kuchner}, M.~J. 2004, ApJ, 612, 1147

\bibitem[{{Lagrange} {et~al.}(2010){Lagrange}, {Bonnefoy}, {Chauvin}, {Apai},
  {Ehrenreich}, {Boccaletti}, {Gratadour}, {Rouan}, {Mouillet}, {Lacour}, \&
  {Kasper}}]{Lagrange2010}
{Lagrange}, A.-M., {Bonnefoy}, M., {Chauvin}, G., {et~al.} 2010, Science, 329,
  57

\bibitem[{{Lagrange} {et~al.}(2009){Lagrange}, {Gratadour}, {Chauvin}, {Fusco},
  {Ehrenreich}, {Mouillet}, {Rousset}, {Rouan}, {Allard}, {Gendron}, {Charton},
  {Mugnier}, {Rabou}, {Montri}, \& {Lacombe}}]{Lagrange2009}
{Lagrange}, A.-M., {Gratadour}, D., {Chauvin}, G., {et~al.} 2009, A\&A, 493,
  L21

\bibitem[{{L{\'e}ger} {et~al.}(2009){L{\'e}ger}, {Rouan}, {Schneider}, {Barge},
  {Fridlund}, {Samuel}, {Ollivier}, {Guenther}, {Deleuil}, {Deeg}, {Auvergne},
  {Alonso}, {Aigrain}, {Alapini}, {Almenara}, {Baglin}, {Barbieri}, {Bruntt},
  {Bord{\'e}}, {Bouchy}, {Cabrera}, {Catala}, {Carone}, {Carpano}, {Csizmadia},
  {Dvorak}, {Erikson}, {Ferraz-Mello}, {Foing}, {Fressin}, {Gandolfi},
  {Gillon}, {Gondoin}, {Grasset}, {Guillot}, {Hatzes}, {H{\'e}brard}, {Jorda},
  {Lammer}, {Llebaria}, {Loeillet}, {Mayor}, {Mazeh}, {Moutou}, {P{\"a}tzold},
  {Pont}, {Queloz}, {Rauer}, {Renner}, {Samadi}, {Shporer}, {Sotin}, {Tingley},
  {Wuchterl}, {Adda}, {Agogu}, {Appourchaux}, {Ballans}, {Baron}, {Beaufort},
  {Bellenger}, {Berlin}, {Bernardi}, {Blouin}, {Baudin}, {Bodin}, {Boisnard},
  {Boit}, {Bonneau}, {Borzeix}, {Briet}, {Buey}, {Butler}, {Cailleau},
  {Cautain}, {Chabaud}, {Chaintreuil}, {Chiavassa}, {Costes}, {Cuna Parrho},
  {de Oliveira Fialho}, {Decaudin}, {Defise}, {Djalal}, {Epstein}, {Exil},
  {Faur{\'e}}, {Fenouillet}, {Gaboriaud}, {Gallic}, {Gamet}, {Gavalda},
  {Grolleau}, {Gruneisen}, {Gueguen}, {Guis}, {Guivarc'h}, {Guterman},
  {Hallouard}, {Hasiba}, {Heuripeau}, {Huntzinger}, {Hustaix}, {Imad},
  {Imbert}, {Johlander}, {Jouret}, {Journoud}, {Karioty}, {Kerjean},
  {Lafaille}, {Lafond}, {Lam-Trong}, {Landiech}, {Lapeyrere}, {Larqu{\'e}},
  {Laudet}, {Lautier}, {Lecann}, {Lefevre}, {Leruyet}, {Levacher}, {Magnan},
  {Mazy}, {Mertens}, {Mesnager}, {Meunier}, {Michel}, {Monjoin}, {Naudet},
  {Nguyen-Kim}, {Orcesi}, {Ottacher}, {Perez}, {Peter}, {Plasson}, {Plesseria},
  {Pontet}, {Pradines}, {Quentin}, {Reynaud}, {Rolland}, {Rollenhagen},
  {Romagnan}, {Russ}, {Schmidt}, {Schwartz}, {Sebbag}, {Sedes}, {Smit},
  {Steller}, {Sunter}, {Surace}, {Tello}, {Tiph{\`e}ne}, {Toulouse}, {Ulmer},
  {Vandermarcq}, {Vergnault}, {Vuillemin}, \& {Zanatta}}]{L'eger2009}
{L{\'e}ger}, A., {Rouan}, D., {Schneider}, J., {et~al.} 2009, A\&A, 506, 287

\bibitem[{{Levine} {et~al.}(2009){Levine}, {Lisman}, {Shaklan}, {Kasting},
  {Traub}, {Alexander}, {Angel}, {Blaurock}, {Brown}, {Brown}, {Burrows},
  {Clampin}, {Cohen}, {Content}, {Dewell}, {Dumont}, {Egerman}, {Ferguson},
  {Ford}, {Greene}, {Guyon}, {Hammel}, {Heap}, {Ho}, {Horner}, {Hunyadi},
  {Irish}, {Jackson}, {Kasdin}, {Kissil}, {Krim}, {Kuchner}, {Kwack}, {Lillie},
  {Lin}, {Liu}, {Marchen}, {Marley}, {Meadows}, {Mosier}, {Mouroulis},
  {Noecker}, {Ohl}, {Oppenheimer}, {Pitman}, {Ridgway}, {Sabatke}, {Seager},
  {Shao}, {Smith}, {Soummer}, {Stapelfeldt}, {Tenerell}, {Trauger}, \&
  {Vanderbei}}]{Levine2009}
{Levine}, M., {Lisman}, D., {Shaklan}, S., {et~al.} 2009, TPF-C Flight Baseline
  Concept, [arXiv:0911.3200]

\bibitem[{{Lunine} {et~al.}(2008){Lunine}, {Fischer}, {Hammel}, {Henning},
  {Hillenbrand}, {Kasting}, {Laughlin}, {Macintosh}, {Marley}, {Melnick},
  {Monet}, {Noecker}, {Peale}, {Quirrenbach}, {Seager}, \& {Winn}}]{Lunine2008}
{Lunine}, J.~I., {Fischer}, D., {Hammel}, H., {et~al.} 2008, Worlds Beyond: A
  Strategy for the Detection and Characterization of Exoplanets,
  [arXiv:0808.2754]

\bibitem[{{Macintosh} {et~al.}(2006){Macintosh}, {Troy}, {Doyon}, {Graham},
  {Baker}, {Bauman}, {Marois}, {Palmer}, {Phillion}, {Poyneer}, {Crossfield},
  {Dumont}, {Levine}, {Shao}, {Serabyn}, {Shelton}, {Vasisht}, {Wallace},
  {Lavigne}, {Valee}, {Rowlands}, {Tam}, \& {Hackett}}]{Macintosh2006a}
{Macintosh}, B., {Troy}, M., {Doyon}, R., {et~al.} 2006, in SPIE Conf. Series,
  Vol. 6272, 62720N

\bibitem[{{Macintosh} {et~al.}(2008){Macintosh}, {Graham}, {Palmer}, {Doyon},
  {Dunn}, {Gavel}, {Larkin}, {Oppenheimer}, {Saddlemyer}, {Sivaramakrishnan},
  {Wallace}, {Bauman}, {Erickson}, {Marois}, {Poyneer}, \&
  {Soummer}}]{Macintosh2008}
{Macintosh}, B.~A., {Graham}, J.~R., {Palmer}, D.~W., {et~al.} 2008, in SPIE
  Conf. Series, Vol. 7015, 701518

\bibitem[{{Marcy} {et~al.}(2005){Marcy}, {Butler}, {Fischer}, {Vogt}, {Wright},
  {Tinney}, \& {Jones}}]{Marcy2005}
{Marcy}, G., {Butler}, R.~P., {Fischer}, D., {et~al.} 2005, PThPS, 158, 24

\bibitem[{{Marley} {et~al.}(1999){Marley}, {Gelino}, {Stephens}, {Lunine}, \&
  {Freedman}}]{Marley1999}
{Marley}, M.~S., {Gelino}, C., {Stephens}, D., {Lunine}, J.~I., \& {Freedman},
  R. 1999, ApJ, 513, 879

\bibitem[{Marois {et~al.}(2008)Marois, Macintosh, Barman, Zuckerman, Song,
  Patience, Lafreni{\`e}re, \& Doyon}]{Marois2008}
Marois, C., Macintosh, B., Barman, T., {et~al.} 2008, Science, 322, 1348

\bibitem[{Marois {et~al.}(2006)Marois, Phillion, \& Macintosh}]{Marois2006a}
Marois, C., Phillion, D.~W., \& Macintosh, B. 2006, in SPIE Conf. Series, Vol.
  6269, 62693M

\bibitem[{{Marois} {et~al.}(2010){Marois}, {Zuckerman}, {Konopacky},
  {Macintosh}, \& {Barman}}]{Marois2010}
{Marois}, C., {Zuckerman}, B., {Konopacky}, Q.~M., {Macintosh}, B., \&
  {Barman}, T. 2010, Nature, 468, 1080

\bibitem[{{Mawet} {et~al.}(2010){Mawet}, {Pueyo}, {Moody}, {Krist}, \&
  {Serabyn}}]{Mawet2010a}
{Mawet}, D., {Pueyo}, L., {Moody}, D., {Krist}, J., \& {Serabyn}, E. 2010, in
  SPIE Conf. Series, Vol. 7739, 773914

\bibitem[{{Mawet} {et~al.}(2005){Mawet}, {Riaud}, {Absil}, \&
  {Surdej}}]{Mawet2005}
{Mawet}, D., {Riaud}, P., {Absil}, O., \& {Surdej}, J. 2005, ApJ, 633, 1191

\bibitem[{{Mayor} {et~al.}(2011){Mayor}, {Marmier}, {Lovis}, {Udry},
  {S{\'e}gransan}, {Pepe}, {Benz}, {Bertaux}, {Bouchy}, {Dumusque}, {Lo Curto},
  {Mordasini}, {Queloz}, \& {Santos}}]{Mayor2011}
{Mayor}, M., {Marmier}, M., {Lovis}, C., {et~al.} 2011, A\&A, submitted

\bibitem[{{Mayor} \& {Queloz}(1995)}]{Mayor1995}
{Mayor}, M. \& {Queloz}, D. 1995, Nature, 378, 355

\bibitem[{{Mohanty} {et~al.}(2007){Mohanty}, {Jayawardhana}, {Hu{\'e}lamo}, \&
  {Mamajek}}]{Mohanty2007}
{Mohanty}, S., {Jayawardhana}, R., {Hu{\'e}lamo}, N., \& {Mamajek}, E. 2007,
  ApJ, 657, 1064

\bibitem[{{Monta{\~n}{\'e}s-Rodriguez}
  {et~al.}(2005){Monta{\~n}{\'e}s-Rodriguez}, {Pall{\'e}}, {Goode}, {Hickey},
  \& {Koonin}}]{Montan'es-Rodriguez2005}
{Monta{\~n}{\'e}s-Rodriguez}, P., {Pall{\'e}}, E., {Goode}, P.~R., {Hickey},
  J., \& {Koonin}, S.~E. 2005, ApJ, 629, 1175

\bibitem[{{Neuh{\"a}user} {et~al.}(2005){Neuh{\"a}user}, {Guenther},
  {Wuchterl}, {Mugrauer}, {Bedalov}, \& {Hauschildt}}]{Neuhauser2005}
{Neuh{\"a}user}, R., {Guenther}, E.~W., {Wuchterl}, G., {et~al.} 2005, A\&A,
  435, L13

\bibitem[{{Patience} {et~al.}(2010){Patience}, {King}, {de Rosa}, \&
  {Marois}}]{Patience2010}
{Patience}, J., {King}, R.~R., {de Rosa}, R.~J., \& {Marois}, C. 2010, A\&A,
  517, A76+

\bibitem[{{Robberto}(2009)}]{Robberto2009}
{Robberto}, M. 2009, NIRCAM Optimal Readout Modes, Tech. Rep. JWST-STScI-001721

\bibitem[{{Schneider} {et~al.}(2008){Schneider}, {Boccaletti}, {Aylward},
  {Baudoz}, {Beuzit}, {Brown}, {Cho}, {Dohlen}, {Ferrari}, {Galicher},
  {Grasset}, {Grenfell}, {Griessmeier}, {Guyon}, {Hough}, {Kasper}, {Keller},
  {Longmore}, {Lopez}, {Martin}, {Mawet}, {Menard}, {Merin}, {Palle}, {Perrin},
  {Pinfield}, {Sein}, {Shore}, {Sotin}, {Sozzetti}, {Stam}, {Surdej},
  {Tamburini}, {Tinetti}, {Udry}, {Verinaud}, \& {Walker}}]{Schneider2008}
{Schneider}, J., {Boccaletti}, A., {Aylward}, A., {et~al.} 2008, ArXiv
  e-prints, [arXiv:0811.2496]

\bibitem[{Schneider {et~al.}(2009)Schneider, Boccaletti, Mawet, Baudoz, Beuzit,
  Doyon, Marley, Stam, Tinetti, Traub, Trauger, Aylward, Cho, Keller, Udry, \&
  the See-coast Team}]{Schneider2009}
Schneider, J., Boccaletti, A., Mawet, D., {et~al.} 2009, Exp. Astron., 23, 357

\bibitem[{{Schneider} {et~al.}(2011){Schneider}, {Dedieu}, {Le Sidaner},
  {Savalle}, \& {Zolotukhin}}]{Schneider2011}
{Schneider}, J., {Dedieu}, C., {Le Sidaner}, P., {Savalle}, R., \&
  {Zolotukhin}, I. 2011, A\&A, 532, A79+

\bibitem[{{Seager} \& {Deming}(2010)}]{Seager2010}
{Seager}, S. \& {Deming}, D. 2010, ARA\&A, 48, 631

\bibitem[{{Seager} {et~al.}(2005){Seager}, {Turner}, {Schafer}, \&
  {Ford}}]{Seager2005}
{Seager}, S., {Turner}, E.~L., {Schafer}, J., \& {Ford}, E.~B. 2005,
  Astrobiology, 5, 372

\bibitem[{{Serabyn} {et~al.}(2010){Serabyn}, {Mawet}, \&
  {Burruss}}]{Serabyn2010}
{Serabyn}, E., {Mawet}, D., \& {Burruss}, R. 2010, Nature, 464, 1018

\bibitem[{{Serabyn} {et~al.}(2011){Serabyn}, {Mawet}, {Wallace}, {Liewer},
  {Trauger}, {Moody}, \& {Kern}}]{Serabyn2011}
{Serabyn}, E., {Mawet}, D., {Wallace}, J.~K., {et~al.} 2011, in SPIE Conf.
  Series, Vol. 8146, 81460L

\bibitem[{{Shaklan} \& {Green}(2006)}]{Shaklan2006}
{Shaklan}, S.~B. \& {Green}, J.~J. 2006, ApOpt, 45, 5143

\bibitem[{{Smith} {et~al.}(2006){Smith}, {Walton}, {Ingley}, {Holland},
  {Cropper}, \& {Pool}}]{Smith2006}
{Smith}, D.~R., {Walton}, D.~M., {Ingley}, R., {et~al.} 2006, in SPIE Conf.
  Series, Vol. 6276, 62760K

\bibitem[{{Soummer} {et~al.}(2007){Soummer}, {Pueyo}, {Sivaramakrishnan}, \&
  {Vanderbei}}]{Soummer2007}
{Soummer}, R., {Pueyo}, L., {Sivaramakrishnan}, A., \& {Vanderbei}, R.~J. 2007,
  Optics Express, 15, 15935

\bibitem[{{Sparks} \& {Ford}(2002)}]{Sparks2002}
{Sparks}, W.~B. \& {Ford}, H.~C. 2002, ApJ, 578, 543

\bibitem[{Stam(2008)}]{Stam2008}
Stam, D.~M. 2008, A\&A, 482, 989

\bibitem[{{Stam} {et~al.}(2004){Stam}, {Hovenier}, \& {Waters}}]{Stam2004}
{Stam}, D.~M., {Hovenier}, J.~W., \& {Waters}, L.~B.~F.~M. 2004, A\&A, 428, 663

\bibitem[{{Traub}(2003)}]{Traub2003}
{Traub}, W.~A. 2003, in ASP Conf. Series, Vol. 294, Scientific Frontiers in
  Research on Extrasolar Planets, ed. D.~{Deming} \& S.~{Seager}, 595--602

\bibitem[{{Traub} \& {Oppenheimer}(2010)}]{Traub2010}
{Traub}, W.~A. \& {Oppenheimer}, B.~R. 2010, Exoplanets, ed. S.~Seager (Tucson,
  AZ: University of Arizona Press), 111--156

\bibitem[{{Trauger} {et~al.}(2010){Trauger}, {Stapelfeldt}, {Traub}, {Krist},
  {Moody}, {Mawet}, {Serabyn}, {Henry}, {Brugarolas}, {Alexander}, {Gappinger},
  {Dawson}, {Mireles}, {Park}, {Pueyo}, {Shaklan}, {Guyon}, {Kasdin},
  {Vanderbei}, {Spergel}, {Belikov}, {Marcy}, {Brown}, {Schneider}, {Woodgate},
  {Egerman}, {Matthews}, {Elias}, {Conturie}, {Vallone}, {Voyer}, {Polidan},
  {Lillie}, {Spittler}, {Lee}, {Hejal}, {Bronowicki}, {Saldivar}, {Ealey}, \&
  {Price}}]{Trauger2010}
{Trauger}, J., {Stapelfeldt}, K., {Traub}, W., {et~al.} 2010, in SPIE Conf.
  Series, Vol. 7731, 773128

\bibitem[{{Trauger} \& {Traub}(2007)}]{Trauger2007}
{Trauger}, J.~T. \& {Traub}, W.~A. 2007, Nature, 446, 771

\bibitem[{{Udry} \& {Santos}(2007)}]{Udry2007}
{Udry}, S. \& {Santos}, N.~C. 2007, ARA\&A, 45, 397

\bibitem[{{Wolstencroft} \& {Raven}(2002)}]{Wolstencroft2002}
{Wolstencroft}, R.~D. \& {Raven}, J.~A. 2002, Icarus, 157, 535

\bibitem[{{Woolf} {et~al.}(2002){Woolf}, {Smith}, {Traub}, \&
  {Jucks}}]{Woolf2002}
{Woolf}, N.~J., {Smith}, P.~S., {Traub}, W.~A., \& {Jucks}, K.~W. 2002, ApJ,
  574, 430

\end{thebibliography}

\end{document}